%% file: main.tex
\pdfoutput=1
\documentclass[sigconf, screen, nonacm]{acmart}

\setcopyright{none}
\usepackage{graphicx}
\usepackage{amsmath}
\usepackage{tikz}
\usepackage{booktabs}
\usepackage{array}
\usetikzlibrary{patterns}

\usepackage{cleveref}
\crefformat{section}{\S#2#1#3}
\crefformat{subsection}{\S#2#1#3}
\crefformat{subsubsection}{\S#2#1#3}

\definecolor{grRed}{HTML}{C8453A}    %
\definecolor{grCoral}{HTML}{D86A3C}  %
\definecolor{grAmber}{HTML}{E3A33B}  %
\definecolor{grSky}{HTML}{4B9CD3}    %
\definecolor{grDeep}{HTML}{2A5CAA}   %
\definecolor{grCyan}{HTML}{20A8C0}   %
\definecolor{grPlum}{HTML}{A6467E}   %
\definecolor{grSlate}{HTML}{5F6B7A}  %

\colorlet{ovlOverGen}{grRed}
\colorlet{ovlSuite}{grDeep}
\colorlet{ovlCluster}{grCyan}

\tikzset{
  ovl/overgenbar/.style={draw=black!65, line width=0.3pt, fill=ovlOverGen!22,
    postaction={draw=none, pattern=north west lines, pattern color=ovlOverGen!85!black}},
  ovl/suitebar/.style={draw=black!65, line width=0.3pt, fill=ovlSuite!26,
    postaction={draw=none, pattern=north east lines, pattern color=ovlSuite!85!black}},
  ovl/clusterbar/.style={draw=black!65, line width=0.3pt, fill=ovlCluster!24,
    postaction={draw=none, pattern=dots, pattern color=ovlCluster!85!black}}
}

\usepackage{xcolor,tabularx}
\usetikzlibrary{arrows.meta,calc,patterns}
\usetikzlibrary{shapes.geometric}

\colorlet{grBlue}{grDeep}
\colorlet{grOrange}{grCoral}
\colorlet{grTeal}{grCyan}
\colorlet{grPurple}{grPlum}

\tikzset{
  grBlock/.style={draw=black!70, fill=white, rounded corners=1.2pt, line width=0.55pt, align=center, inner sep=3pt, outer sep=0pt, font=\sffamily\fontsize{8.5}{10}\selectfont},
  grInput/.style={grBlock, dashed},
  grNative/.style={grBlock, draw=grBlue!85!black, fill=grBlue!8},
  grTyped/.style={grBlock, draw=grPurple!85!black, fill=grPurple!10},
  grMemory/.style={grBlock, draw=grTeal!85!black, fill=grTeal!8},
  grSystem/.style={grBlock, draw=grOrange!75!black, fill=grOrange!10},
  grArrow/.style={draw=black!75, line width=0.6pt, -{Latex[length=1.8mm]}},
  grBoth/.style={draw=black!75, line width=0.6pt, <->, >=Latex},
  grTitle/.style={anchor=west, font=\sffamily\fontsize{9}{11}\selectfont\bfseries},
  grNote/.style={align=center, font=\sffamily\fontsize{8}{9.5}\selectfont}
}

\usepackage{tikz}
\usetikzlibrary{arrows.meta,patterns}

\colorlet{grcBlue}{grDeep}
\colorlet{grcOrange}{grRed}
\colorlet{grcTeal}{grCyan}
\colorlet{grcPurple}{grPlum}
\colorlet{grcGold}{grAmber}
\colorlet{grcSky}{grSky}
\colorlet{grcPink}{grPlum}
\colorlet{grcGrey}{grSlate}
\colorlet{grcRed}{grRed}

\tikzset{
  grcFigure/.style={x=1cm,y=1cm,font=\sffamily\fontsize{8}{9}\selectfont},
  grcTitle/.style={font=\sffamily\fontsize{8}{9}\selectfont\bfseries,inner sep=1pt},
  grcNote/.style={align=center,inner sep=1pt},
  grcNumber/.style={font=\sffamily\fontsize{7}{8}\selectfont,inner sep=0pt},
  grcBlock/.style={draw=black!65,fill=white,rounded corners=1pt,align=center,inner sep=2pt},
  grcInput/.style={grcBlock,dashed},
  grcNative/.style={grcBlock,draw=grcBlue!80!black,fill=grcBlue!8},
  grcMemory/.style={grcBlock,draw=grcTeal!75!black,fill=grcTeal!8},
  grcSystem/.style={grcBlock,draw=grcOrange!80!black,fill=grcOrange!8},
  grcTyped/.style={grcBlock,draw=grcPurple!80!black,fill=grcPurple!9},
  grcArrow/.style={-{Latex[length=1.2mm]},draw=black!75,line width=.6pt},
  grcBoth/.style={{Latex[length=1.2mm]}-{Latex[length=1.2mm]},draw=black!75,line width=.6pt},
  grcSuite/.style={draw=grcBlue!80!black,line width=.25pt,fill=grcBlue!26,postaction={draw=none,pattern=north east lines,pattern color=grcBlue!85!black}},
  grcCluster/.style={draw=grcTeal!75!black,line width=.25pt,fill=grcTeal!24,postaction={draw=none,pattern=dots,pattern color=grcTeal!85!black}},
  grcOvergen/.style={draw=grcOrange!80!black,line width=.25pt,fill=grcOrange!22,postaction={draw=none,pattern=north west lines,pattern color=grcOrange!85!black}},
  grcVzero/.style={draw=black!65,line width=.25pt,fill=grcRed!24,postaction={draw=none,pattern=dots,pattern color=grcRed!85!black}},
  grcVone/.style={draw=black!65,line width=.25pt,fill=grcGold!30,postaction={draw=none,pattern=horizontal lines,pattern color=grcGold!85!black}},
  grcVtwo/.style={draw=black!65,line width=.25pt,fill=grcSky!26,postaction={draw=none,pattern=north east lines,pattern color=grcSky!85!black}},
  grcVthree/.style={draw=black!65,line width=.25pt,fill=grcBlue!26,postaction={draw=none,pattern=crosshatch,pattern color=grcBlue!85!black}},
  grcSearch/.style={draw=black!60,line width=.2pt,fill=grcGold!30,postaction={draw=none,pattern=north east lines,pattern color=grcGold!85!black}},
  grcRTL/.style={draw=black!60,line width=.2pt,fill=grcRed!24,postaction={draw=none,pattern=crosshatch,pattern color=grcRed!85!black}},
  grcSimulation/.style={draw=black!60,line width=.2pt,fill=grcTeal!26,postaction={draw=none,pattern=dots,pattern color=grcTeal!85!black}},
  grcImplementation/.style={draw=black!60,line width=.2pt,fill=grcBlue!26,postaction={draw=none,pattern=north west lines,pattern color=grcBlue!85!black}}
}


\begin{document}

\title{Bottom-Up FPGA Reconfigurable Accelerator Overlay Generation}

\author{Mohamed Bouaziz}
\affiliation{%
  \institution{King Abdullah University of Science and Technology (KAUST)}
  \city{Thuwal}
  \country{Saudi Arabia}}
\author{Suhaib A. Fahmy}
\affiliation{%
  \institution{King Abdullah University of Science and Technology (KAUST)}
  \city{Thuwal}
  \country{Saudi Arabia}}
\renewcommand{\shortauthors}{Bouaziz and Fahmy}

\begin{abstract}
Programmable FPGA overlays run new workloads without re-syn\-the\-sis, yet they pay twice. An overlay adds a second layer of programmability on top of the FPGA's own overhead. Moreover, overlay generators usually start from a regular template, so they spend logic on switches and links that the workloads never use, and they close timing at low clocks. We present a flow that instead grows an overlay bottom-up from the dataflow graphs of the workloads it must run. The flow clusters workloads by structural similarity, ranks shared slots with spectral descriptors of each operation's surroundings, and adds only the units and links a workload lacks. It builds every unit on native hard blocks, with their internal pipeline registers, dual-ported memories and double-pumped DSPs, and it uses fixed-function engines where wider data types exceed the overlay's 16-bit lanes. Workloads then switch by configuration alone. On an AMD VCK5000, across 17 workloads from three suites and against OverGen rebuilt for the same board, our images run at 250--400\,MHz instead of 35--50\,MHz. As a result, they are 9.52--10.33$\times$ faster in geometric mean, and up to 36.97$\times$, while using 19.4--22.1$\times$ less of the FPGA and 9.67--10.51$\times$ less estimated board energy.
\end{abstract}

\maketitle

\input{Sections/1-Introduction}
\input{Sections/2-Background}
\input{Sections/3-Overlay-Generation}
\input{Sections/4-Automatic-Flow}

\input{Sections/5-Experimental-Results}

\input{Sections/6-Related-Work}
\input{Sections/7-Conclusion}

\begin{acks}
Claude AI helped with parts of the coding and tweaking figures and tables. ChatGPT helped improve clarity.
\end{acks}

\bibliographystyle{ACM-Reference-Format}
\bibliography{refs}

\end{document}

%% file: Sections/1-Introduction.tex
\section{Introduction}
\label{sec:introduction}

\input{Figures/flow-overview}

Spatial accelerators execute computations through interconnected functional units (FUs), in contrast to the von Neumann execution model, which runs instructions from a stored program in memory. This exposes parallelism and makes data movement through the interconnects an explicit design parameter. Stream-Dataflow~\cite{nowatzki2017streamdataflow}, Plasticine~\cite{prabhakar2017plasticine}, and Spatial~\cite{koeplinger2018spatial} illustrate how this can be achieved by organizing architectures and their programming models around the workload structures. However, their effectiveness depends on the match between the workload's requirements and the available computation, communication, and storage resources.

Field-programmable gate arrays (FPGAs) offer a way to adapt these resources to a workload at a finer level of granularity, as they come with a bit-reconfigurable fabric that allows designers to specialize both computation and data movement. However, they require design either at the register-transfer-level (RTL) or via High-level synthesis (HLS). The former requires substantial engineering effort, and the latter raises the programming abstraction at the cost of searching for efficient implementation parameters~\cite{cong2011hls,cong2018autoaccel,sohrabizadeh2022autodse}. Besides, the designed accelerator can only run the underlying workload. Thus, even the slightest change in the workload algorithm requires altering the generated hardware, which must undergo the time-consuming FPGA synthesis and physical implementation. This makes design iteration costly and in-depth explorations unfeasible.

For this purpose, \emph{FPGA overlays} have been introduced. An overlay is a programmable architecture implemented on the FPGA once, while a compiler only performs a fast mapping of workloads to it~\cite{coole2010intermediate,brant2012zuma,li2019overlaysurvey,jain2022jit}. It has shown effectiveness in many applications, such as neural-network and language-model inference~\cite{yu2020opu,abdelfattah2018dla,qian2025autocompiler,wu2025flightopu,he2026skipopu}, image and signal processing~\cite{choudhury2023flowpix,bai2025dspopu}, programmable packet parsing~\cite{mashreghi2023parser,mashreghi2024prismparser}, and privacy-preserving computation~\cite{fang2017secure,yang2025ola}. Yet, it presents a central design problem: the overlay must maintain sufficient flexibility to execute workloads with as little hardware overhead as possible to preserve fast compilation and scalability.

This overhead is added on top of the fact that an FPGA already needs roughly 20--35$\times$ the area, 3--4$\times$ the delay and 14$\times$ the dynamic power of an ASIC for the same logic~\cite{kuon2007gap}, and an overlay adds a second layer of programmability on top, often achieving a lower clock. 
However, architectures mapped to FPGAs can be efficiently pipelined thanks to their capable fabrics. Besides, building the overlay from the device's hard-block primitives and leveraging their micro-architectural features, such as internal registers and native operations, helps recover much of this loss.

Architecture-generation frameworks, more generally than just FPGA overlays, address this problem through workload-driven specialization. DSAGEN~\cite{weng2020dsagen} explores architecture description graphs, OverGen~\cite{liu2022overgen} specializes FPGA overlays using workload and memory information, and REVAMP refines computational and supporting resources for an application domain~\cite{bandara2022revamp}.  Other approaches specialize programmable processor families, streaming custom instructions, or dataflow templates~\cite{boutros2021specializing,ma2021pdlfgpu,drewes2024template,sun2025labidus,silva2025hpcgra,duarte2017scratch,cong2018autoaccel}, with static and profile-guided analysis helping automate architectural choices~\cite{ma2024specializedoverlays}. 
However, each search is expressed through a chosen representation, a set of legal transformations, and an initial hardware organization that is not necessarily efficient. These choices determine the cost and selection of the architecture being searched. For instance, a central problem is that many frameworks start from a rectangular-shaped hardware layout and optimize for workloads' DFGs that generally come in a converging/reduction shape. This makes optimizing these architectures to fit workload requirements compound a non-matching layout. They also overlook two FPGA-specific axes. The first is that FUs and how well they map onto hard blocks set the upper bound on clock frequency and, hence, on throughput. The second is the routing between them, which sets a lower bound on area, with a double effect: the overlay's reconfigurable switches require multiplexers and crossbars built from logic, and the FPGA itself spends a large share of its area connecting that logic~\cite{kuon2007gap}. 

A complementary direction constructs shared hardware from workload graphs. RADISH and APEX generate configurable computational structures from application graphs~\cite{moreano2005merging,willsey2019radish,melchert2023apex}. AS-Overlays mines recurring graph patterns across applications to specialize processing-element cores within a retained interconnect organization~\cite{mbongue2019asoverlays}. DeCO further demonstrates that a DSP-based overlay can be shaped around the stage-wise requirements of workload DFGs rather than a uniform rectangular hardware organization~\cite{jain2016deco}. \textbf{These results motivate a more specific question: how should workloads be clustered, corresponding operations identified, and shared hardware constructed when the implementation must respect the capabilities and execution constraints of native FPGA resources, in both compute and routing?} \emph{To address this multi-level question, we introduce a bottom-up overlay generation flow}.

The first issue therefore is deciding which workloads should share an overlay. An application domain or benchmark suite label does not specify the structure of a DFG. For example, two workloads from the experiments, \texttt{fir} and \texttt{fft} can both be classified as signal processing despite their different dependency patterns. The workloads considered in \S~\ref{sec:evaluation} contain such differences. \textbf{To solve this issue, we address the extraction of structural descriptors using workload similarity clustering so that workload clustering depends on shared operations instead of semantic labels alone.}

The second issue is the construction of the shared architecture overlay. This architecture should be based on what operations the workloads need and what their interconnection patterns looks like, neighbor-to-neighbor and beyond. This guarantees that the end result will not just approximate what compute, interconnect, and memory components are needed in isolation. Instead, these should help grow the architecture by adding components in patterns, based on an analysis of each workload's requirements. \textbf{For this, our flow uses spectral graph analysis to guide the growth process based on detected patterns that connect operations.} The spectral analysis helps provide information about an operation's structural context beyond its neighborhood. Thus, the architecture grows by reusing compatible resources and adding resources or connections when the selected correspondence cannot accommodate a workload. Therefore, the interconnect grows only where the workloads need it, which addresses the routing axis.

The third issue is realizing this graph efficiently on the FPGA. Conventionally, arithmetic operations are assigned to digital signal processing (DSP) hard blocks, and arrays to hard block memories (BRAMs/URAMs). These hard blocks are optimized for these purposes and are much more efficient than the general-purpose resources of an FPGA. However, synthesizing operators to these blocks does not necessarily capture their full micro-architectural features or constraints. Prior work exploited native DSP operations, internal pipelining, and runtime sharing, blocks cascading and DSP-memory organization, and programmable computation around distributed memory blocks~\cite{ronak2016mapping,ronak2016sharing,cheah2014idea,jain2015efficient,jain2016throughput,jain2016deco,jain2022jit,shi2020ftdl,basalama2020spar2,panahi2021memorycentric,kabir2025davinci}. \textbf{Our flow makes supported operations, timing, and storage interfaces explicit through native templates, and memory and streaming plans. It also leverages the AMD Versal hard blocks as they provide more capabilities than previous generations~\cite{amd2022am004,amd2026am007}.}

Our flow combines these solutions in the automatic flow shown in Figure~\ref{fig:flow-overview}. An MLIR-assisted frontend prepares programs for workload DFG extraction used for workload clustering and architecture overlay construction. Each cluster defines the workloads that will share an overlay, and the spectral analysis guides the overlay's growth in compute, memory, and interconnect. We build the overlay using operations supported by FPGA hard blocks and account for memory layout and pipeline timing. These choices guide RTL generation and the configurations used to run each workload. The resulting overlay can execute the corresponding cluster of workloads using their generated configurations, as well as unseen workloads whose DFGs are subsets of those in the cluster. This allows workload changes without having to repeat FPGA synthesis and physical implementation. The resulting \emph{image} runs each clustered workload as a \emph{mode} selected by its configuration.

To summarize, our contributions are as follows:
\begin{itemize}
    \item A workload-clustering method based on structure similarity rather than domain and benchmark-suite labels.
    \item An architecture overlay growth method based on spectral analysis that constructs shared computation and connectivity using the computational and memory patterns present in the workload clusters.
    \item Native hard block resource templates and explicit storage and execution descriptions that leverage the micro-architectural features to allow programmable operators, pipelining, and memory interfacing.
    \item An MLIR-assisted prototype and evaluation framework that distinguishes architecture construction, workload configuration, and implementation variants.
\end{itemize}

We evaluate our flow against OverGen~\cite{liu2022overgen}, checking every output word-for-word. We demonstrate that our images are 9.52$\times$ faster in geometric mean with one image per suite and 10.33$\times$ with similarity clusters, while using 19.4--22.1$\times$ less fabric and 9.67--10.51$\times$ less estimated board energy.

%% file: Figures/flow-overview.tex
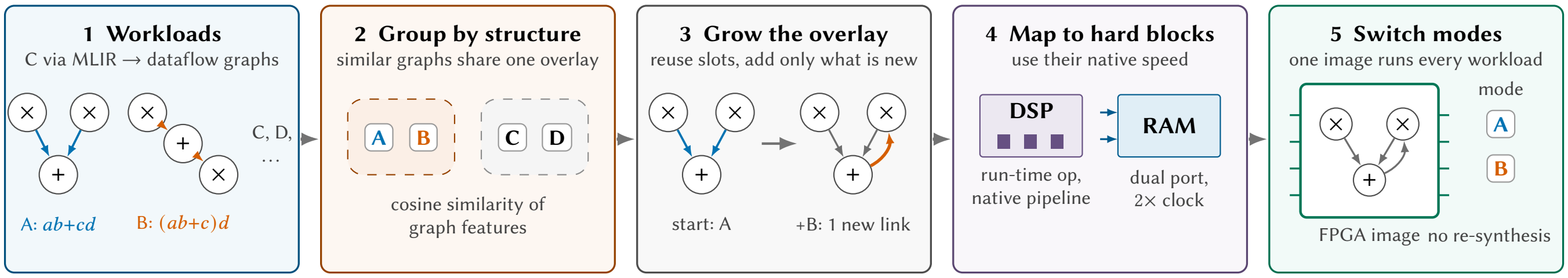
\begin{figure*}[t]
  \centering
  \definecolor{grInputBlue}{HTML}{0072B2}
  \definecolor{grGroupOrange}{HTML}{D55E00}
  \definecolor{grNativePurple}{HTML}{8064A2}
  \definecolor{grModeTeal}{HTML}{009E73}
  \definecolor{grGrey}{HTML}{6E6E6E}
  \resizebox{\textwidth}{!}{%
  \begin{tikzpicture}[
    font=\sffamily\fontsize{8}{9}\selectfont,
    stepArrow/.style={-{Latex[length=2mm]},draw=black!55,line width=1pt},
    panel/.style={rounded corners=3pt,line width=.6pt},
    title/.style={font=\sffamily\fontsize{7.5}{9}\selectfont\bfseries,anchor=north},
    note/.style={font=\sffamily\fontsize{6.3}{7.3}\selectfont,align=center,text=black!75},
    op/.style={circle,draw=black!70,fill=white,minimum size=4.3mm,inner sep=0pt,font=\sffamily\fontsize{7.5}{8}\selectfont},
    dot/.style={circle,inner sep=0pt,minimum size=1.7mm},
    eA/.style={-{Latex[length=1.3mm]},draw=grInputBlue,line width=.7pt},
    eB/.style={-{Latex[length=1.3mm]},draw=grGroupOrange,line width=.7pt},
    eN/.style={-{Latex[length=1.3mm]},draw=grGrey,line width=.6pt},
    tok/.style={draw=black!40,rounded corners=2pt,fill=white,inner sep=2pt,font=\sffamily\bfseries\fontsize{7}{8}\selectfont}
  ]
    \foreach \x/\c in {0/grInputBlue,3.55/grGroupOrange,7.1/grGrey,10.65/grNativePurple,14.2/grModeTeal} {
      \draw[panel,draw=\c!70!black,fill=\c!5] (\x,0) rectangle (\x+3.3,3.0);
    }
    \foreach \x in {3.3,6.85,10.4,13.95} { \draw[stepArrow] (\x+.02,1.5) -- (\x+.23,1.5); }

    \node[title] at (1.65,2.9) {1\enspace Workloads};
    \node[title] at (5.2,2.9) {2\enspace Group by structure};
    \node[title] at (8.75,2.9) {3\enspace Grow the overlay};
    \node[title] at (12.3,2.9) {4\enspace Map to hard blocks};
    \node[title] at (15.85,2.9) {5\enspace Switch modes};

    \node[note] at (1.65,2.38) {C via MLIR $\to$ dataflow graphs};
    \begin{scope}[shift={(.25,.55)}]
      \node[op] (a1) at (0,1.25) {$\times$}; \node[op] (a2) at (.7,1.25) {$\times$}; \node[op] (a3) at (.35,.55) {$+$};
      \draw[eA] (a1) -- (a3); \draw[eA] (a2) -- (a3);
      \node[note,text=grInputBlue] at (.35,0) {A: $ab{+}cd$};
      \node[op] (b1) at (1.35,1.25) {$\times$}; \node[op] (b2) at (1.75,.9) {$+$}; \node[op] (b3) at (2.15,.55) {$\times$};
      \draw[eB] (b1) -- (b2); \draw[eB] (b2) -- (b3);
      \node[note,text=grGroupOrange] at (1.75,0) {B: $(ab{+}c)d$};
      \node[note] at (2.75,.9) {C, D,\\\ldots};
    \end{scope}

    \node[note] at (5.2,2.38) {similar graphs share one overlay};
    \draw[grGroupOrange!70!black,dashed,rounded corners=5pt,fill=grGroupOrange!10] (3.85,1.1) rectangle (5.05,1.95);
    \draw[black!45,dashed,rounded corners=5pt,fill=black!4] (5.35,1.1) rectangle (6.55,1.95);
    \node[tok,text=grInputBlue] at (4.2,1.52) {A}; \node[tok,text=grGroupOrange] at (4.7,1.52) {B};
    \node[tok] at (5.7,1.52) {C}; \node[tok] at (6.2,1.52) {D};
    \node[note] at (5.2,.62) {cosine similarity of\\graph features};

    \node[note] at (8.75,2.38) {reuse slots, add only what is new};
    \begin{scope}[shift={(7.45,.75)}]
      \node[op] (g1) at (0,1.05) {$\times$}; \node[op] (g2) at (.75,1.05) {$\times$}; \node[op] (g3) at (.375,.35) {$+$};
      \draw[eA] (g1) -- (g3); \draw[eA] (g2) -- (g3);
      \node[note] at (.375,-.2) {start: A};
      \draw[stepArrow,line width=.8pt] (1.05,.7) -- (1.45,.7);
      \node[op] (h1) at (1.7,1.05) {$\times$}; \node[op] (h2) at (2.45,1.05) {$\times$}; \node[op] (h3) at (2.075,.35) {$+$};
      \draw[eN] (h1) -- (h3); \draw[eN] (h2) -- (h3);
      \draw[eB,line width=1pt] (h3) to[bend right=35] (h2);
      \node[note] at (2.075,-.2) {+B: 1 new link};
    \end{scope}

    \node[note] at (12.3,2.38) {use their native speed};
    \begin{scope}[shift={(10.95,.55)}]
      \draw[draw=grNativePurple!80!black,fill=grNativePurple!15,rounded corners=1pt] (0,.75) rectangle (1.15,1.45);
      \node[font=\sffamily\bfseries\fontsize{7.5}{8}\selectfont] at (.575,1.27) {DSP};
      \foreach \xx in {.2,.5,.8} { \fill[grNativePurple!80!black] (\xx,.84) rectangle (\xx+.14,1.0); }
      \node[note] at (.575,.4) {run-time op,\\native pipeline};
      \draw[draw=grInputBlue!80!black,fill=grInputBlue!12,rounded corners=1pt] (1.55,.75) rectangle (2.7,1.45);
      \node[font=\sffamily\bfseries\fontsize{7.5}{8}\selectfont] at (2.125,1.1) {RAM};
      \draw[-{Latex[length=1.2mm]},grInputBlue!80!black,line width=.6pt] (1.35,1.25) -- (1.55,1.25);
      \draw[-{Latex[length=1.2mm]},grInputBlue!80!black,line width=.6pt] (1.35,.95) -- (1.55,.95);
      \node[note] at (2.125,.4) {dual port,\\$2\times$ clock};
    \end{scope}

    \node[note] at (15.85,2.38) {one image runs every workload};
    \begin{scope}[shift={(14.55,.62)}]
      \draw[draw=grModeTeal!75!black,fill=white,rounded corners=2pt,line width=.8pt] (0,0) rectangle (1.55,1.5);
      \foreach \yy in {.25,.55,.85,1.15} { \draw[grModeTeal!75!black,line width=.6pt] (-.12,\yy) -- (0,\yy); \draw[grModeTeal!75!black,line width=.6pt] (1.55,\yy) -- (1.67,\yy); }
      \node[op,minimum size=3.6mm] (m1) at (.4,1.05) {$\times$}; \node[op,minimum size=3.6mm] (m2) at (1.15,1.05) {$\times$}; \node[op,minimum size=3.6mm] (m3) at (.775,.42) {$+$};
      \draw[eN] (m1) -- (m3); \draw[eN] (m2) -- (m3); \draw[eN] (m3) to[bend right=35] (m2);
      \node[note] at (.775,-.22) {FPGA image};
      \node[tok,text=grInputBlue] at (2.25,1.05) {A};
      \node[tok,text=grGroupOrange] at (2.25,.55) {B};
      \node[note] at (2.25,1.45) {mode};
      \node[note] at (2.12,-.22) {no re-synthesis};
    \end{scope}

  \end{tikzpicture}%
  }
  \caption{Overview of our flow: group, grow, map to hard blocks, and switch modes.}
  \Description{Five panels. Workloads become dataflow graphs; a reduction A and a chain B are shown. Structurally similar graphs are grouped, so A and B share an overlay while C and D form another group. The overlay grows from A's graph; adding B reuses all three slots and adds a single link. Slots are built on native DSP blocks with their internal pipeline and on memories used with both ports and a doubled clock. One implemented FPGA image then runs A or B by selecting a mode, with no re-synthesis.}
  \label{fig:flow-overview}
\end{figure*}

%% file: Sections/2-Background.tex
\section{Background}
\label{sec:background}

In this section, we introduce the concepts used by our flow. The first is the native FPGA resources that make an overlay both programmable and fast. The second is the similarity measure that decides which workloads share an overlay. The third is the spectral analysis that reveals the structure each cluster's overlay must provide.

\subsection{FPGA Resources and Implementation}
\label{sec:background-fpga}

An FPGA combines general-purpose configurable logic, such as lookup tables (LUTs) and flip-flops (FFs), hard blocks, such as digital signal processing (DSP) blocks for arithmetic and block RAMs/UltraRAMs (BRAMs/URAMs), and programmable routing resources to implement an accelerator design.

The hard blocks can run in many modes and can be pipelined through internal registers to enable higher clock rates. For instance, a DSP can run different arithmetic operations on different data types~\cite{amd2022am004}, and a BRAM/URAM can store data in different sizes and organize its read and write scheme in several ways~\cite{amd2026am007}. These modes can be configured at runtime, enabling overlay programmability.

On the other hand, communication with external memory can add latency. For this, a burst transfers a contiguous block of words in a single request to reduce the communication overhead. The data can also be reused, which increases latency if not kept on-chip. For this, a scratchpad memory block holds explicitly loaded data on-chip. 

Using these resources requires significant engineering effort to design the accelerator in RTL and a time-consuming hardware implementation. HLS and hardware-oriented languages automate part of the translation from a higher-level description, but the resulting hardware still requires the time-consuming implementation~\cite{koeplinger2018spatial,sohrabizadeh2022autodse}.
Therefore, A change to the datapath or its pipeline structure is expensive. Overlays address this cost by implementing a reusable programmable architecture.

\subsection{FPGA Overlays and Workload Mapping}
\label{sec:background-overlays}

\input{Figures/fpga-overlay-background}

An overlay implements a programmable architecture on the FPGA. Typically, it is composed of configurable functional units (FUs), storage, and selectable connections. Mapping a workload assigns its operations to compatible units and arranges the movement using these overlay components without having to handle the low-level FPGA bit-configurable components~\cite{liu2022overgen,jain2015efficient}.

This creates two levels of configuration.
The FPGA image establishes the hardware, including its available functional units, routing structures, and primitive attributes. A workload configuration selects functions and paths within that hardware, as illustrated in Figure~\ref{fig:fpga-overlay-background}. For example, the shared compute graph in Figure~\ref{fig:flow-overview} can support a reduction or a chain by changing operation and operand selections.
In our current native generator, configurations for the selected workloads are incorporated during hardware construction,
and invocation selects one of those workload modes.

The benefit of reusing an overlay depends on the cost of the flexibility it contains. Supporting another workload may require an additional operation, a new operand source, more storage, or a different memory interface. Each such requirement can increase the resources or control needed by the overlay. Conversely, compatible workloads may share much of their implementation. Therefore, the workload set should be selected before constructing the overlay architecture. Thus, the clustering method must identify useful opportunities for sharing based on the workloads' compute and memory access patterns.

\subsection{Comparing DFGs with Cosine Similarity}
\label{sec:background-cosine}

In order to cluster workloads based on their compute and memory access patterns. A workload's DFG describes both the operations it needs and the dependencies between them. When considering workloads for a common overlay, we first compare their computational requirements. This identifies workloads with similar demands for operation types before examining how they are connected.
Figure~\ref{fig:background-dfgs} shows two DFGs that compute $p=ab$ and $q=pc$. The path graph produces $z=q+d$, while the reuse graph produces $y=q+p$. Both require the same operators but differ in dependencies.

\input{Figures/background-dfgs}

We represent a workload by a feature vector $\mathbf{f}_G\in\mathbb{R}_{\geq 0}^{m}$, where each coordinate counts a particular property across all workloads. For instance, for a vector containing only multiplication and addition counts results in the two graphs having
\(
  \mathbf{f}_{\mathrm{path}}
  =\mathbf{f}_{\mathrm{reuse}}
  =(2,1)^{\mathsf T}.
\)
To detect this similarity, we define the cosine similarity as
\begin{equation}
  s(\mathbf{f},\mathbf{g})
  =
  \frac{\mathbf{f}^{\mathsf T}\mathbf{g}}
       {\|\mathbf{f}\|_2\|\mathbf{g}\|_2}
  =\cos\theta \in [-1,1],
  \label{eq:background-cosine}
\end{equation}
where $\cos\theta$ measures much the vectors differ and $\|\mathbf{f}\|_2=\sqrt{\sum_k f_k^2}$.
For nonnegative features, a score of 1 means that the vectors have the same feature proportions. A score of 0 means that they have no nonzero coordinates in common. For instance, the $\mathbf{f}_{\mathrm{path}} $ and $ \mathbf{f}_{\mathrm{reuse}}$ vectors have similarity 1. Thus, the comparison helps distinguish workloads with similar components from those with differing ones. This helps cluster workloads that guide an overlay construction. 

Generally, our flow also includes resource-kind and dependency features in its vectors, as described in \S~\ref{sec:generation-clustering}. Such features capture aggregate interaction patterns. However, they still represent the overall workload and do not convey information about the interactions that occur. Thus, feature similarity is used to cluster similar workloads. Once a cluster is created, we use spectral analysis to describe individual operations and their surroundings. This enables distinguishing between two decisions: which workloads to cluster together and what patterns should be present for workloads within one cluster to be represented.

\subsection{Graph Signals and Fourier Analysis}
\label{sec:background-gft}

\input{Figures/graph-fourier-example}

The operator count does not describe the extra connection in $G_{\mathrm{reuse}}$. 
Therefore, the operation's neighbourhood should be analyzed. For this, we tag vertex properties (such as the type of operation), propagate the tags through graph connections using the heat diffusion model~\cite{chung1997heat}, and record the resulting values at each vertex. Applying the model at several scales captures local and broader interaction patterns. The graph Fourier transform (GFT) expresses these interactions in terms of patterns~\cite{shuman2013emerging}.
For this, we use \emph{graph signals} that assign a numerical value to each vertex to describe a property. For instance, an addition tag assigns one at the addition vertex and zero at the others. Thus, in vertex order $(v_1,v_2,v_3)$, both $G_{path}$ and $G_{reuse}$ in Figure~\ref{fig:background-dfgs} graphs have the same initial signal $\mathbf{x}=(0,0,1)^{\mathsf T}$. 

For this example, we keep the compute connections and treat them as undirected during analysis. This gives a three-vertex path for $G_{\mathrm{path}}$ and a triangle for $G_{\mathrm{reuse}}$. The \emph{adjacency matrix} $A$ of an analysis graph $H$ with $n$ vertices has $A_{ij}=1$ when vertices $i$ and $j$ are connected and zero otherwise. This means that $A_{ij}=A_{ji}$ because the analysis graph is undirected. As the \emph{degree} of a vertex is its number of neighbours, we also use the \emph{degree matrix} $D$, which places these counts on its diagonal and has zero entries elsewhere, so $D_{ii}=\sum_j A_{ij}$. This leads to the combinatorial Laplacian
  \(L=D-A\)
that expresses the local differences.
Its action on a signal $\mathbf{x}$ at vertex $i$ is
\(
  (L\mathbf{x})_i
  =\sum_j A_{ij}(x_i-x_j).
\)

$G_{path}$ has vertex degrees $(1,2,1)$. Adding the reuse edge between $v_1$ and $v_3$ in $G_{reuse}$ gives every vertex degree two. The corresponding Laplacians are as in Eq.~\ref{eq:background-laplacians}. This added connection changes the matrix even though the operation-count vector and initial addition tag are unchanged.

\begin{equation}
  L_{\mathrm{path}}
  =
  \begin{bmatrix}
     1 & -1 &  0 \\
    -1 &  2 & -1 \\
     0 & -1 &  1
  \end{bmatrix},
  \qquad
  L_{\mathrm{reuse}}
  =
  \begin{bmatrix}
     2 & -1 & -1 \\
    -1 &  2 & -1 \\
    -1 & -1 &  2
  \end{bmatrix}.
  \label{eq:background-laplacians}
\end{equation}

Although very similar, the two graphs allow propagating values differently. To extract the propagation patterns, we express the signal as a combination of patterns associated with these connections. Since $L$ is symmetric, its eigenvectors can be chosen orthogonal and of unit length. An eigenvector $u_k$ satisfies $Lu_k=\lambda_k u_k$, so applying $L$ to the pattern is equivalent to scaling it by $\lambda_k$. Collecting the eigenvectors as the columns of $U$ gives
  \(L=U\Lambda U^{\mathsf T}\) and
  \(U^{\mathsf T}U=I\),
where $\Lambda$ is diagonal and contains the nonnegative eigenvalues $(\lambda_k)_{0\leq i<n}$. These patterns form a basis for the vertex signal components~\cite{hammond2011wavelets}. Figure~\ref{fig:graph-fourier-example} shows $G_{path}$'s constant pattern $u_0$, endpoint-varying pattern $u_1$, and middle-versus-endpoints pattern $u_2$.

Besides, the eigenvalue $\lambda_k=u_k^{\mathsf T}Lu_k
      =\frac{1}{2}\sum_{i,j}
        A_{ij}\bigl(u_k(i)-u_k(j)\bigr)^2$ measures how much a pattern varies across connected vertices. 
$G_{path}$'s three patterns have eigenvalues $0$, $1$, and $3$. This means that the constant pattern has no variation across edges, while the other patterns vary increasingly strongly.

In order to determine how much each pattern contributes to the tag signal, the GFT is used $\widehat{\mathbf{x}}=U^{\mathsf T}\mathbf{x}$, and its inverse $\mathbf{x}=U\widehat{\mathbf{x}}$ reconstructs the signal.
Each coefficient $\widehat{x}_k=u_k^{\mathsf T}\mathbf{x}$ is the contribution of pattern $u_k$~\cite{shuman2013emerging}. For the addition tag $\mathbf{x}=(0,0,1)^{\mathsf T}$ on the $G_{path}$ DFG,
\(
  \widehat{\mathbf{x}}
  =
  \left(
    \frac{1}{\sqrt{3}},
    -\frac{1}{\sqrt{2}},
    \frac{1}{\sqrt{6}}
  \right)^{\mathsf T}.
\)
This means that all three patterns have contributions.

The heat diffusion model propagates this property (i.e. the addition) by attenuating each coefficient and reconstructing the vertex values:
\begin{equation}
  \mathbf{x}_t
  =
  e^{-rL}\mathbf{x}
  =
  Ue^{-r\Lambda}U^{\mathsf T}\mathbf{x},
  \qquad r\geq 0.
  \label{eq:background-heat}
\end{equation}
Coefficient $k$ is multiplied by $e^{-r\lambda_k}$, so patterns with greater variation are attenuated more strongly. This reduces the signal's overall variation across edges as information about the addition propagates through the graph.

At the illustrative scale $r=\ln 2$, the path's components at $\lambda_1=1$ and $\lambda_2=3$ are multiplied by $1/2$ and $1/8$, respectively. Reconstruction gives
\(
  \mathbf{x}_{\mathrm{path},r}
  =
  \left(
    \frac{5}{48},
    \frac{7}{24},
    \frac{29}{48}
  \right)^{\mathsf T}.
\)
This means that the first multiplication receives less of the addition tag than the second because it is separated from the addition by another operation.

For the $G_{reuse}$ graph, the same eigenvectors form a valid basis, but their eigenvalues are $0$, $3$, and $3$. Thus, the extra edge changes the attenuation of $u_1$ from $1/2$ to $1/8$ at the same scale. Applying the model to the same initial tag gives
\(
  \mathbf{x}_{\mathrm{reuse},r}
  =
  \left(
    \frac{7}{24},
    \frac{7}{24},
    \frac{5}{12}
  \right)^{\mathsf T}.
\)
The two multiplication vertices now receive equal values because both connect directly to the addition and occupy symmetric positions in this undirected analysis graph. Figure~\ref{fig:graph-fourier-example} shows how the additional connection changes the neighbourhood description at $v_1$, although its operation type and the overall operation count remain the same.

Repeating the calculation for different vertex properties and diffusion scales produces a description of each vertex's surroundings. Smaller scales keep more local detail, while larger scales mix information over broader parts of the graph~\cite{shuman2013emerging,hammond2011wavelets}. We use these reconstructed vertex values as descriptors.

The example shows how connectivity affects the tags that are propagated. For architecture construction, our flow uses tags for compatible resource kinds and labelled input and output ports, as described in \S~\ref{sec:generation-spectral}. This representation preserves the roles and operand positions required by the generator.

%% file: Figures/fpga-overlay-background.tex
\begin{figure}[t]
  \centering
  \definecolor{grLogicBlue}{HTML}{0072B2}
  \definecolor{grDSPPurple}{HTML}{8064A2}
  \definecolor{grMemoryTeal}{HTML}{009E73}
  \definecolor{grModeOrange}{HTML}{D55E00}
  \resizebox{\columnwidth}{!}{%
  \begin{tikzpicture}[
    font=\sffamily\fontsize{8}{9}\selectfont,
    flow/.style={-{Latex[length=1.5mm]},draw=black!75,line width=.65pt},
    buildflow/.style={flow,draw=grLogicBlue},
    modeflow/.style={flow,draw=grModeOrange,dashed},
    panel/.style={draw=black!55,fill=black!1,rounded corners=2pt,line width=.6pt},
    title/.style={font=\sffamily\fontsize{9}{10}\selectfont\bfseries},
    subhead/.style={font=\sffamily\fontsize{8}{9}\selectfont\bfseries},
    note/.style={align=center,inner sep=1pt},
    block/.style={draw,fill=white,align=center,minimum height=4.3mm,inner sep=1.5pt},
    reg/.style={draw,fill=white,minimum width=4mm,minimum height=4.3mm,inner sep=0pt},
    stage/.style={draw,rounded corners=1.5pt,align=center,minimum height=4.8mm,inner sep=2pt}
  ]

    \draw[panel] (8.2,0.35) rectangle (16.8,2.1);

    \node[subhead,align=left,anchor=west,inner sep=1pt] at (8.4,1.75) {Build\\hardware};

    \node[stage,draw=grLogicBlue!75!black,fill=grLogicBlue!7,minimum width=1.1cm] (rtl) at (10.55,1.75) {Overlay RTL};
    \node[stage,draw=grLogicBlue!75!black,fill=grLogicBlue!7,minimum width=2cm] (impl) at (12.7,1.75) {Synthesis, P\&R};
    \node[stage,draw=grLogicBlue!75!black,fill=grLogicBlue!7,minimum width=1.45cm] (image) at (14.9,1.75) {FPGA image};

    \draw[buildflow] (rtl) -- (impl);
    \draw[buildflow] (impl) -- (image);

    \draw[dashed,black!35] (8.4,1.35) -- (16.6,1.35);

    \node[subhead,align=left,anchor=west,inner sep=1pt] at (8.4,1) {Invoke\\workload};

    \node[stage,draw=grModeOrange!80!black,fill=grModeOrange!7,minimum width=1.8cm] (mode) at (10.7,1) {Select workload};
    \node[stage,draw=grMemoryTeal!75!black,fill=grMemoryTeal!7,minimum width=3cm] (overlay) at (14.9,1) {Implemented overlay};

    \draw[modeflow] (mode.east) -- node[above,note] {controls} (overlay.west);
    \draw[buildflow] (image.south) -- (image.south |- overlay.north);

    \node[note] at (12.85,0.5) {Fixed resources, with controls selected by workload mode.};

  \end{tikzpicture}%
  }

  \caption{Our deployment flow.}
  \Description{The figure shows that hardware synthesis, placement, routing, and FPGA image generation are separated from workload invocation. A workload mode selects controls within the overlay that are implemented.}
  \label{fig:fpga-overlay-background}
\end{figure}
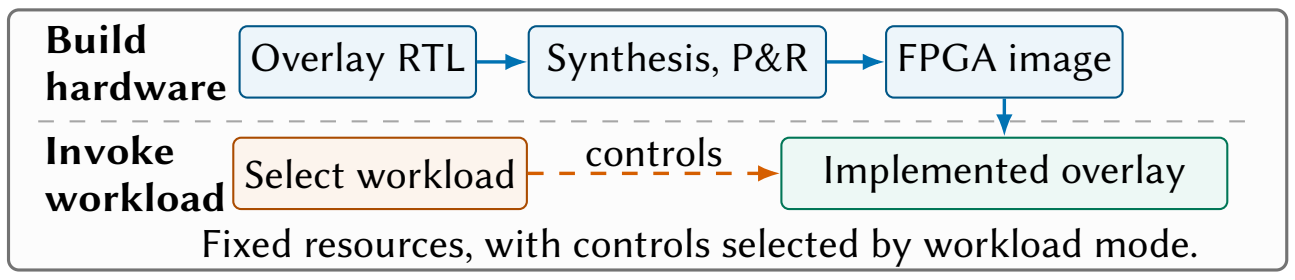

%% file: Figures/background-dfgs.tex
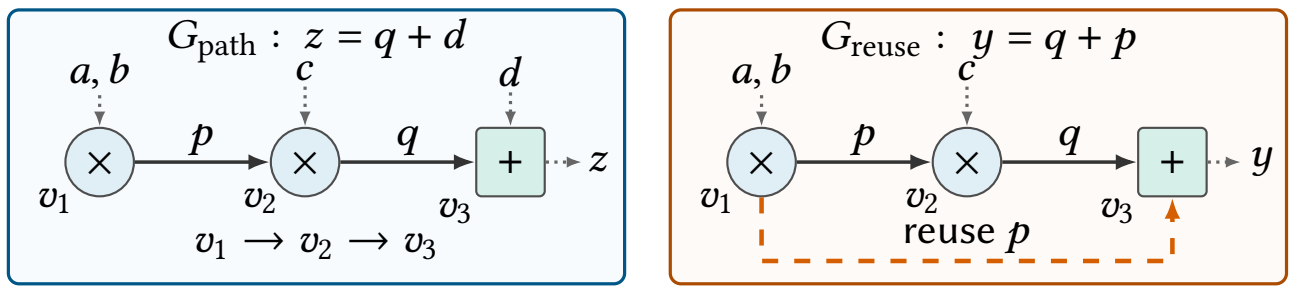
\begin{figure}[t]
  \centering
  \definecolor{dfgBlue}{HTML}{0072B2}%
  \definecolor{dfgOrange}{HTML}{D55E00}%
  \definecolor{dfgTeal}{HTML}{009E73}%
  \resizebox{\columnwidth}{!}{%
  \begin{tikzpicture}[
    font=\sffamily\fontsize{8}{9}\selectfont,
    flow/.style={-{Latex[length=1.4mm]},draw=black!80,line width=.65pt},
    external/.style={-{Latex[length=1.mm]},draw=black!60,densely dotted,line width=.6pt},
    reuse/.style={flow,draw=dfgOrange,dashed,line width=.85pt},
    panel/.style={rounded corners=2pt,line width=.6pt},
    head/.style={font=\sffamily\fontsize{8}{9}\selectfont\bfseries,inner sep=1pt},
    note/.style={inner sep=1pt},
    vertex label/.style={font=\sffamily\fontsize{8}{9}\selectfont,inner sep=0pt},
    op/.style={draw=black!70,minimum size=4.5mm,inner sep=0pt,label distance=1pt},
    mul/.style={op,circle,fill=dfgBlue!12},
    add/.style={op,rectangle,rounded corners=1pt,fill=dfgTeal!16}
  ]

    \draw[panel,draw=dfgBlue!75!black,fill=dfgBlue!3] (0,1.2) rectangle (4.05,3);
    \draw[panel,draw=dfgOrange!80!black,fill=dfgOrange!3] (4.35,1.2) rectangle (8.4,3);
    \node[head] at (2.03,2.8) {$G_{\mathrm{path}}:\ z=q+d$};
    \node[head] at (6.38,2.8) {$G_{\mathrm{reuse}}:\ y=q+p$};

    \node[mul,label={[vertex label]below left:$v_1$}] (p1) at (.60,2) {$\times$};
    \node[mul,label={[vertex label]below left:$v_2$}] (p2) at (1.95,2) {$\times$};
    \node[add,label={[vertex label]below left:$v_3$}] (p3) at (3.30,2) {$+$};

    \draw[flow] (p1) -- node[above,inner sep=1pt] {$p$} (p2);
    \draw[flow] (p2) -- node[above,inner sep=1pt] {$q$} (p3);

    \node[note] (pab) at (.60,2.60) {$a,b$};
    \node[note] (pc) at (1.95,2.60) {$c$};
    \node[note] (pd) at (3.30,2.60) {$d$};
    \node[note] (pz) at (3.88,2) {$z$};

    \draw[external] (pab) -- (p1);
    \draw[external] (pc) -- (p2);
    \draw[external] (pd) -- (p3);
    \draw[external] (p3) -- (pz);

    \node[note] at (2.025,1.45) {$v_1\rightarrow v_2\rightarrow v_3$};

    \node[mul,label={[vertex label]below left:$v_1$}] (r1) at (4.95,2) {$\times$};
    \node[mul,label={[vertex label]below left:$v_2$}] (r2) at (6.30,2) {$\times$};
    \node[add,label={[vertex label]below left:$v_3$}] (r3) at (7.65,2) {$+$};

    \draw[flow] (r1) -- node[above,inner sep=1pt] {$p$} (r2);
    \draw[flow] (r2) -- node[above,inner sep=1pt] {$q$} (r3);
    \draw[reuse] (r1.south) -- (4.95,1.35) -- node[above,inner sep=1pt] {reuse $p$} (7.65,1.35) -- (r3.south);

    \node[note] (rab) at (4.95,2.60) {$a,b$};
    \node[note] (rc) at (6.30,2.60) {$c$};
    \node[note] (ry) at (8.23,2) {$y$};

    \draw[external] (rab) -- (r1);
    \draw[external] (rc) -- (r2);
    \draw[external] (r3) -- (ry);

  \end{tikzpicture}%
  }
  \caption{DFGs of \texttt{path} and \texttt{reuse} workloads.}
  \Description{Both DFGs compute p equals a times b and q equals p times c, using two multiplications and one addition. The path adds an external d to q. The reuse graph adds p to q, introducing a direct edge from the first multiplication to the addition. Both operation-count vectors are (2,1) and have cosine similarity one. Multiplications are blue circles and additions are green squares.}
  \label{fig:background-dfgs}
\end{figure}

%% file: Figures/graph-fourier-example.tex
\begin{figure}[t]
  \centering
  \definecolor{gfBlue}{HTML}{0072B2}%
  \definecolor{gfOrange}{HTML}{D55E00}%
  \definecolor{gfTeal}{HTML}{009E73}%
  \definecolor{gfPurple}{HTML}{8064A2}%
  \resizebox{\columnwidth}{!}{%
  \begin{tikzpicture}[
    font=\sffamily\fontsize{8}{9}\selectfont,
    panel/.style={rounded corners=2pt,line width=.6pt},
    head/.style={font=\sffamily\fontsize{8}{9}\selectfont\bfseries,inner sep=1pt},
    note/.style={inner sep=1pt},
    signal/.style={circle,draw=black!65,minimum size=5.5mm,inner sep=0pt},
    addsignal/.style={signal,rectangle,rounded corners=1pt},
    link/.style={draw=black!75,line width=.65pt}
  ]

    \node[head] at (4.2,4.75) {Fourier modes of $G_{\mathrm{path}}$};

    \draw[panel,draw=gfBlue!80!black,fill=gfBlue!4] (0,2.7) rectangle (2.7,4.45);
    \draw[panel,draw=gfOrange!80!black,fill=gfOrange!4] (2.85,2.7) rectangle (5.55,4.45);
    \draw[panel,draw=gfTeal!80!black,fill=gfTeal!4] (5.7,2.7) rectangle (8.4,4.45);

    \node[head] at (1.35,4.22) {$\lambda_0=0$};
    \node[head] at (4.2,4.22) {$\lambda_1=1$};
    \node[head] at (7.05,4.22) {$\lambda_2=3$};

    \node[note] at (1.35,3.87) {$u_0=(1,1,1)^{\mathsf T}/\sqrt{3}$};
    \node[note] at (4.2,3.87) {$u_1=(1,0,-1)^{\mathsf T}/\sqrt{2}$};
    \node[note] at (7.05,3.87) {$u_2=(1,-2,1)^{\mathsf T}/\sqrt{6}$};

    \foreach \offset in {0,2.85,5.7} {
      \draw[black!35] (\offset+.42,3.3) -- (\offset+2.28,3.3);
      \node[note] at (\offset+.2,3.3) {$0$};

      \foreach \x/\vertex in {.55/v_1,1.35/v_2,2.15/v_3} {
        \fill[black!50] (\offset+\x,3.3) circle (.8pt);
        \node[note] at (\offset+\x,2.9) {$\vertex$};
      }
    }

    \foreach \x in {.55,1.35,2.15} {
      \draw[gfBlue,line width=.8pt] (\x,3.3) -- (\x,{3.3+.5/sqrt(6)});
      \fill[gfBlue] (\x,{3.3+.5/sqrt(6)}) circle (1.6pt);
    }

    \foreach \x/\value in {3.4/1,4.2/0,5/-1} {
      \pgfmathsetmacro{\stemHeight}{3.3+.5*\value/(sqrt(2)*sqrt(2))}
      \draw[gfOrange,line width=.8pt] (\x,3.3) -- (\x,\stemHeight);
      \fill[gfOrange] (\x-.055,\stemHeight-.055) rectangle (\x+.055,\stemHeight+.055);
    }

    \foreach \x/\value in {6.25/0.5,7.05/-1,7.85/0.5} {
      \pgfmathsetmacro{\stemHeight}{3.3+.5*\value/sqrt(12)}
      \draw[gfTeal!85!black,line width=.8pt] (\x,3.3) -- (\x,\stemHeight);
      \fill[gfTeal!85!black] (\x,\stemHeight+.085) -- (\x-.06,\stemHeight-.04) -- (\x+.06,\stemHeight-.04) -- cycle;
    }

    \node[note] at (4.2,2.5) {Addition tag $\mathbf{x}=(0,0,1)^{\mathsf T}$; heat diffusion model at $t=\ln 2$};

    \draw[panel,draw=gfBlue!80!black,fill=gfBlue!3] (0,0.8) rectangle (4.05,2.29);
    \draw[panel,draw=gfOrange!80!black,fill=gfOrange!3] (4.35,0.8) rectangle (8.4,2.29);

    \node[head] at (2.03,2.1) {$G_{\mathrm{path}}:\ \lambda=(0,1,3)$};
    \node[head] at (6.38,2.1) {$G_{\mathrm{reuse}}:\ \lambda=(0,3,3)$};

    \node[signal,fill=gfPurple!12] (p1) at (0.65,1.6) {$\tfrac{5}{48}$};
    \node[signal,fill=gfPurple!30] (p2) at (2.03,1.6) {$\tfrac{7}{24}$};
    \node[addsignal,fill=gfPurple!55] (p3) at (3.4,1.6) {$\tfrac{29}{48}$};

    \draw[link] (p1) -- (p2);
    \draw[link] (p2) -- (p3);

    \node[signal,fill=gfPurple!30] (r1) at (5,1.6) {$\tfrac{7}{24}$};
    \node[signal,fill=gfPurple!30] (r2) at (6.38,1.6) {$\tfrac{7}{24}$};
    \node[addsignal,fill=gfPurple!42] (r3) at (7.75,1.6) {$\tfrac{5}{12}$};

    \draw[link] (r1) -- (r2);
    \draw[link] (r2) -- (r3);
    \draw[link,draw=gfOrange,dashed,line width=.85pt] (r1.south) -- (5,1) -- (7.75,1) -- (r3.south);

    \foreach \x/\vertex in {.65/v_1,2.025/v_2,3.4/v_3,5/v_1,6.375/v_2,7.75/v_3} {
      \node[note] at (\x-.27,1.2) {$\vertex$};
    }

  \end{tikzpicture}%
  }
  \caption{Fourier modes and addition tags after applying the heat diffusion model.}
  \Description{The top row shows the three orthonormal path modes with eigenvalues zero, one, and three, using circle, square, and triangle stem markers on a common scale. Both graphs start with addition tag (0,0,1). At diffusion scale ln(2), the path gives (5/48,7/24,29/48), while the reuse graph gives (7/24,7/24,5/12). The reuse graph has eigenvalues zero, three, and three. The lower undirected links show the connections used for this illustrative analysis.}
  \label{fig:graph-fourier-example}
\end{figure}
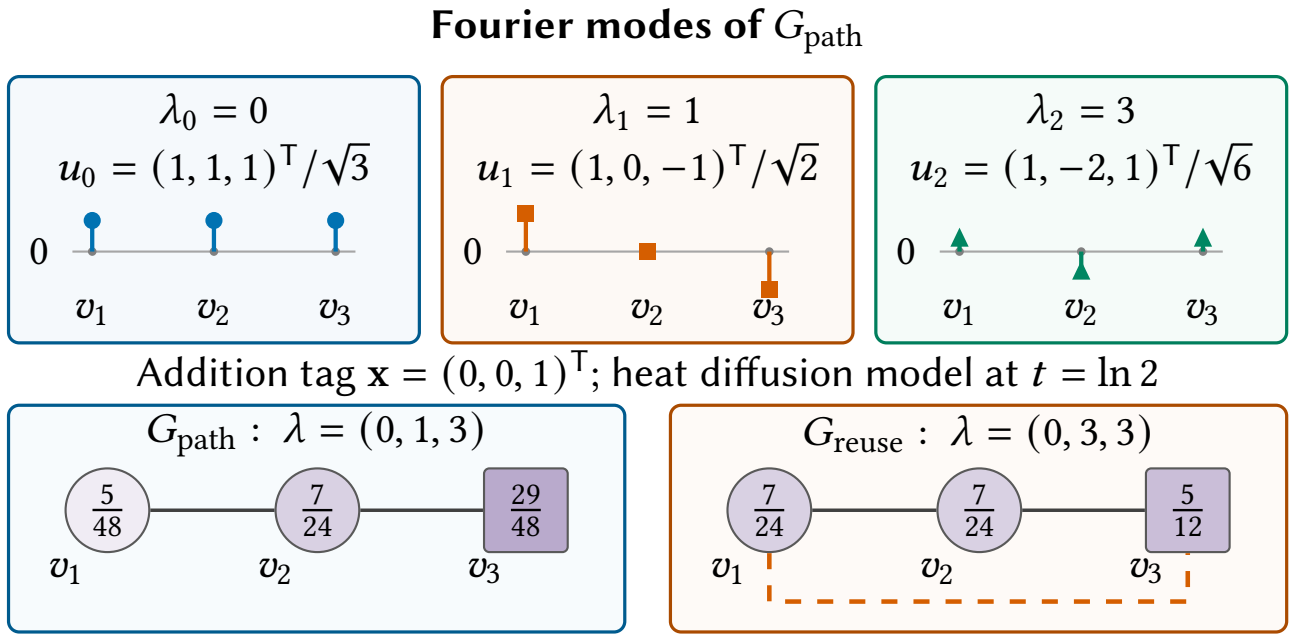

%% file: Sections/3-Overlay-Generation.tex
\section{Overlay Generation}
\label{sec:overlay-generation}

\input{Figures/overlay-shapes}

Our flow clusters workloads with similar structure, maps their operations onto shared hardware, and adds the required connections.
Figure~\ref{fig:overlay-shapes} draws the same workload (\texttt{bgr2grey}~\S\ref{sec:evaluation}) on both overlays. OverGen~\cite{liu2022overgen} routes it through 22 of its 35 switches and 65 of its 349 links, whereas our overlay uses 16 of its 43 links and 6 of its 8 selectors. This shows that not only does our flow grow the necessary FUs and links based on the workload, but it also does the same for switching/operand-selection, which has been demonstrated to consume a lot of the FPGA fabric~\cite{lysecky2005firm,brant2012zuma,vaishnav2018survey}. The overlay is programmed by selecting the FU operations and operand-selector paths. This eliminates a lot of unneeded logic across the whole workloads' set (\S\ref{sec:eval-resources}) and shrinks the time-to-run (\S\ref{sec:eval-generation}).

\subsection{Supported Graphs and Execution Model}
\label{sec:generation-model}

Here, we define the terminology, the supported graph representation, and the execution model. 

A workload is a finite directed acyclic graph $G_w=(V_w,E_w)$. Vertices specify operations; an edge $(u,v,p)$ supplies the result of $u$ to operand $p$ of $v$. Memory vertices name bounded arrays and their accesses. A \emph{frame} evaluates this graph once; a run executes a bounded sequence of frames. This form expresses the control.

An overlay contains hardware \emph{slots} $S$ and connections $E_A$ of an architecture $A$. A slot's \emph{kind} $\kappa(s)$ specifies the operations it supports. The mapping $\mu_w:V_w\to S$ is injective: each workload uses a different slot for each vertex, while different workloads may share slots.
We call one implemented FPGA design an \emph{image}, which holds a single overlay. Producing it requires synthesis and physical implementation. Each workload the overlay supports is a \emph{mode} of that image, selected at runtime by writing its \emph{configuration}, the FU operations, operand selectors and memory bindings, into control registers. Thus, switching workloads costs a register write, whereas supporting a workload outside the image requires \emph{reimplementation}, i.e., a new image.

\subsection{Functional Units from Native DSP Paths}
\label{sec:generation-compute}

\input{Figures/native-programmable-fu}
\input{Tables/programmable-fu-ops}

Our general FUs drive four Versal DSP58s~\cite{amd2022am004} with one opcode, treating a 64-bit word as four 16-bit lanes. Each block pairs a signed multiplier with a 58-bit arithmetic unit as shown in Figure~\ref{fig:native-programmable-fu}; its operation is chosen at runtime, whereas the mode (integer, floating-point) and pipeline are chosen at design time. For wider types, dedicated units are used instead: a 64-bit integer product is built from three 32-bit ones, and 32- and 64-bit floating-point runs on vendor IP built from DSP58s in integer mode. Besides programmable FUs, a grown overlay may contain fixed-op units, each performing one operation the FU does not provide, such as division by a constant or a signed maximum.

In most designs, DSP blocks can run at twice the clock rate of the surrounding fabric logic~\cite{ronak2017multipumping}, leaving them idle for part of each clock period. For this, our flow enables \emph{double pumping}, which runs DSP at twice the logic clock frequency, doubling the work per cycle. As Figure~\ref{fig:native-programmable-fu} shows, an in-kernel mixed-mode clock manager (MMCM)~\cite{amd_pg065} generates that clock, so two lanes share one DSP.

\subsection{Memory Semantics and Physical Allocation}
\label{sec:generation-memory}

Each array has a fixed size in 64-bit words, and a word holds one 64-bit integer or float, two 32-bit floats, or four 16-bit integers. Most follow affine access, in which a frame advances by a fixed stride from a starting offset. Gather and scatter accesses instead point directly to a whole word. The affine writers that share an output are checked to ensure they never collide. In addition, reads and writes keep program order, and unwritten outputs stay zero.

The words are stored in banked memory components, with consecutive words mapped to consecutive banks. Each pass across the banks fills the next row, and a bank that outgrows one memory block stacks another memory block. The compiler tries banking over \{1,2,4,8\} banks, favoring the least memory, then the fewest rounds lost to bank conflicts, then the fewest blocks and banks. In addition, compatible arrays that never overlap in time share storage. Figure~\ref{fig:native-memory} shows the native memory types~\cite{amd2026am007} and a window that keeps words for reuse; its overlap depends on the executor.

Like DSPs (\S\ref{sec:generation-compute}), BRAM and UltraRAM can run faster than the surrounding logic clock. For this, our flow can enable \emph{dual-porting}, which uses both ports of each BRAM and UltraRAM to serve two accesses per bank per cycle instead of one, and double pumping doubles this to four accesses per bank per cycle.
\input{Figures/native-memory}
\subsection{Structural Workload clustering}
\label{sec:generation-clustering}

\input{Tables/cluster-features}

Our flow clusters workloads by cosine similarity (\S~\ref{sec:background-cosine}), so that each cluster tends to need the same units and interactions. The feature vectors count opcodes, slot kinds and arities, directed connections by operand port, and graph size and depth (Table~\ref{tab:cluster-features}).

Each workload starts in its own cluster. The candidates are the 32 most similar pairs plus each workload's three nearest neighbors. We visit candidates in descending similarity. For a pair $(i,j)$, let $\mathcal C_i$ and $\mathcal C_j$ be the clusters currently holding the two workloads, and $Q(\mathcal C)$ the cost of an overlay supporting every workload in $\mathcal C$. The merged cluster has at most four workloads and costs less than the two clusters separately

For native workloads, each proposed cluster gets one trial architecture $\mathcal A_{\mathcal C}$, built by the neighbor-count method of \S~\ref{sec:generation-growth}, and $Q(\mathcal C)=C(\mathcal A_{\mathcal C})$, the heuristic score of Eq.~\ref{eq:architecture-cost}. 
A final pass assigns each remaining workload to its remaining neighbors, in the same order and under the same test.

\subsection{Spectral Matching and Incremental Growth}
\label{sec:generation-matching}

Once workloads are clustered, each cluster's overlay is grown one workload at a time. For that, we study how each operation is described by its surroundings and then match it to an existing overlay slot or grow a new one.

\subsubsection{Spectral Descriptors with Operand Roles}
\label{sec:generation-spectral}

\input{Figures/port-expansion}

After clustering workloads, operation matching is performed to determine which slots and kinds (as defined in \S~\ref{sec:generation-model}) are required for the overlay. In practice, a slot’s kind is fixed at design time, for example, an FU in a given DSP mode, while the operation it performs is chosen per workload mode at runtime. Matching needs to describe how operations connect. Figure~\ref{fig:port-expansion} illustrates how matching expands each operation into a centre vertex and one vertex per port, and each dependency into a vertex that records its word format. After that, we deduce $K$ distinct tags (as defined in \S~\ref{sec:background-gft}) that identify resource kinds, port roles, and word formats on the expanded graph $G$.

To capture how interaction patterns are represented in the graphs, we use spectral analysis (\S~\ref{sec:background-gft}). We treat each tag as a heat source that diffuses along its edges~\cite{shuman2013emerging,hammond2011wavelets} (Section~\ref{sec:background-gft}). With the matrix $X_G$ marking each vertex's tag and $L_G=U_G\Lambda_GU_G^{\mathsf T}$ the Laplacian of the expanded graph, the heat after time $t$ is
\begin{equation}
  Z_G(t)=U_Ge^{-t\Lambda_G}U_G^{\mathsf T}X_G.
  \label{eq:label-heat}
\end{equation}
The row of $Z_G(t)$ at an operation $v$ describes the \emph{diffused heat} of tags that has reached it: at $t=0.1$ it comes mostly from $v$ and its own ports, and by $t=2.0$ it reflects the operations around it. Reading it at $t\in\{0.1,0.5,2.0\}$ gives $v$ a fingerprint $\phi_G(v)$ of its surroundings at three distances (Figure~\ref{fig:spectral-assignment}a). The same construction gives $\phi_{\mathcal A}(s)$ for each slot $s$ of the overlay architecture $\mathcal A$. 

The cost of placing $v$ in an existing slot $s$ is the difference between their neighbor counts, which estimates the connections added or left unused. This quantity is increased by $\lambda\,d_{\mathrm H}(v,s)$, where $d_{\mathrm H}(v,s)$ is the mean squared difference between $\phi_G(v)$ and $\phi_{\mathcal A}(s)$, and $\lambda > 0$. Adding the $\lambda\,d_{\mathrm H}(v,s)$ nudges the selection towards considering far neighboring patterns and not only neighboring connections during the selection. In addition, the factor $\lambda$ is picked so that $\lambda\,d_{\mathrm H}(v,s) < 1$, so that it stays below the equivalent of one connection. This matching method selects only from slots whose wiring fits equally well, favoring the one whose surroundings most closely resemble $v$'s. This is enhanced by the incremental assignment discussed in \S~\ref{sec:generation-growth}.

\subsubsection{Incremental Assignment and Graph Growth}
\label{sec:generation-growth}

\input{Figures/spectral-assignment}

Another descriptor $\mathbf q_v$ counts $v$'s incoming and outgoing edges separately by the neighbor's kind and the destination operand port. Thus, we define $d_{\mathrm D}(v,s)$ as the distance $\|\mathbf q_v-\mathbf q_s\|_1$ for a slot $s$ on $\mathcal A$. The candidates are the existing slots, plus one new slot for each vertex of $G$. A vertex $v$ of kind $\kappa(v)$ takes an existing slot $s$ of the same kind at cost: $$c_{\mathrm{old}}(v,s)=32\,(d_{\mathrm D}(v,s)+4\,d_{\mathrm H}(v,s))$$, or a new slot at the cost: $$c_{\mathrm{new}}(v)=\rho_{\kappa(v)}+32\,\|\mathbf q_v\|_1$$, where $\rho$ is 64 for a constant, 1024 for an FU, 4096 for a URAM load and 2048 for any other kind. Each cost also receives a random offset below $10^{-8}$, so that equal costs maintain the same order in every run. Finally, the Hungarian assignment method~\cite{kuhn1955hungarian} finds the injective, compatible mapping $\mu_G^\star$ that minimizes $\sum_{v\in V_G}c(v,\mu(v))$ (Figure~\ref{fig:spectral-assignment}b). Unused new slots are removed, and the connections become $E_{\mathcal A}=\bigcup_w\{(\mu_w(u),\mu_w(v),p):(u,v,p)\in E_w\}$.

\subsection{Candidate Selection}
\label{sec:generation-selection}

Once all workloads are mapped, the architecture is scored as a whole, including storage and connections. Let $n_{\mathrm C}$ and $n_{\mathrm K}$ count compute and constant slots, and $n_{\mathrm B}$ and $n_{\mathrm U}$ the planned BRAM and URAM blocks, sized for at most 512 frames per mode and taking each resource's maximum across modes. In the connection union, $k_{s,p}$ counts the distinct sources of operand $p$ of slot $s$, and $f_s$ the distinct slot--port pairs that $s$ drives. The terms
\begin{equation}
  \begin{aligned}
    R_{\mathrm{bits}}&=64(|S|-n_{\mathrm{store}}), &
    M_{\mathrm{bits}}&=64\textstyle\sum_{s,p}\max(k_{s,p}-1,0),\\
    D_{\mathrm{mux}}&=\max_{s,p:k_{s,p}>0}\lceil\log_2k_{s,p}\rceil, &
    F_{\mathrm{ex}}&=\textstyle\sum_s\max(f_s-4,0)
  \end{aligned}
  \label{eq:structural-cost-terms}
\end{equation}
estimate result-register bits (the $n_{\mathrm{store}}$ store and scatter slots hold no result), multiplexer bits, multiplexer depth (zero when empty) and fanout beyond four. Under spectral construction, two candidates compete. Each builds the architecture by adding workloads one at a time. The first uses the order in which the workloads were given to the flow, while the second uses a shuffled order. Since workloads are treated one at a time, this can lead to different architectures. 
Each candidate is then augmented by moving an operation to a new slot, to one of up to eight slots of its kind that its workload does not use, or by swapping with a same-kind operation of its workload. The selected candidate is the one that lowers $C$ in Eq.~\ref{eq:architecture-cost}.
\begin{equation}
  \begin{aligned}
    C(\mathcal A)={}&1024n_{\mathrm C}+2048n_{\mathrm B}+4096n_{\mathrm U}+64n_{\mathrm K}\\
      &+R_{\mathrm{bits}}+M_{\mathrm{bits}}+64D_{\mathrm{mux}}+64F_{\mathrm{ex}}.
  \end{aligned}
  \label{eq:architecture-cost}
\end{equation}

%% file: Figures/overlay-shapes.tex
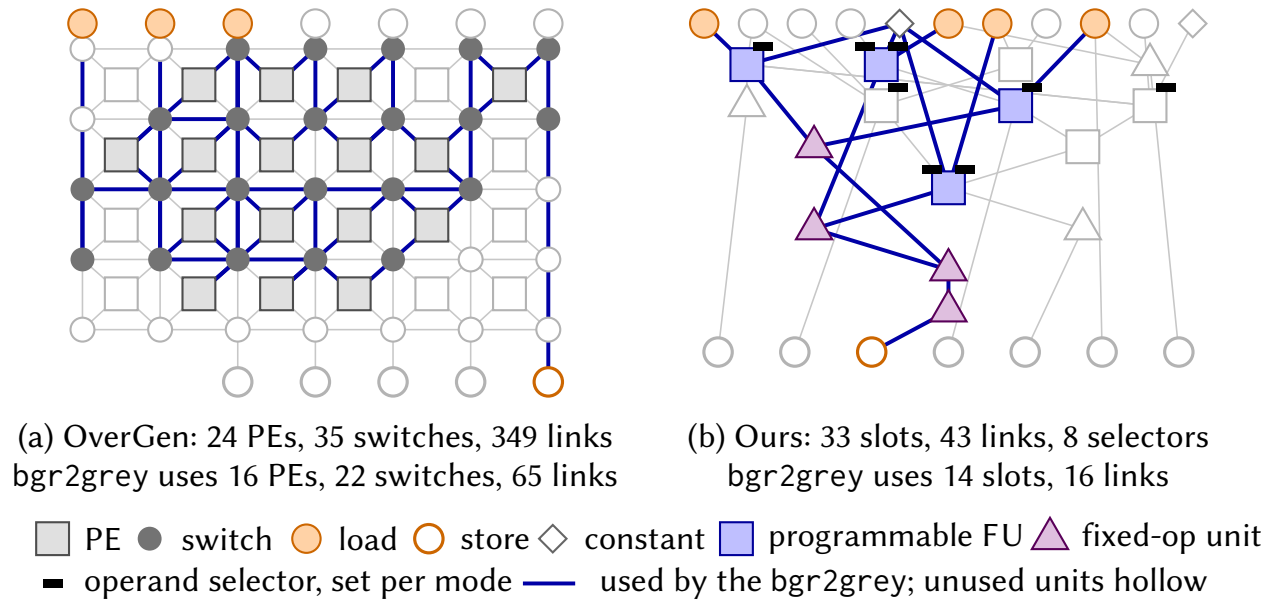
\begin{figure}[t]
  \centering
  \resizebox{\columnwidth}{!}{%
  \begin{tikzpicture}[font=\sffamily\fontsize{6.5}{7.5}\selectfont,
    edgeA/.style={black!35,line width=.3pt},
    edgeB/.style={black!22,line width=.3pt},
    edgeOn/.style={blue!65!black,line width=.75pt},
    idle/.style={fill=white,draw=black!30,text opacity=.3},
    swn/.style={circle,fill=black!55,inner sep=0pt,minimum size=1.6mm},
    pen/.style={rectangle,draw=black!70,fill=black!12,inner sep=0pt,minimum size=2.2mm},
    fun/.style={rectangle,draw=blue!60!black,fill=blue!25,inner sep=0pt,minimum size=2.2mm},
    spn/.style={regular polygon,regular polygon sides=3,draw=violet!70!black,fill=violet!25,inner sep=0pt,minimum size=2.8mm},
    ldn/.style={circle,draw=orange!80!black,fill=orange!35,inner sep=0pt,minimum size=1.9mm},
    stn/.style={circle,draw=orange!80!black,fill=white,line width=.6pt,inner sep=0pt,minimum size=1.9mm},
    cnn/.style={diamond,draw=black!60,fill=white,inner sep=0pt,minimum size=1.9mm},
    seln/.style={rectangle,fill=black,inner sep=0pt,minimum width=1.3mm,minimum height=.6mm}]

    \begin{scope}
      \draw[edgeB] (0.30,0.15) -- (3.42,0.15);
      \draw[edgeB] (0.30,0.62) -- (3.42,0.62);
      \draw[edgeB] (0.30,1.09) -- (3.42,1.09);
      \draw[edgeB] (0.30,1.56) -- (3.42,1.56);
      \draw[edgeB] (0.30,2.03) -- (3.42,2.03);
      \draw[edgeB] (0.30,0.15) -- (0.30,2.03);
      \draw[edgeB] (0.82,0.15) -- (0.82,2.03);
      \draw[edgeB] (1.34,0.15) -- (1.34,2.03);
      \draw[edgeB] (1.86,0.15) -- (1.86,2.03);
      \draw[edgeB] (2.38,0.15) -- (2.38,2.03);
      \draw[edgeB] (2.90,0.15) -- (2.90,2.03);
      \draw[edgeB] (3.42,0.15) -- (3.42,2.03);
      \draw[edgeB] (0.56,0.39) -- (0.30,0.15);
      \draw[edgeB] (0.56,0.39) -- (0.82,0.15);
      \draw[edgeB] (0.56,0.39) -- (0.30,0.62);
      \draw[edgeB] (0.56,0.39) -- (0.82,0.62);
      \draw[edgeB] (0.56,0.85) -- (0.30,0.62);
      \draw[edgeB] (0.56,0.85) -- (0.82,0.62);
      \draw[edgeB] (0.56,0.85) -- (0.30,1.09);
      \draw[edgeB] (0.56,0.85) -- (0.82,1.09);
      \draw[edgeB] (0.56,1.32) -- (0.30,1.09);
      \draw[edgeB] (0.56,1.32) -- (0.82,1.09);
      \draw[edgeB] (0.56,1.32) -- (0.30,1.56);
      \draw[edgeB] (0.56,1.32) -- (0.82,1.56);
      \draw[edgeB] (0.56,1.79) -- (0.30,1.56);
      \draw[edgeB] (0.56,1.79) -- (0.82,1.56);
      \draw[edgeB] (0.56,1.79) -- (0.30,2.03);
      \draw[edgeB] (0.56,1.79) -- (0.82,2.03);
      \draw[edgeB] (1.08,0.39) -- (0.82,0.15);
      \draw[edgeB] (1.08,0.39) -- (1.34,0.15);
      \draw[edgeB] (1.08,0.39) -- (0.82,0.62);
      \draw[edgeB] (1.08,0.39) -- (1.34,0.62);
      \draw[edgeB] (1.08,0.85) -- (0.82,0.62);
      \draw[edgeB] (1.08,0.85) -- (1.34,0.62);
      \draw[edgeB] (1.08,0.85) -- (0.82,1.09);
      \draw[edgeB] (1.08,0.85) -- (1.34,1.09);
      \draw[edgeB] (1.08,1.32) -- (0.82,1.09);
      \draw[edgeB] (1.08,1.32) -- (1.34,1.09);
      \draw[edgeB] (1.08,1.32) -- (0.82,1.56);
      \draw[edgeB] (1.08,1.32) -- (1.34,1.56);
      \draw[edgeB] (1.08,1.79) -- (0.82,1.56);
      \draw[edgeB] (1.08,1.79) -- (1.34,1.56);
      \draw[edgeB] (1.08,1.79) -- (0.82,2.03);
      \draw[edgeB] (1.08,1.79) -- (1.34,2.03);
      \draw[edgeB] (1.60,0.39) -- (1.34,0.15);
      \draw[edgeB] (1.60,0.39) -- (1.86,0.15);
      \draw[edgeB] (1.60,0.39) -- (1.34,0.62);
      \draw[edgeB] (1.60,0.39) -- (1.86,0.62);
      \draw[edgeB] (1.60,0.85) -- (1.34,0.62);
      \draw[edgeB] (1.60,0.85) -- (1.86,0.62);
      \draw[edgeB] (1.60,0.85) -- (1.34,1.09);
      \draw[edgeB] (1.60,0.85) -- (1.86,1.09);
      \draw[edgeB] (1.60,1.32) -- (1.34,1.09);
      \draw[edgeB] (1.60,1.32) -- (1.86,1.09);
      \draw[edgeB] (1.60,1.32) -- (1.34,1.56);
      \draw[edgeB] (1.60,1.32) -- (1.86,1.56);
      \draw[edgeB] (1.60,1.79) -- (1.34,1.56);
      \draw[edgeB] (1.60,1.79) -- (1.86,1.56);
      \draw[edgeB] (1.60,1.79) -- (1.34,2.03);
      \draw[edgeB] (1.60,1.79) -- (1.86,2.03);
      \draw[edgeB] (2.12,0.39) -- (1.86,0.15);
      \draw[edgeB] (2.12,0.39) -- (2.38,0.15);
      \draw[edgeB] (2.12,0.39) -- (1.86,0.62);
      \draw[edgeB] (2.12,0.39) -- (2.38,0.62);
      \draw[edgeB] (2.12,0.85) -- (1.86,0.62);
      \draw[edgeB] (2.12,0.85) -- (2.38,0.62);
      \draw[edgeB] (2.12,0.85) -- (1.86,1.09);
      \draw[edgeB] (2.12,0.85) -- (2.38,1.09);
      \draw[edgeB] (2.12,1.32) -- (1.86,1.09);
      \draw[edgeB] (2.12,1.32) -- (2.38,1.09);
      \draw[edgeB] (2.12,1.32) -- (1.86,1.56);
      \draw[edgeB] (2.12,1.32) -- (2.38,1.56);
      \draw[edgeB] (2.12,1.79) -- (1.86,1.56);
      \draw[edgeB] (2.12,1.79) -- (2.38,1.56);
      \draw[edgeB] (2.12,1.79) -- (1.86,2.03);
      \draw[edgeB] (2.12,1.79) -- (2.38,2.03);
      \draw[edgeB] (2.64,0.39) -- (2.38,0.15);
      \draw[edgeB] (2.64,0.39) -- (2.90,0.15);
      \draw[edgeB] (2.64,0.39) -- (2.38,0.62);
      \draw[edgeB] (2.64,0.39) -- (2.90,0.62);
      \draw[edgeB] (2.64,0.85) -- (2.38,0.62);
      \draw[edgeB] (2.64,0.85) -- (2.90,0.62);
      \draw[edgeB] (2.64,0.85) -- (2.38,1.09);
      \draw[edgeB] (2.64,0.85) -- (2.90,1.09);
      \draw[edgeB] (2.64,1.32) -- (2.38,1.09);
      \draw[edgeB] (2.64,1.32) -- (2.90,1.09);
      \draw[edgeB] (2.64,1.32) -- (2.38,1.56);
      \draw[edgeB] (2.64,1.32) -- (2.90,1.56);
      \draw[edgeB] (2.64,1.79) -- (2.38,1.56);
      \draw[edgeB] (2.64,1.79) -- (2.90,1.56);
      \draw[edgeB] (2.64,1.79) -- (2.38,2.03);
      \draw[edgeB] (2.64,1.79) -- (2.90,2.03);
      \draw[edgeB] (3.16,0.39) -- (2.90,0.15);
      \draw[edgeB] (3.16,0.39) -- (3.42,0.15);
      \draw[edgeB] (3.16,0.39) -- (2.90,0.62);
      \draw[edgeB] (3.16,0.39) -- (3.42,0.62);
      \draw[edgeB] (3.16,0.85) -- (2.90,0.62);
      \draw[edgeB] (3.16,0.85) -- (3.42,0.62);
      \draw[edgeB] (3.16,0.85) -- (2.90,1.09);
      \draw[edgeB] (3.16,0.85) -- (3.42,1.09);
      \draw[edgeB] (3.16,1.32) -- (2.90,1.09);
      \draw[edgeB] (3.16,1.32) -- (3.42,1.09);
      \draw[edgeB] (3.16,1.32) -- (2.90,1.56);
      \draw[edgeB] (3.16,1.32) -- (3.42,1.56);
      \draw[edgeB] (3.16,1.79) -- (2.90,1.56);
      \draw[edgeB] (3.16,1.79) -- (3.42,1.56);
      \draw[edgeB] (3.16,1.79) -- (2.90,2.03);
      \draw[edgeB] (3.16,1.79) -- (3.42,2.03);
      \draw[edgeB] (0.30,2.2) -- (0.30,2.03);
      \draw[edgeB] (0.82,2.2) -- (0.82,2.03);
      \draw[edgeB] (1.34,2.2) -- (1.34,2.03);
      \draw[edgeB] (1.86,2.2) -- (1.86,2.03);
      \draw[edgeB] (2.38,2.2) -- (2.38,2.03);
      \draw[edgeB] (2.90,2.2) -- (2.90,2.03);
      \draw[edgeB] (3.42,2.2) -- (3.42,2.03);
      \draw[edgeB] (3.42,-.2) -- (3.42,0.15);
      \draw[edgeB] (2.90,-.2) -- (2.90,0.15);
      \draw[edgeB] (2.38,-.2) -- (2.38,0.15);
      \draw[edgeB] (1.86,-.2) -- (1.86,0.15);
      \draw[edgeB] (1.34,-.2) -- (1.34,0.15);
      \draw[edgeOn] (0.30,2.20) -- (0.30,0.62);
      \draw[edgeOn] (0.82,2.20) -- (0.82,0.62);
      \draw[edgeOn] (1.34,2.20) -- (1.34,0.62);
      \draw[edgeOn] (1.34,2.20) -- (1.34,1.09);
      \draw[edgeOn] (1.34,2.20) -- (1.34,1.56);
      \draw[edgeOn] (1.34,2.20) -- (1.34,2.03);
      \draw[edgeOn] (3.16,1.79) -- (3.42,2.03);
      \draw[edgeOn] (2.12,0.85) -- (2.38,1.09);
      \draw[edgeOn] (2.12,0.39) -- (2.38,0.62);
      \draw[edgeOn] (1.60,1.79) -- (1.86,2.03);
      \draw[edgeOn] (1.60,1.32) -- (1.86,1.56);
      \draw[edgeOn] (1.60,0.85) -- (1.86,1.09);
      \draw[edgeOn] (1.60,0.39) -- (1.86,0.62);
      \draw[edgeOn] (1.08,1.79) -- (1.34,2.03);
      \draw[edgeOn] (1.08,1.32) -- (1.34,1.56);
      \draw[edgeOn] (1.08,0.85) -- (1.34,1.09);
      \draw[edgeOn] (1.08,0.39) -- (1.34,0.62);
      \draw[edgeOn] (0.56,1.32) -- (0.82,1.56);
      \draw[edgeOn] (2.64,1.32) -- (2.90,1.56);
      \draw[edgeOn] (2.64,0.85) -- (2.90,1.09);
      \draw[edgeOn] (2.12,1.79) -- (2.38,2.03);
      \draw[edgeOn] (2.12,1.32) -- (2.38,1.56);
      \draw[edgeOn] (3.42,2.03) -- (3.42,1.56);
      \draw[edgeOn] (3.42,1.56) -- (3.42,-0.20);
      \draw[edgeOn] (2.38,2.03) -- (2.38,1.56);
      \draw[edgeOn] (2.38,1.56) -- (2.12,1.32);
      \draw[edgeOn] (2.38,1.56) -- (2.64,1.32);
      \draw[edgeOn] (2.38,1.09) -- (2.64,0.85);
      \draw[edgeOn] (2.38,1.09) -- (2.90,1.09);
      \draw[edgeOn] (2.38,0.62) -- (2.64,0.85);
      \draw[edgeOn] (1.86,2.03) -- (1.86,1.56);
      \draw[edgeOn] (1.86,1.56) -- (2.12,1.32);
      \draw[edgeOn] (1.86,1.56) -- (2.12,1.79);
      \draw[edgeOn] (1.86,1.09) -- (1.86,0.62);
      \draw[edgeOn] (1.86,1.09) -- (2.38,1.09);
      \draw[edgeOn] (1.86,0.62) -- (1.60,0.39);
      \draw[edgeOn] (1.86,0.62) -- (2.12,0.39);
      \draw[edgeOn] (1.86,0.62) -- (2.12,0.85);
      \draw[edgeOn] (1.34,2.03) -- (1.34,1.56);
      \draw[edgeOn] (1.34,2.03) -- (1.60,1.79);
      \draw[edgeOn] (1.34,1.56) -- (0.82,1.56);
      \draw[edgeOn] (1.34,1.56) -- (1.08,1.32);
      \draw[edgeOn] (1.34,1.56) -- (1.60,1.32);
      \draw[edgeOn] (1.34,1.09) -- (1.34,0.62);
      \draw[edgeOn] (1.34,1.09) -- (1.60,0.85);
      \draw[edgeOn] (1.34,1.09) -- (1.86,1.09);
      \draw[edgeOn] (1.34,0.62) -- (1.08,0.39);
      \draw[edgeOn] (1.34,0.62) -- (1.08,0.85);
      \draw[edgeOn] (1.34,0.62) -- (1.86,0.62);
      \draw[edgeOn] (1.34,0.62) -- (1.60,0.85);
      \draw[edgeOn] (0.82,1.56) -- (0.56,1.32);
      \draw[edgeOn] (0.82,1.56) -- (1.08,1.79);
      \draw[edgeOn] (0.82,1.09) -- (0.56,1.32);
      \draw[edgeOn] (0.82,1.09) -- (1.34,1.09);
      \draw[edgeOn] (0.82,1.09) -- (1.08,1.32);
      \draw[edgeOn] (0.82,0.62) -- (0.82,1.09);
      \draw[edgeOn] (0.82,0.62) -- (1.34,0.62);
      \draw[edgeOn] (0.82,0.62) -- (1.08,0.85);
      \draw[edgeOn] (0.30,1.09) -- (0.82,1.09);
      \draw[edgeOn] (0.30,0.62) -- (0.30,1.09);
      \draw[edgeOn] (2.90,2.03) -- (3.16,1.79);
      \draw[edgeOn] (2.90,1.56) -- (2.90,2.03);
      \draw[edgeOn] (2.90,1.56) -- (3.16,1.79);
      \draw[edgeOn] (2.90,1.09) -- (2.64,1.32);
      \draw[edgeOn] (2.90,1.09) -- (2.90,1.56);
      \node[swn,idle] at (0.30,0.15) {};
      \node[swn] at (0.30,0.62) {};
      \node[swn] at (0.30,1.09) {};
      \node[swn,idle] at (0.30,1.56) {};
      \node[swn,idle] at (0.30,2.03) {};
      \node[swn,idle] at (0.82,0.15) {};
      \node[swn] at (0.82,0.62) {};
      \node[swn] at (0.82,1.09) {};
      \node[swn] at (0.82,1.56) {};
      \node[swn,idle] at (0.82,2.03) {};
      \node[swn,idle] at (1.34,0.15) {};
      \node[swn] at (1.34,0.62) {};
      \node[swn] at (1.34,1.09) {};
      \node[swn] at (1.34,1.56) {};
      \node[swn] at (1.34,2.03) {};
      \node[swn,idle] at (1.86,0.15) {};
      \node[swn] at (1.86,0.62) {};
      \node[swn] at (1.86,1.09) {};
      \node[swn] at (1.86,1.56) {};
      \node[swn] at (1.86,2.03) {};
      \node[swn,idle] at (2.38,0.15) {};
      \node[swn] at (2.38,0.62) {};
      \node[swn] at (2.38,1.09) {};
      \node[swn] at (2.38,1.56) {};
      \node[swn] at (2.38,2.03) {};
      \node[swn,idle] at (2.90,0.15) {};
      \node[swn,idle] at (2.90,0.62) {};
      \node[swn] at (2.90,1.09) {};
      \node[swn] at (2.90,1.56) {};
      \node[swn] at (2.90,2.03) {};
      \node[swn,idle] at (3.42,0.15) {};
      \node[swn,idle] at (3.42,0.62) {};
      \node[swn,idle] at (3.42,1.09) {};
      \node[swn] at (3.42,1.56) {};
      \node[swn] at (3.42,2.03) {};
      \node[pen,idle] at (0.56,0.39) {};
      \node[pen,idle] at (0.56,0.85) {};
      \node[pen] at (0.56,1.32) {};
      \node[pen,idle] at (0.56,1.79) {};
      \node[pen] at (1.08,0.39) {};
      \node[pen] at (1.08,0.85) {};
      \node[pen] at (1.08,1.32) {};
      \node[pen] at (1.08,1.79) {};
      \node[pen] at (1.60,0.39) {};
      \node[pen] at (1.60,0.85) {};
      \node[pen] at (1.60,1.32) {};
      \node[pen] at (1.60,1.79) {};
      \node[pen] at (2.12,0.39) {};
      \node[pen] at (2.12,0.85) {};
      \node[pen] at (2.12,1.32) {};
      \node[pen] at (2.12,1.79) {};
      \node[pen,idle] at (2.64,0.39) {};
      \node[pen] at (2.64,0.85) {};
      \node[pen] at (2.64,1.32) {};
      \node[pen,idle] at (2.64,1.79) {};
      \node[pen,idle] at (3.16,0.39) {};
      \node[pen,idle] at (3.16,0.85) {};
      \node[pen,idle] at (3.16,1.32) {};
      \node[pen] at (3.16,1.79) {};
      \node[ldn] at (0.30,2.2) {};
      \node[ldn] at (0.82,2.2) {};
      \node[ldn] at (1.34,2.2) {};
      \node[ldn,idle] at (1.86,2.2) {};
      \node[ldn,idle] at (2.38,2.2) {};
      \node[ldn,idle] at (2.90,2.2) {};
      \node[ldn,idle] at (3.42,2.2) {};
      \node[stn] at (3.42,-.2) {};
      \node[stn,idle] at (2.90,-.2) {};
      \node[stn,idle] at (2.38,-.2) {};
      \node[stn,idle] at (1.86,-.2) {};
      \node[stn,idle] at (1.34,-.2) {};
      \node[anchor=north,align=center] at (1.86,-.36) {(a) OverGen: 24 PEs, 35 switches, 349 links\\\texttt{bgr2grey} uses 16 PEs, 22 switches, 65 links};
    \end{scope}

    \begin{scope}[shift={(4.3,0)}]
      \draw[edgeB] (1.35,1.65) -- (0.771,0.0);
      \draw[edgeB] (1.35,1.65) -- (1.8,1.1);
      \draw[edgeB] (0.818,2.2) -- (1.35,1.65);
      \draw[edgeB] (0.45,1.925) -- (0.45,1.65);
      \draw[edgeB] (0.45,1.925) -- (3.15,1.65);
      \draw[edgeB] (0.45,1.65) -- (0.257,0.0);
      \draw[edgeB] (3.109,2.2) -- (3.15,1.65);
      \draw[edgeB] (2.782,2.2) -- (3.15,1.925);
      \draw[edgeB] (2.782,2.2) -- (2.829,0.0);
      \draw[edgeB] (1.35,1.925) -- (2.25,1.65);
      \draw[edgeB] (1.8,1.1) -- (2.7,0.825);
      \draw[edgeB] (2.25,1.65) -- (2.7,1.375);
      \draw[edgeB] (2.25,1.65) -- (1.8,0.0);
      \draw[edgeB] (2.7,0.825) -- (2.314,0.0);
      \draw[edgeB] (1.8,2.2) -- (1.35,1.925);
      \draw[edgeB] (1.8,2.2) -- (3.15,1.925);
      \draw[edgeB] (2.455,2.2) -- (2.25,1.925);
      \draw[edgeB] (2.127,2.2) -- (2.25,1.925);
      \draw[edgeB] (0.491,2.2) -- (1.35,1.65);
      \draw[edgeB] (0.491,2.2) -- (0.45,1.925);
      \draw[edgeB] (1.145,2.2) -- (1.35,1.925);
      \draw[edgeB] (2.7,1.375) -- (1.8,1.1);
      \draw[edgeB] (3.15,1.65) -- (0.45,1.925);
      \draw[edgeB] (3.15,1.65) -- (2.7,1.375);
      \draw[edgeB] (2.25,1.925) -- (1.35,1.65);
      \draw[edgeB] (3.15,1.925) -- (3.343,0.0);
      \draw[edgeB] (3.436,2.2) -- (3.15,1.65);
      \draw[edgeOn] (0.164,2.2) -- (0.45,1.925);
      \draw[edgeOn] (0.45,1.925) -- (0.9,1.375);
      \draw[edgeOn] (1.473,2.2) -- (0.45,1.925);
      \draw[edgeOn] (1.473,2.2) -- (1.35,1.925);
      \draw[edgeOn] (1.473,2.2) -- (1.8,1.1);
      \draw[edgeOn] (1.473,2.2) -- (2.25,1.65);
      \draw[edgeOn] (1.8,0.55) -- (1.8,0.275);
      \draw[edgeOn] (1.8,0.275) -- (1.286,0.0);
      \draw[edgeOn] (0.9,0.825) -- (1.8,0.55);
      \draw[edgeOn] (0.9,1.375) -- (1.8,0.55);
      \draw[edgeOn] (2.782,2.2) -- (2.25,1.65);
      \draw[edgeOn] (1.35,1.925) -- (0.9,0.825);
      \draw[edgeOn] (1.8,1.1) -- (0.9,0.825);
      \draw[edgeOn] (2.25,1.65) -- (0.9,1.375);
      \draw[edgeOn] (1.8,2.2) -- (1.35,1.925);
      \draw[edgeOn] (2.127,2.2) -- (1.8,1.1);
      \node[ldn] at (0.164,2.2) {};
      \node[stn,idle] at (0.771,0.0) {};
      \node[fun,idle] at (1.35,1.65) {};
      \node[ldn,idle] at (0.818,2.2) {};
      \node[fun] at (0.45,1.925) {};
      \node[cnn] at (1.473,2.2) {};
      \node[spn,idle] at (0.45,1.65) {};
      \node[spn] at (1.8,0.55) {};
      \node[spn] at (1.8,0.275) {};
      \node[spn] at (0.9,0.825) {};
      \node[spn] at (0.9,1.375) {};
      \node[ldn,idle] at (3.109,2.2) {};
      \node[ldn] at (2.782,2.2) {};
      \node[fun] at (1.35,1.925) {};
      \node[fun] at (1.8,1.1) {};
      \node[fun] at (2.25,1.65) {};
      \node[spn,idle] at (2.7,0.825) {};
      \node[ldn] at (1.8,2.2) {};
      \node[ldn,idle] at (2.455,2.2) {};
      \node[ldn] at (2.127,2.2) {};
      \node[ldn,idle] at (0.491,2.2) {};
      \node[ldn,idle] at (1.145,2.2) {};
      \node[fun,idle] at (2.7,1.375) {};
      \node[fun,idle] at (3.15,1.65) {};
      \node[fun,idle] at (2.25,1.925) {};
      \node[spn,idle] at (3.15,1.925) {};
      \node[stn,idle] at (2.314,0.0) {};
      \node[stn,idle] at (0.257,0.0) {};
      \node[stn,idle] at (3.343,0.0) {};
      \node[cnn,idle] at (3.436,2.2) {};
      \node[stn,idle] at (1.8,0.0) {};
      \node[stn] at (1.286,0.0) {};
      \node[stn,idle] at (2.829,0.0) {};
      \node[seln] at (1.910,1.220) {};
      \node[seln] at (3.260,1.770) {};
      \node[seln] at (0.560,2.045) {};
      \node[seln] at (1.460,2.045) {};
      \node[seln] at (2.360,1.770) {};
      \node[seln] at (1.240,2.045) {};
      \node[seln] at (1.690,1.220) {};
      \node[seln] at (1.460,1.770) {};

      \node[anchor=north,align=center] at (1.8,-.36) {(b) Ours: 33 slots, 43 links, 8 selectors\\\texttt{bgr2grey} uses 14 slots, 16 links};
    \end{scope}

    \begin{scope}[shift={(0,-1.25)}]
      \node[pen] at (.1,0) {}; \node[anchor=west] at (.2,0) {PE};
      \node[swn] at (.75,0) {}; \node[anchor=west] at (.85,0) {switch};
      \node[ldn] at (1.8,0) {}; \node[anchor=west] at (1.9,0) {load};
      \node[stn] at (2.62,0) {}; \node[anchor=west] at (2.72,0) {store};
      \node[cnn] at (3.45,0) {}; \node[anchor=west] at (3.55,0) {constant};
      \node[fun] at (4.68,0) {}; \node[anchor=west] at (4.78,0) {programmable FU};
      \node[spn] at (6.78,0) {}; \node[anchor=west] at (6.88,0) {fixed-op unit};
      \node[seln] at (.1,-.3) {}; \node[anchor=west] at (.2,-.3) {operand selector, set per mode};
      \draw[edgeOn] (3.25,-.3) -- (3.6,-.3); \node[anchor=west] at (3.65,-.3) {used by the \texttt{bgr2grey}; unused units hollow};
    \end{scope}

  \end{tikzpicture}%
  }
  \caption{OverGen (a) vs.\ our overlay (b), one mode in bold.}
  \label{fig:overlay-shapes}
\end{figure}

%% file: Figures/native-programmable-fu.tex
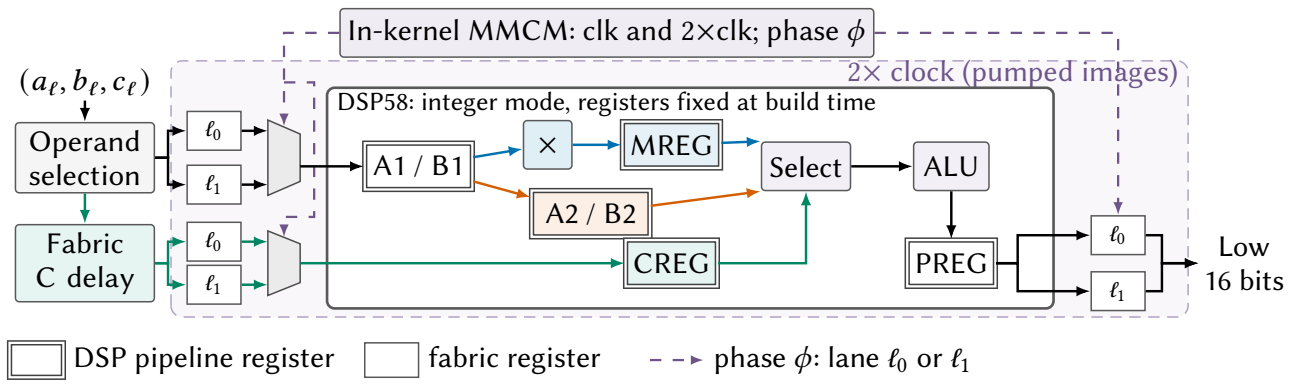
\begin{figure}[t]
  \centering
  \definecolor{fuBlue}{HTML}{0072B2}
  \definecolor{fuOrange}{HTML}{D55E00}
  \definecolor{fuTeal}{HTML}{009E73}
  \definecolor{fuPurple}{HTML}{8064A2}
  \resizebox{\columnwidth}{!}{%
  \begin{tikzpicture}[
    font=\sffamily\fontsize{8}{9}\selectfont,
    flow/.style={-{Latex[length=1.3mm]},line width=.65pt},
    control/.style={-{Latex[length=1.2mm]},draw=fuPurple!85!black,dashed,line width=.6pt},
    block/.style={draw=black!65,rounded corners=1pt,fill=black!4,align=center,minimum height=4.5mm,inner sep=2pt},
    reg/.style={draw=black!75,double,double distance=.6pt,fill=white,align=center,minimum height=4.3mm,inner sep=2pt},
    freg/.style={draw=black!65,fill=white,align=center,minimum height=3.6mm,minimum width=5mm,inner sep=1pt,font=\sffamily\fontsize{6.5}{7}\selectfont},
    mux/.style={trapezium,trapezium angle=70,shape border rotate=270,draw=black!70,fill=black!8,minimum width=6.5mm,minimum height=2.6mm,inner sep=0pt},
    note/.style={align=center,inner sep=1pt}
  ]

    \node[note] (args) at (.75,2.2) {$(a_\ell,b_\ell,c_\ell)$};
    \node[block,text width=1.15cm] (operands) at (.75,1.5) {Operand\\selection};
    \node[block,text width=1.15cm,fill=fuTeal!10] (cdelay) at (.75,.52) {Fabric\\C delay};
    \draw[flow] (args) -- (operands);
    \draw[flow,draw=fuTeal!85!black] (operands) -- (cdelay);

    \path[fill=fuPurple!5,draw=fuPurple!60,densely dashed,rounded corners=3pt] (1.55,.02) rectangle (11.0,2.4);
    \node[note,anchor=north east,text=fuPurple!85!black] at (10.95,2.45) {$2\times$ clock (pumped images)};
    \node[block,fill=fuPurple!12] (mmcm) at (5.6,2.66) {In-kernel MMCM: clk and $2\times$clk; phase $\phi$};

    \node[freg] (ra0) at (1.95,1.75) {$\ell_0$};
    \node[freg] (ra1) at (1.95,1.25) {$\ell_1$};
    \draw[draw=black!70,fill=black!8] (2.45,1.85) -- (2.75,1.72) -- (2.75,1.28) -- (2.45,1.15) -- cycle;
    \node[freg] (rc0) at (1.95,.72) {$\ell_0$};
    \node[freg] (rc1) at (1.95,.32) {$\ell_1$};
    \draw[draw=black!70,fill=black!8] (2.45,.82) -- (2.75,.70) -- (2.75,.34) -- (2.45,.22) -- cycle;
    \draw[flow] (operands.east) -- ++(.12,0) |- (ra0.west);
    \draw[flow] (operands.east) -- ++(.12,0) |- (ra1.west);
    \draw[flow,draw=fuTeal!85!black] (cdelay.east) -- ++(.12,0) |- (rc0.west);
    \draw[flow,draw=fuTeal!85!black] (cdelay.east) -- ++(.12,0) |- (rc1.west);
    \draw[flow] (ra0.east) -- (2.45,1.75);
    \draw[flow] (ra1.east) -- (2.45,1.25);
    \draw[flow,draw=fuTeal!85!black] (rc0.east) -- (2.45,.72);
    \draw[flow,draw=fuTeal!85!black] (rc1.east) -- (2.45,.32);
    \draw[control] (mmcm.west) -| (2.6,1.79);
    \draw[control] (2.6,2.2) -- (2.88,2.2) -- (2.88,.95) -- (2.6,.95) -- (2.6,.76);

    \draw[draw=black!65,rounded corners=2pt,line width=.7pt,fill=white] (3.0,.1) rectangle (9.75,2.15);
    \node[note,anchor=north west,font=\sffamily\fontsize{7}{8}\selectfont] at (3.05,2.15) {DSP58: integer mode, registers fixed at build time};

    \node[reg,minimum width=1cm] (ab1) at (3.85,1.42) {A1 / B1};
    \node[block,fill=fuBlue!12,minimum width=4.5mm] (mul) at (5.05,1.62) {$\times$};
    \node[reg,fill=fuBlue!10] (mreg) at (6.2,1.62) {MREG};
    \node[reg,fill=fuOrange!10,minimum width=1.1cm] (ab2) at (5.45,.98) {A2 / B2};
    \node[reg,fill=fuTeal!10] (creg) at (6.2,.52) {CREG};
    \node[block,fill=fuPurple!10] (select) at (7.45,1.42) {Select};
    \node[block,fill=fuPurple!10,minimum width=7mm] (alu) at (8.8,1.42) {ALU};
    \node[reg] (preg) at (8.8,.52) {PREG};

    \draw[flow] (2.75,1.42) -- (ab1.west);
    \draw[flow,draw=fuBlue] (ab1) -- (mul);
    \draw[flow,draw=fuBlue] (mul) -- (mreg);
    \draw[flow,draw=fuBlue] (mreg) -- (select.north west);
    \draw[flow,draw=fuOrange] (ab1) -- (ab2);
    \draw[flow,draw=fuOrange] (ab2) -- (select.south west);
    \draw[flow,draw=fuTeal!85!black] (2.75,.52) -- (creg.west);
    \draw[flow,draw=fuTeal!85!black] (creg) -- (7.45,.52) -- (select.south);
    \draw[flow] (select) -- (alu);
    \draw[flow] (alu) -- (preg);

    \node[freg] (r0) at (10.35,.78) {$\ell_0$};
    \node[freg] (r1) at (10.35,.26) {$\ell_1$};
    \draw[flow] (preg.east) -- ++(.2,0) |- (r0.west);
    \draw[flow] (preg.east) -- ++(.2,0) |- (r1.west);
    \node[note,anchor=west] (out) at (11.1,.52) {Low\\16 bits};
    \draw[flow] (r0.east) -- ++(.15,0) |- (out.west);
    \draw[flow] (r1.east) -- ++(.15,0) |- (out.west);
    \draw[control] (mmcm.east) -| (r0.north);

    \begin{scope}[shift={(0,-.38)}]
      \node[reg,minimum width=5mm,minimum height=3.2mm,inner sep=0pt] at (.3,0) {};
      \node[anchor=west,note] at (.6,0) {DSP pipeline register};
      \node[freg,minimum height=3.2mm,inner sep=0pt] at (3.6,0) {};
      \node[anchor=west,note] at (3.9,0) {fabric register};
      \draw[control] (6.0,0) -- (6.5,0); \node[anchor=west,note] at (6.55,0) {phase $\phi$: lane $\ell_0$ or $\ell_1$};
    \end{scope}

  \end{tikzpicture}%
  }
  \caption{DSP58~\cite{amd2022am004} pipeline registers and double pumping.}
  \Description{Operand selection and a fabric C delay feed one DSP58. Inside it, registers A1/B1, A2/B2, MREG, CREG and PREG are enabled at build time and form the native pipeline. In pumped images the DSP runs on a doubled clock from an in-kernel MMCM: the operands of two lanes are captured by fabric registers, a 2:1 multiplexer driven by the phase signal feeds one lane per half slow cycle, and two phase-enabled registers return the two results to the slow clock, whose low 16 bits form each lane's result.}
  \label{fig:native-programmable-fu}
\end{figure}

%% file: Tables/programmable-fu-ops.tex
\begin{table}[t]
  \centering
  \footnotesize
  \setlength{\tabcolsep}{3pt}
  \caption{Operations by data type.}
  \label{tab:programmable-fu-ops}
  \begin{tabular*}{\columnwidth}{@{\extracolsep{\fill}}lll@{}}
    \toprule
    Unit & Operations & Result \\
    \midrule
    i16$\times$4 FU       & MADD, MUL, ADD, SUB      & $ab{+}c$, $ab$, $a{\pm}b$, mod $2^{16}$ \\
                          & AND, OR, XOR, PASS       & bitwise; $a$ \\
    i16$\times$4 fixed-op & MAX, DIV$_k$             & signed $\max(a,b)$; $\operatorname{trunc}(a/k)$ \\
                          & HSUM, JOIN               & lane sums; packed low halves \\
    i64 engines           & ADD, MUL, LT             & $a{+}b$, $ab$ mod $2^{64}$; signed $a{<}b$ \\
    f64 FP cores          & ADD, SUB, MUL, DIV, SQRT & each rounded to nearest even \\
    f32 FP cores          & ADD, SUB, MUL, NEG       & rounded; $-a$ exact \\
    \bottomrule
  \end{tabular*}
\end{table}

%% file: Figures/native-memory.tex
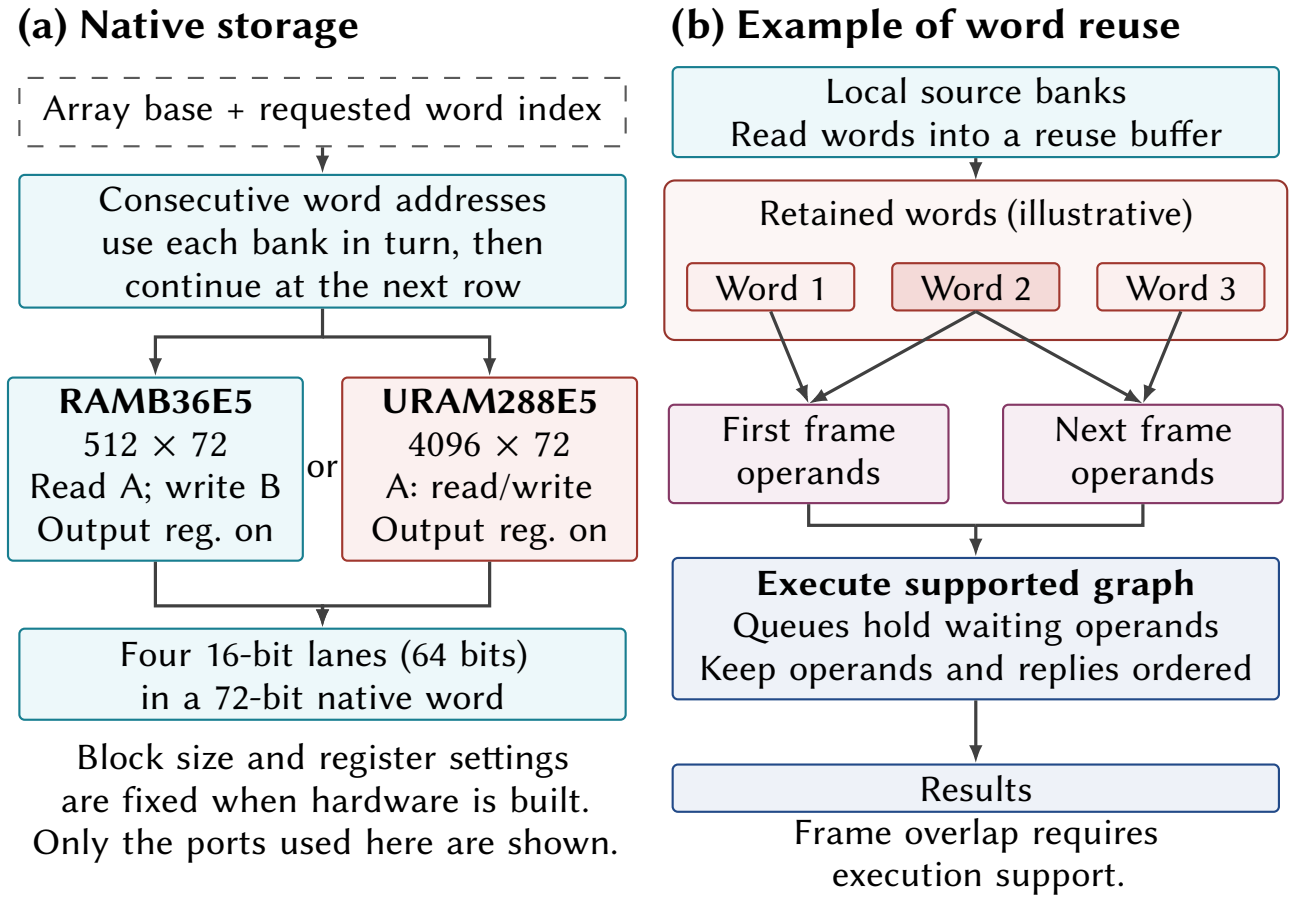
\begin{figure}[t]
  \centering
  \resizebox{\columnwidth}{!}{%
  \begin{tikzpicture}[grcFigure,font=\sffamily\fontsize{7}{8}\selectfont]

    \node[grcTitle,anchor=west] at (0,5.05) {(a) Native storage};
    \node[grcTitle,anchor=west] at (4.3,5.05) {(b) Example of word reuse};

    \node[grcInput,text width=3.85cm,minimum height=4.5mm] (logical) at (2.05,4.5) {Array base + requested word index};
    \node[grcMemory,text width=3.85cm] (mapping) at (2.05,3.65) {Consecutive word addresses\\use each bank in turn, then\\continue at the next row};

    \node[grcMemory,text width=1.8cm] (bram) at (.95,2.15) {\textbf{RAMB36E5}\\$512\times72$\\Read A; write B\\Output reg. on};
    \node[grcSystem,text width=1.8cm] (uram) at (3.15,2.15) {\textbf{URAM288E5}\\$4096\times72$\\A: read/write\\Output reg. on};
    \node[grcNote] at (2.05,2.15) {or};

    \node[grcMemory,text width=3.85cm] (word) at (2.05,.8) {Four 16-bit lanes (64 bits)\\in a 72-bit native word};

    \draw[grcArrow] (logical) -- (mapping);
    \draw[grcArrow] (mapping.south) -- (2.05,3.02) -| (bram.north);
    \draw[grcArrow] (2.05,3.02) -| (uram.north);
    \draw[grcArrow] (bram.south) -- (.95,1.25) -| (word.north);
    \draw[grcArrow] (uram.south) -- (3.15,1.25) -| (word.north);

    \node[grcNote,text width=4cm] at (2.05,-.05) {Block size and register settings\\are fixed when hardware is built.\\Only the ports used here are shown.};

    \node[grcMemory,text width=3.85cm] (source) at (6.35,4.5) {Local source banks\\Read words into a reuse buffer};

    \draw[draw=grcOrange!75!black,fill=grcOrange!5,rounded corners=2pt] (4.3,3.0) rectangle (8.4,4.05);
    \node[grcNote] at (6.35,3.82) {Retained words (illustrative)};

    \node[grcSystem,minimum width=1.1cm] (w1) at (5.0,3.35) {Word 1};
    \node[grcSystem,minimum width=1.1cm,fill=grcOrange!20] (w2) at (6.35,3.35) {Word 2};
    \node[grcSystem,minimum width=1.1cm] (w3) at (7.7,3.35) {Word 3};

    \draw[grcArrow] (source.south) -- (6.35,4.05);

    \node[grcTyped,text width=1.7cm] (first) at (5.25,2.25) {First frame\\operands};
    \node[grcTyped,text width=1.7cm] (next) at (7.45,2.25) {Next frame\\operands};

    \draw[grcArrow] (w1.south) -- (first.north);
    \draw[grcArrow] (w2.south) -- (first.north);
    \draw[grcArrow] (w2.south) -- (next.north);
    \draw[grcArrow] (w3.south) -- (next.north);

    \node[grcNative,text width=3.85cm] (execute) at (6.35,1.1) {\textbf{Execute supported graph}\\Queues hold waiting operands\\Keep operands and replies ordered};

    \draw[draw=black!75,line width=.6pt] (first.south) -- (5.25,1.78) -- (6.35,1.78);
    \draw[draw=black!75,line width=.6pt] (next.south) -- (7.45,1.78) -- (6.35,1.78);
    \draw[grcArrow] (6.35,1.78) -- (execute.north);

    \node[grcNative,text width=3.85cm] (output) at (6.35,.05) {Results};
    \draw[grcArrow] (execute) -- (output);

    \node[grcNote,text width=4cm] at (6.35,-.40) {Frame overlap requires execution support.};

  \end{tikzpicture}%
  }
  \caption{Native memory using RAMB36E5/URAM288E5~\cite{amd2026am007}.}
  \Description{The left panel distributes consecutive word addresses across banks and rows, using alternative BRAM or UltraRAM configurations with enabled output registers. Four 16-bit lanes occupy 64 bits of a 72-bit native word. The right panel illustrates three retained words without specifying buffer capacity. Word 2 supplies both the first frame and the next frame, illustrating reuse without another read from the source banks. Supported graph execution retains waiting operands in queues and preserves operand and reply ordering. Frame overlap requires support from the execution backend.}
  \label{fig:native-memory}
\end{figure}

%% file: Tables/cluster-features.tex
\begin{table}[t]
  \centering
  \caption{Features of the native workload descriptor.}
  \label{tab:cluster-features}
  \small
  \begin{tabular}{@{}p{0.2\columnwidth}p{0.74\columnwidth}@{}}
    \toprule
    Family & Counts \\
    \midrule
    Operations & Each opcode. \\
    Kinds & Each FU, memory, constant or specialised kind. \\
    \#Ops. & Operations by operand count (e.g.\ 1 for PASS, 2 for ADD, 3 for MADD). \\
    Edges & Each class $(\kappa(u),\kappa(v),p)$: source kind, destination kind, operand port. \\
    Shape & Vertices, edges, DAG depth and maximum fan-out. \\
    \bottomrule
  \end{tabular}
\end{table}

%% file: Figures/port-expansion.tex
\begin{figure}[t]
  \centering
  \definecolor{peBlue}{HTML}{0072B2}
  \definecolor{peOrange}{HTML}{D55E00}
  \definecolor{peTeal}{HTML}{009E73}
  \definecolor{pePurple}{HTML}{8064A2}
  \resizebox{\columnwidth}{!}{%
  \begin{tikzpicture}[
    font=\sffamily\fontsize{8}{9}\selectfont,
    panel/.style={draw=black!50,rounded corners=2pt,line width=.6pt},
    head/.style={font=\sffamily\fontsize{8}{9}\selectfont\bfseries,inner sep=1pt},
    flow/.style={-{Latex[length=1.4mm]},draw=peBlue,line width=.7pt},
    link/.style={draw=black!65,line width=.6pt},
    center/.style={draw=peBlue!80!black,fill=peBlue!12,rounded corners=2pt,align=center,minimum width=6.5mm,minimum height=6mm,inner sep=1.5pt},
    port/.style={draw=peTeal!70!black,fill=peTeal!7,rounded corners=1pt,minimum width=5.4mm,minimum height=3.8mm,inner sep=1pt},
    outport/.style={port,draw=peOrange!80!black,fill=peOrange!10},
    edgevertex/.style={draw=pePurple!80!black,fill=pePurple!10,align=center,minimum width=9mm,minimum height=6mm,inner sep=1.5pt},
    note/.style={align=center,inner sep=1pt}
  ]

    \draw[panel] (0,0) rectangle (2.15,2.85);
    \node[head] at (1.075,2.6) {(a) DFG edge};
    \node[center] (originalu) at (.9,2.0) {$u$\\MUL};
    \node[center] (originalv) at (.9,1.0) {$v$\\ADD};
    \draw[flow] (originalu) -- node[right,inner sep=2pt] {port $b$} (originalv);
    \node[note] at (1.075,.35) {Opcodes\\in mode};

    \draw[panel] (2.3,0) rectangle (8.4,2.85);
    \node[head] at (5.35,2.6) {(b) Undirected port expansion};

    \node[center] (cu) at (3.6,1.45) {FU\\$c_u$};
    \node[port] (ua) at (2.75,2.15) {in $a$};
    \node[port] (ub) at (2.75,1.45) {in $b$};
    \node[port] (uc) at (2.75,.9) {in $c$};
    \node[outport] (uo) at (4.4,1.45) {out};

    \node[edgevertex] (edge) at (5.35,1.45) {word\\format};

    \node[port] (vb) at (6.3,1.45) {in $b$};
    \node[center] (cv) at (7.15,1.45) {FU\\$c_v$};
    \node[port] (va) at (7.15,2.15) {in $a$};
    \node[port] (vc) at (7.15,.9) {in $c$};
    \node[outport] (vo) at (8.05,1.45) {out};

    \draw[link] (ua) -- (cu);
    \draw[link] (ub) -- (cu);
    \draw[link] (uc) -- (cu);
    \draw[link] (cu) -- (uo);
    \draw[link,draw=pePurple!85!black,line width=.8pt] (uo) -- (edge);
    \draw[link,draw=pePurple!85!black,line width=.8pt] (edge) -- (vb);
    \draw[link] (vb) -- (cv);
    \draw[link] (va) -- (cv);
    \draw[link] (vc) -- (cv);
    \draw[link] (cv) -- (vo);

    \node[note,text width=5.8cm] at (5.35,.35) {Labels encode kind, role, port, and format.\\Vertex IDs are not features.};

  \end{tikzpicture}%
  }
  \caption{Port expansion of one DFG edge preserves endpoint roles and operand positions in an undirected graph.}
  \Description{A multiplication u supplies operand b of addition v. The undirected auxiliary graph replaces both operations with compatible FU centers. Each center connects to input a, b, c and output vertices. A word-format vertex connects the source output to destination input b. Matching labels exclude vertex identities; original opcodes remain in workload modes.}
  \label{fig:port-expansion}
\end{figure}
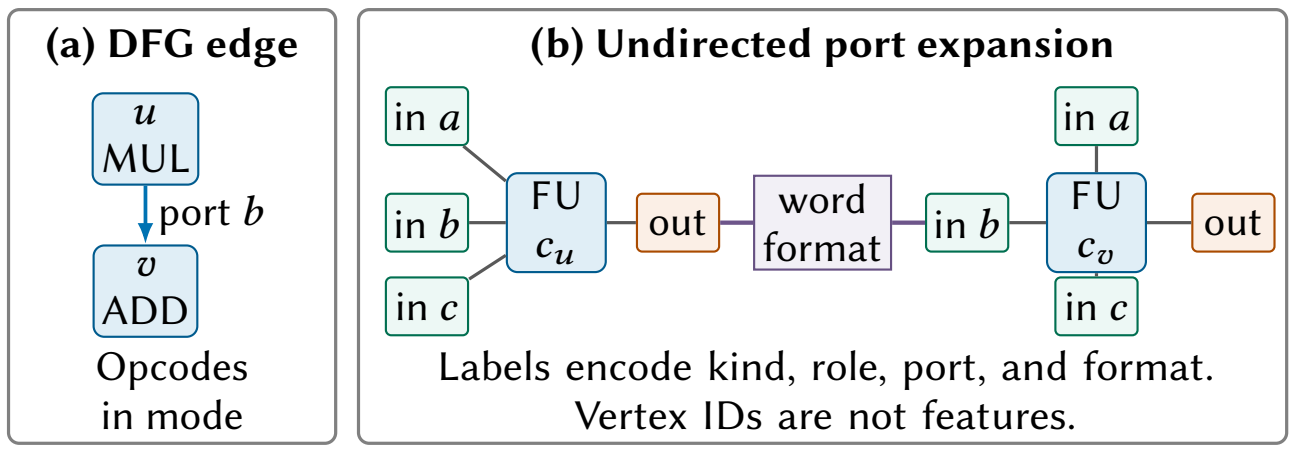

%% file: Figures/spectral-assignment.tex
\begin{figure}[t]
  \centering
  \resizebox{\columnwidth}{!}{%
  \begin{tikzpicture}[grcFigure,font=\sffamily\fontsize{7}{8}\selectfont]

    \node[grcTitle,anchor=west] at (0,3.0) {(a) Fingerprint of $v$};
    \node[grcTitle,anchor=west] at (4.3,3.0) {(b) Assigning $G$ (left) to $\mathcal A$ (right)};

    \foreach \rx/\ry in {1.82/1.0,1.2/0.72,0.72/0.44} {
      \draw[draw=grcOrange!75!black,dashed,fill=grcOrange!6] (1.9,1.75) ellipse ({\rx} and {\ry});
    }
    \foreach \t/\ly in {0.1/1.31,0.5/1.03,2.0/0.75} {
      \node[grcNote,fill=white,inner sep=.4pt] at (1.9,\ly) {$t=\t$};
    }
    \node[grcMemory,minimum width=.55cm] (ldA) at (0.42,2.22) {ld};
    \node[grcMemory,minimum width=.55cm] (ldB) at (0.42,1.28) {ld};
    \node[grcMemory,minimum width=.55cm] (stA) at (3.38,1.75) {st};
    \node[grcSystem,minimum width=.55cm] (v) at (1.9,1.75) {$v$};
    \foreach \n/\x/\y/\l in {pa/1.3/2.02/a,pb/1.3/1.48/b,po/2.5/1.75/o} {
      \node[circle,draw=black!60,fill=white,inner sep=.6pt,font=\sffamily\fontsize{5}{5}\selectfont] (\n) at (\x,\y) {\l};
    }
    \foreach \n/\x/\y in {ea/0.86/2.12,eb/0.86/1.38,eo/2.95/1.75} {
      \node[circle,fill=black!55,inner sep=.9pt] (\n) at (\x,\y) {};
    }
    \draw[black!55] (ldA) -- (ea) -- (pa) -- (v) -- (pb) -- (eb) -- (ldB);
    \draw[black!55] (v) -- (po) -- (eo) -- (stA);
    \node[grcNote] at (1.9,0.4) {$\phi_G(v)$: mix of tags reaching $v$ at each $t$};

    \node[grcMemory,minimum width=.7cm] (gl) at (4.75,2.55) {ld};
    \node[grcSystem,minimum width=.7cm] (gm) at (4.75,2.05) {$\times$};
    \node[grcSystem,minimum width=.7cm] (ga) at (4.75,1.55) {$+$};
    \node[grcMemory,minimum width=.7cm] (gs) at (4.75,1.05) {st};
    \node[grcMemory,minimum width=1.05cm] (al) at (7.65,2.55) {ld};
    \node[grcSystem,minimum width=1.05cm] (af) at (7.65,2.05) {FU};
    \node[grcSystem,minimum width=1.05cm,dashed] (an) at (7.65,1.55) {new FU};
    \node[grcMemory,minimum width=1.05cm] (as) at (7.65,1.05) {st};
    \draw[black!40] (gl) -- (gm) -- (ga) -- (gs);
    \draw[grcArrow] (gl) -- (al);
    \draw[grcArrow] (gm) -- node[grcNote,above,inner sep=1pt] {$c_{\mathrm{old}}$} (af);
    \draw[grcArrow,dashed] (ga) -- node[grcNote,above,inner sep=1pt] {$c_{\mathrm{new}}$} (an);
    \draw[grcArrow] (gs) -- (as);
    \node[grcNote,text width=4cm] at (6.2,0.48) {Hungarian: lowest total cost,\\one slot per operation};

  \end{tikzpicture}%
  }
  \caption{Diffusion fingerprints (a), incremental assign. (b).}
  \Description{Panel a shows an operation v with its two input ports, its output port and the connection vertices to two loads and a store, inside three nested dashed rings labelled t equals 0.1, 0.5 and 2.0. Panel b shows an incoming workload with a load, a multiply, an add and a store. The load, multiply and store map to an existing load, FU and store slot; the add maps to a new FU slot.}
  \label{fig:spectral-assignment}
\end{figure}
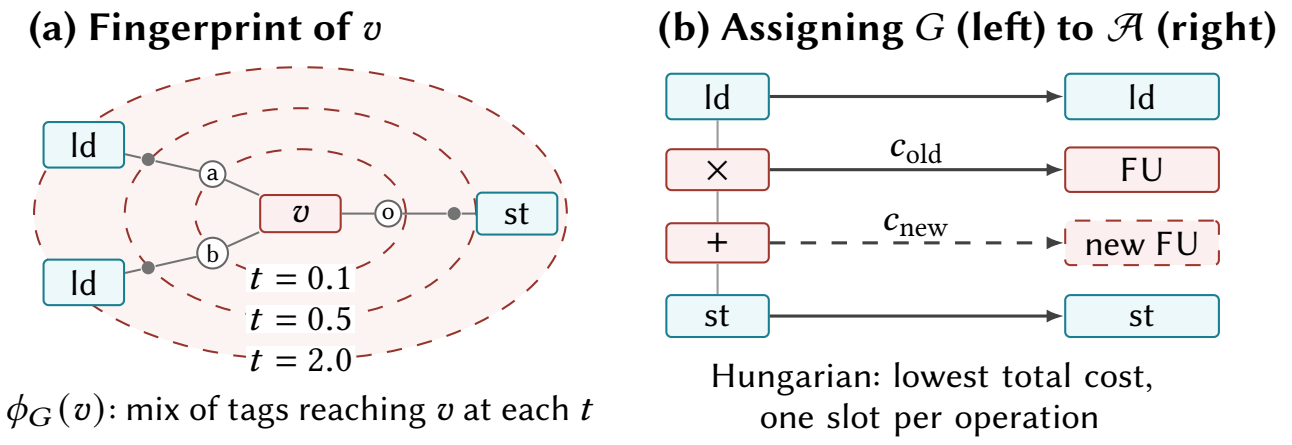

%% file: Sections/4-Automatic-Flow.tex
\section{The Automatic Flow}
\label{sec:automatic-flow}

Our flow compiles supported programs into a graph-based overlay and separately builds architectures from fixed-function engines or micro-programs. An \emph{engine} is an RTL implementation with its own controller; a \emph{microprogram} lists memory and arithmetic operations for an executor. These inputs follow distinct paths in Figure~\ref{fig:automatic-flow}.

\input{Figures/automatic-flow.tex}

\subsection{Source Normalization and Graph Extraction}
\label{sec:flow-frontend}

\input{Tables/compiler-stages.tex}

Our front end builds on MLIR, using its built-in families of intermediate representation (IR) called \emph{dialects} and types~\cite{lattner2021mlir,llvmMLIRLangRef}. ClangIR lifts C into MLIR~\cite{llvmClangIR}, and a restricted adapter rewrites it into the dialects of Table~\ref{tab:compiler-stages}. The result is enhanced by promoting local storage to values, simplifying operations and removing repeated expressions~\cite{llvmMLIRPasses}. Then, an extractor emits the finalized DFGs, which are clustered, matched and grown into an overlay (\S\ref{sec:generation-clustering}--\ref{sec:generation-matching}) or lowered into typed pipelines (\S\ref{sec:flow-lowering}).

The frontend accepts a single-block function with one result and one to eight arguments of the supported types (\S~\ref{sec:generation-compute}), turning each argument and the result into a stream. Unsupported control flow, calls, pointers, types and conversions are rejected. C leaves signed overflow undefined, so we give no guarantee for programs whose signed arithmetic overflows, even though our hardware simply keeps the low bits. The C frontend reads data only through its arguments and writes only its result. Besides, any other memory access must be expressed as an explicit DFG.

\subsection{Architecture Construction and Lowering}
\label{sec:flow-lowering}

This stage turns graphs into RTL along one of two paths. 
The native path builds a programmable overlay that several workloads share by switching modes at runtime, which requires every unit to work on a common word format, namely four 16-bit lanes. Graphs with other types take the typed path, trading sharing for a fixed pipeline built around the graph's own types. In this case, workloads can share fixed operators if they satisfy the rules of \S~\ref{sec:generation-compute}.

On the native path, the generator \emph{constructs} the overlay as in \S~\ref{sec:overlay-generation}: it emits the matched units, operand selectors, connections, and memory bindings. Then it loads every FU in the selected mode, which remains fixed until the run ends.

On the typed path, the generator instead \emph{lowers} each graph to a fixed pipeline of operators on scalar or 4-lane \texttt{i8}, \texttt{i32} or \texttt{f32} values. Before lowering, a rewrite turns four signed-byte products of the same two input vectors, summed by an \texttt{i32} adder tree, into a single dot-product unit, which two cascaded DSP blocks compute natively~\cite{amd2022am004}.

\subsection{Pipeline Latency and Execution Control}
\label{sec:flow-pipeline}

Each DSP and BRAM contains optional internal registers, such as the ones in Figure~\ref{fig:native-programmable-fu}. Enabling them shortens the clock period. The price is latency. Each enabled register delays the arrival by one cycle; thus, operands of the same operation must still arrive in the same cycle.

The native path keeps the registers enabled and lets a controller absorb their latency. Its reference controller issues an operation once its predecessors have completed and its PE can accept operands. The results stay available until consumed, and frames run one after another. Figure~\ref{fig:pipeline-contracts}(a) shows the capture and completion steps this adds around the three native stages of a DSP block.

\input{Figures/pipeline-contracts.tex}

The typed pipeline hides latency entirely by planning arrivals in advance and accepting new work each cycle. It passes \emph{tokens}, each carrying one invocation's arguments and mode, with all inputs entering at stage zero. 
This also means that an operation waits for its slowest operand, and each faster operand is delayed by the difference in their arrival times. 
Figure~\ref{fig:pipeline-contracts}(b) illustrates the alignment.

A result leaves the pipeline as soon as the next stage is ready to accept it. If it is busy, a shared enable signal freezes the whole pipeline until it is ready. Otherwise, a new token enters every cycle.

\subsection{Explicit Storage and FIFO Composition}
\label{sec:flow-storage}

A pipeline that accepts a token every cycle is only as fast as the data that feeds it. However, access to external memory is slow. Therefore, storage is declared explicitly in the source, and the generator builds each buffer from BRAM/URAM memory primitives with known timing~\cite{amd2026ug1353}. A static one-dimensional array declaration selects a FIFO, RAM, or delay line, and the computation may not access the array in any other way. Reads return after one cycle from a FIFO, two from a BRAM, and three from a URAM. Delay lines share the pipeline's enable, and reset clears their valid bits.

The data source and the pipeline rarely run in step, so an optional FIFO buffers it at the input. The data is written whenever it is available and read whenever the pipeline can accept a token. Each FIFO word holds all the arguments of one invocation, so a single read supplies a complete token. The FIFO delivers each value one cycle after it is requested and keeps it visible for only that cycle. If the pipeline is stalled in that cycle, the value would be lost, so the adapter holds it until the pipeline takes it and only then requests the next one. This preserves all values but can slow the input to below one token per cycle.

\subsection{Bulk Transfers and Pipelined Memory}
\label{sec:flow-implementation}

Off-chip memory is mostly the slowest component in a design. A DDR round trip costs tens of cycles, and an engine that waits for each reply before issuing the next spends most of its time idle. The memory-wrapper generator removes that wait by staging each kernel's data on-chip. From an existing build, its recorded memory addresses and its array layouts, it derives per-mode \emph{transfer regions}: address ranges copied through a private URAM \emph{scratchpad}. Each run then proceeds in three phases, a bulk load, the computation and a bulk store (Figure~\ref{fig:buffered-memory-flow}), with loads moving up to 256 words per burst and keeping up to 16 requests in flight.

\input{Figures/buffered-memory-flow.tex}

During the computation, the engines no longer see DDR at all. The scratchpad accepts a request every cycle and returns replies in order, so an engine can keep issuing while earlier requests are pending, provided it respects the dependencies between them. FIFOs absorb stalls within each phase. Beyond hiding latency, staging turns scattered accesses into large transfers and keeps reused data close to the compute.

\subsection{Execution Backends and Local Reuse}
\label{sec:flow-scheduling}

The bottlenecks of workloads are different. Some are shaped by irregular loops and data reuse, some by long floating-point latencies, and some by the cost of running one frame at a time. No single way to execute operations serves them all, so the flow offers several \emph{backends}. Each specifies how operations execute and how they access memory. The grown overlay works on 16-bit lanes, so for workloads that need 64-bit or floating-point types, its compute is replaced by fixed-function engines or a micro-programmed executor, while the image around them grows in the same way. All three below are distinct from the typed static pipeline.

\subsubsection{Fixed-Function Engines}
\label{sec:flow-mach}

A fixed-function engine is hand-written RTL for one workload, with its loop schedule and data reuse built in. Re-deriving these from C would only lose quality, so system generation takes such engines as they are.

\subsubsection{Micro-Programmed Executor}
\label{sec:flow-dsp}

Floating-point operators take many cycles, so a single loop iteration would leave them idle most of the time. This backend fills that time by interleaving up to 16 independent iterations, each keeping its own \emph{context} of 16 registers, while operands shared by the group load only once. Loads, multiplication/division, and addition are issued on separate paths. Each path keeps its own order and issues one operation per cycle once its operands and unit are ready. A value with one writer and one reader can pass straight through a result queue, and multiplication and addition are rounded separately.

The generator derives the issue tables from the microprograms, while each workload keeps its own loop controller. A register-readiness table is checked one cycle ahead and corrected for writes landing in the current cycle. Besides, memory hazards are respected. A load never passes an earlier store to the same element (RAW), a store never passes the load of its own element (WAR), and stores to the same element keep their order (WAW); only stores to different elements may reorder.

\subsubsection{Grown-Overlay Execution}
\label{sec:flow-vision}

The reference controller of \S~\ref{sec:flow-pipeline} finishes one frame before starting the next, leaving much of the graph idle. This executor allows successive frames to overlap while respecting strict rules. In active mode, every node fires exactly once per frame and consumes its operands, and every value has exactly one consumer. Under these rules, operands and load responses arrive in frame order, so frames remain separate without tagging each value, and queues hold values as producers and consumers drift apart.

%% file: Figures/automatic-flow.tex
\begin{figure}[t]
  \centering
  \resizebox{\columnwidth}{!}{%
  \begin{tikzpicture}[
    x=1cm,y=1cm,
    grBlock/.append style={font=\sffamily\fontsize{7}{8}\selectfont,inner sep=1.5pt},
    grTitle/.append style={font=\sffamily\fontsize{7}{8.5}\selectfont\bfseries},
    grNote/.append style={font=\sffamily\fontsize{6.5}{7.5}\selectfont,inner sep=1pt},
    grArrow/.append style={-{Latex[length=1.1mm]}},
    half/.style={minimum width=2.1cm,text width=2.04cm,minimum height=6.2mm},
    wide/.style={minimum width=3.7cm,text width=3.62cm,minimum height=6.2mm}
  ]

    \path[pattern=north east lines,pattern color=grBlue] (0,.42) rectangle (.16,.6);
    \node[grTitle,anchor=west] at (.2,.51) {(a) Compiler-generated compute};
    \path[pattern=dots,pattern color=grOrange] (4.6,.42) rectangle (4.76,.6);
    \node[grTitle,anchor=west] at (4.8,.51) {(b) System generation};

    \node[grInput,half] (dfgin) at (1.05,0) {Native DFGs};
    \node[grInput,half] (c) at (3.4,0) {Supported C};
    \node[grNative,half] (admit) at (1.05,-.75) {Validate\\ops, memory};
    \node[grNative,half] (graphs) at (3.4,-.75) {ClangIR, MLIR\\$\to$ source DFGs};
    \node[grNative,half] (growth) at (1.05,-1.5) {\textbf{Graph growth}\\modes, banks};
    \node[grTyped,half] (typed) at (3.4,-1.5) {\textbf{Typed lowering}\\sharing, alignment};
    \node[grNative,half] (nativeRTL) at (1.05,-2.25) {Programmable\\graph RTL};
    \node[grTyped,half] (typedRTL) at (3.4,-2.25) {Pipeline RTL\\+ optional FIFO};

    \node[grSystem,wide] (groups) at (6.45,0) {\textbf{Select workload set}\\given, or cosine groups and costs};
    \node[grInput,wide] (defs) at (6.45,-.75) {Engines, microprograms,\\admitted graph RTL from (a)};
    \node[grSystem,wide] (backends) at (6.45,-1.5) {\textbf{Backend and memory wrapper}\\composition, layouts, adapters};
    \node[grSystem,wide] (systemRTL) at (6.45,-2.25) {\textbf{System RTL} for the set};

    \node[grBlock,minimum width=8.3cm,text width=8.2cm] (build) at (4.15,-3.05) {Functional checks, RTL simulation $\to$ FPGA implementation:\\image, timing and resource reports};
    \node[grNote] at (4.15,-3.55) {Dashed: supplied inputs. Solid: transformations or generated outputs.};

    \draw[grArrow] (dfgin) -- (admit);
    \draw[grArrow] (admit) -- (growth);
    \draw[grArrow] (growth) -- (nativeRTL);
    \draw[grArrow] (c) -- (graphs);
    \draw[grArrow] (graphs) -- (typed);
    \draw[grArrow] (graphs.south west) -- (growth.north east);
    \draw[grArrow] (typed) -- (typedRTL);
    \draw[grArrow] (groups.east) -- (8.44,0) -- (8.44,-1.5) -- (backends.east);
    \draw[grArrow] (defs) -- (backends);
    \draw[grArrow] (backends) -- (systemRTL);
    \foreach \n in {nativeRTL,typedRTL,systemRTL} {
      \draw[grArrow] (\n.south) -- (\n.south |- build.north);
    }

  \end{tikzpicture}%
  }
  \caption{Compiler and system-generation flow.}
  \label{fig:automatic-flow}
\end{figure}
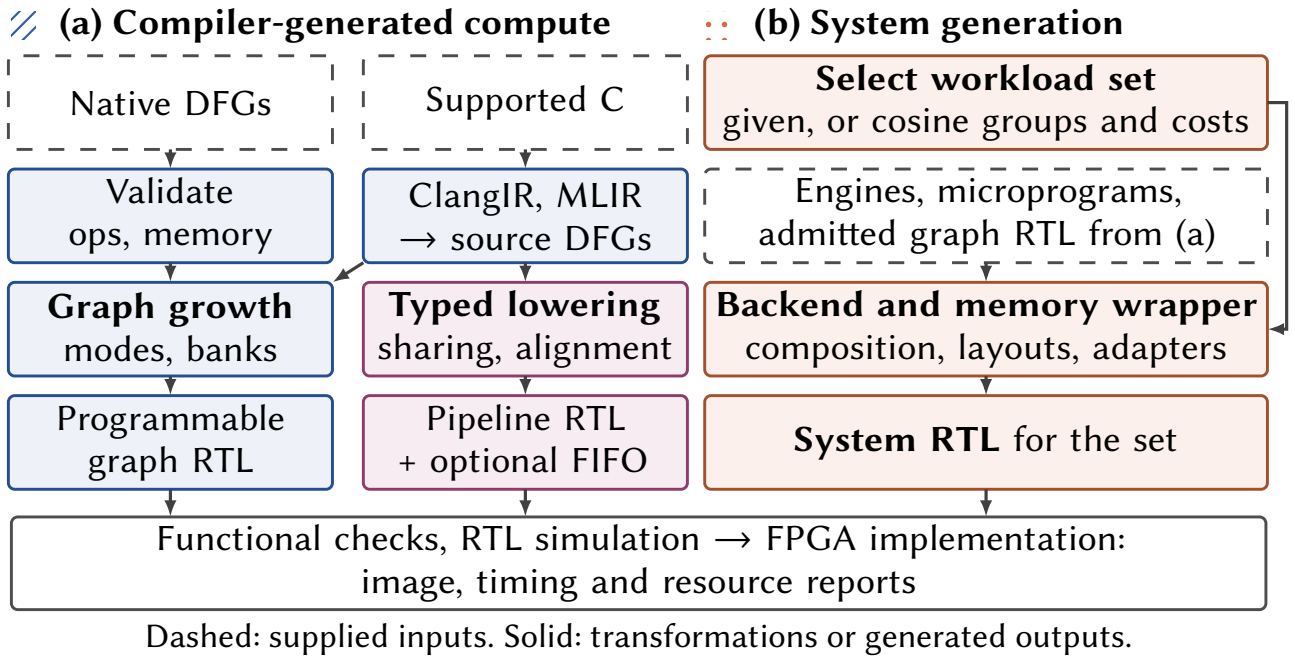

%% file: Tables/compiler-stages.tex
\begin{table}[t]
\centering
\caption{Compiler and system-generation stages.}
\label{tab:compiler-stages}
\begingroup
\footnotesize
\setlength{\tabcolsep}{3pt}
\renewcommand{\arraystretch}{1.08}

\begin{tabularx}{\columnwidth}{@{}>{\raggedright\arraybackslash}p{0.25\columnwidth}>{\raggedright\arraybackslash}X@{}}
\toprule
Stage & Main operation and output \\
\midrule

\multicolumn{2}{@{}l}{\textbf{Compiler-generated compute}} \\
\addlinespace[2pt]

Frontend & Translate supported C to CIR, then normalize it in MLIR. \\

DFG admission & Validate extracted source DFGs or supplied native graphs, including memory interfaces. \\

Native growth & Construct and verify shared native graphs; emit programmable RTL, modes, and memory bindings. \\

Typed lowering & Share compatible fixed-operation units and align data, mode, and valid signals. \\

Storage & Emit declared RAMs, FIFOs, and delays; optionally buffer the typed compute input. \\

\midrule
\multicolumn{2}{@{}l}{\textbf{System generation}} \\
\addlinespace[2pt]

Workload set & Use supplied membership or group typed workload records by similarity and estimated cost. \\

Backends & Compose supplied engines, derive program issue tables, or generate restricted graph overlap and queues. \\

Memory wrapper & Add burst transfers, FIFOs, and scratchpad storage with sequential load, compute, and store phases. \\

\midrule
Checks and build & Run path-specific functional and RTL checks, then implement modules or images and report timing and resources. \\

\bottomrule
\end{tabularx}
\endgroup
\end{table}

%% file: Figures/pipeline-contracts.tex
\begin{figure}[t]
  \centering
  \definecolor{pcBlue}{HTML}{0072B2}
  \definecolor{pcPurple}{HTML}{8064A2}
  \definecolor{pcTeal}{HTML}{009E73}
  \resizebox{\columnwidth}{!}{%
  \begin{tikzpicture}[
    font=\sffamily\fontsize{8}{9}\selectfont,
    head/.style={anchor=west,font=\sffamily\fontsize{8}{9}\selectfont\bfseries,inner sep=1pt},
    note/.style={align=center,inner sep=1pt},
    small/.style={note,font=\sffamily\fontsize{7}{8}\selectfont},
    block/.style={draw=black!65,rounded corners=1pt,fill=white,align=center,minimum height=3mm,inner sep=1.5pt},
    native/.style={block,draw=pcBlue!85!black,fill=pcBlue!9},
    typed/.style={block,draw=pcPurple!85!black,fill=pcPurple!12,minimum width=8.5mm},
    delay/.style={block,draw=pcTeal!75!black,fill=pcTeal!9,minimum height=4mm},
    flow/.style={-{Latex[length=1.3mm]},draw=black!75,line width=.65pt}
  ]

    \node[head] at (0,3.9) {(a) Reference PE/controller: unstalled};
    \foreach \x/\index in {.7/0,2.1/1,3.5/2,4.9/3,6.3/4,7.7/5} {
      \node[note] at (\x,3.6) {$E_{\index}$};
      \draw[black!25] (\x,3.48) -- (\x,2.7);
    }
    \node[native,text width=1.12cm] (accept) at (.7,3.05) {Capture\\operands};
    \node[native,text width=3.65cm] (stages) at (3.5,3.05) {3 native\\registered stages};
    \node[native,text width=1.08cm] (capture) at (6.3,3.05) {Latch;\\done};
    \node[native,text width=1.22cm] (depend) at (7.7,3.05) {Next PE\\may issue};
    \draw[flow] (accept) -- (stages);
    \draw[flow] (stages) -- (capture);
    \draw[flow] (capture) -- (depend);

    \node[head] at (0,2.5) {(b) Typed \texttt{i32} pipeline: $z=(ab+c)d$};
    \draw[draw=pcPurple!80!black,dashed,rounded corners=2pt,line width=.6pt] (0.7,0.4) rectangle (7.6,2.25);
    \foreach \x/\stage in {.25/0,2.85/3,5.55/5,8.15/8} {
      \node[small] at (\x,2.1) {$s=\stage$};
    }
    \node[note] (ab) at (0.25,1.7) {$a,b$};
    \node[typed] (m1) at (1.5,1.7) {$\times$\\$\ell=3$};
    \node[typed] (add) at (4.2,1.7) {$+$\\$\ell=2$};
    \node[typed] (m2) at (6.9,1.7) {$\times$\\$\ell=3$};
    \node[note] (z) at (8.15,1.7) {$z$};
    \draw[flow] (ab) -- (m1);
    \draw[flow] (m1) -- (add);
    \draw[flow] (add) -- (m2);
    \draw[flow] (m2) -- (z);

    \node[note] (cin) at (0.25,1.1) {$c$};
    \node[note] (din) at (0.25,0.65) {$d$};
    \node[delay,text width=58pt] (cdelay) at (1.95,1.1) {Delay 3 stages};
    \node[delay,text width=2.8cm] (ddelay) at (3.05,0.65) {Delay 5 stages};
    \draw[flow] (cin) -- (cdelay);
    \draw[flow] (din) -- (ddelay);
    \draw[flow] (cdelay) -- (4.2,1.1) -- (add.south);
    \draw[flow] (ddelay) -- (6.9,0.65) -- (m2.south);

  \end{tikzpicture}%
  }
  \caption{Reference-controller timing (a), op. pipelining (b)}
  \Description{The reference PE captures operands at E0, runs three native stages at E1 through E3, captures the result and marks done at E4, and permits dependent issue at E5 when ready. The typed chain has operator latencies three, two, and three, with delays of three stages for c and five for d. Operators, delays, mode predicates, and valid bits share a clock enable. A blocked valid output freezes all state. Typed latency counts enabled edges, including acceptance.}
  \label{fig:pipeline-contracts}
\end{figure}
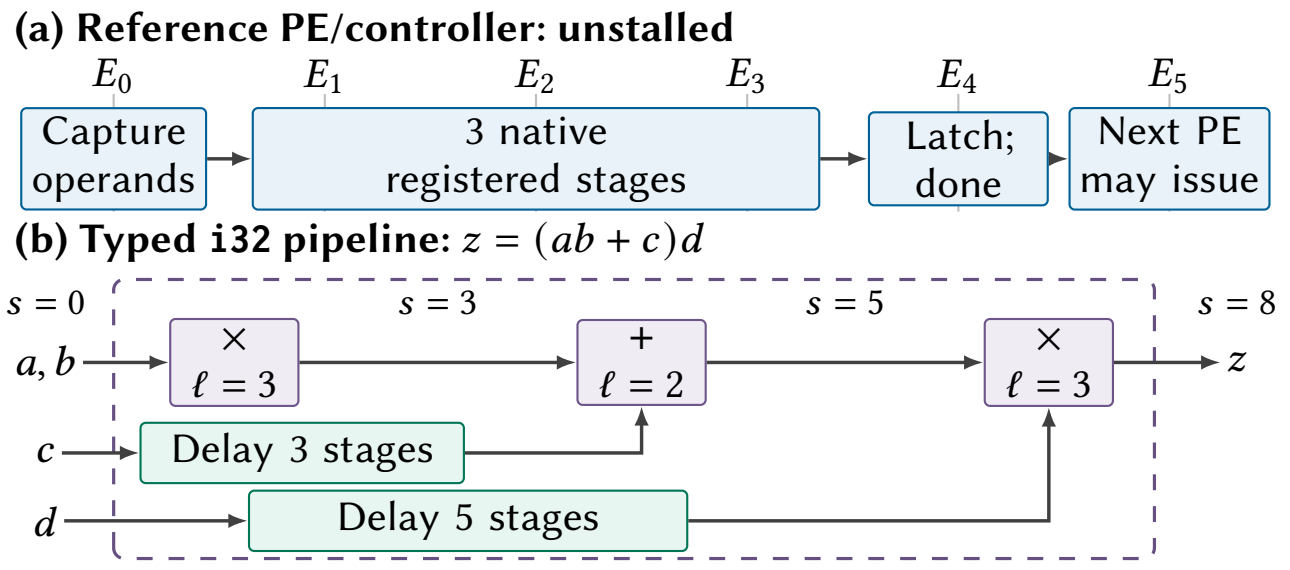

%% file: Figures/buffered-memory-flow.tex
\begin{figure}[t]
  \centering
  \resizebox{\columnwidth}{!}{%
  \begin{tikzpicture}[grcFigure,font=\sffamily\fontsize{7}{8}\selectfont]

    \node[grcBlock,minimum width=.65cm,minimum height=2.95cm] (ddr) at (.45,1.9) {};
    \node[grcNote,rotate=90] at (.45,1.9) {External DDR};

    \node[grcMemory,minimum width=.65cm,minimum height=2.95cm] (scratch) at (7.95,1.9) {};
    \node[grcNote,rotate=90] at (7.95,1.9) {UltraRAM scratchpad};

    \node[grcNative,text width=2.15cm] (read) at (2.25,3.05) {\textbf{1 Burst reads}\\$\leq$16 $\times$ 256 words};
    \node[grcMemory,text width=1.48cm] (infifo) at (4.65,3.05) {Input FIFO\\$512\times64$};
    \node[grcNative,text width=1.18cm] (loader) at (6.45,3.05) {Loader};

    \draw[grcArrow] (.775,3.05) -- (read);
    \draw[grcArrow] (read) -- (infifo);
    \draw[grcArrow] (infifo) -- (loader);
    \draw[grcArrow] (loader) -- (7.625,3.05);

    \node[grcSystem,text width=2.05cm] (engine) at (2.25,2.3) {\textbf{2a} Engine or\\executor};
    \node[grcSystem,text width=3cm] (server) at (5.7,2.3) {Pipelined scratchpad server\\1 request/cycle, in order};

    \draw[grcBoth] (engine) -- (server);
    \draw[grcBoth] (server) -- (7.625,2.3);

    \node[grcTyped,text width=2.05cm] (vision) at (2.25,1.55) {\textbf{2b} Graph core\\frame overlap};
    \node[grcMemory,text width=1.48cm] (banks) at (4.65,1.55) {Local banks\\opt.\ window};
    \node[grcTyped,text width=1.18cm] (tile) at (6.45,1.55) {Tile\\fill/fetch};

    \draw[grcBoth] (vision) -- (banks);
    \draw[grcBoth] (banks) -- (tile);
    \draw[grcBoth] (tile) -- (7.625,1.55);

    \node[grcNative,text width=2.05cm] (write) at (2.25,.8) {\textbf{3 Burst writes}};
    \node[grcMemory,text width=1.48cm] (outfifo) at (4.65,.8) {Output FIFO\\$512\times64$};
    \node[grcNative,text width=1.18cm] (reader) at (6.45,.8) {Output\\reader};

    \draw[grcArrow] (7.625,.8) -- (reader);
    \draw[grcArrow] (reader) -- (outfifo);
    \draw[grcArrow] (outfifo) -- (write);
    \draw[grcArrow] (write) -- (.775,.8);

    \node[grcNote] at (4.2,.08) {Phases 1--3 run in sequence; FIFOs decouple transfers within a phase.\\The mode selects the transfer regions and the compute configuration.};

  \end{tikzpicture}%
  }
  \caption{Memory phases around one scratchpad.}
  \Description{Bulk reads move 64-bit words from DDR through a 512 by 64-bit input FIFO into one private UltraRAM scratchpad, with at most 16 bursts in flight and 256 words per burst. Compute uses either an engine or microprogram executor with an ordered pipelined scratchpad server, or a restricted graph core with local banks and streamed tile transfers. Bulk stores reverse the direction through an output reader, a 512 by 64-bit FIFO, and burst writes. The three phases do not overlap.}
  \label{fig:buffered-memory-flow}
\end{figure}

%% file: Sections/5-Experimental-Results.tex
\section{Experimental Results}
\label{sec:evaluation}

\input{Figures/overgen-comparison.tex}

We compare our flow with OverGen~\cite{liu2022overgen} on Versal, using common inputs and reference outputs. We measure speedup, resource usage, image-generation cost, clustering effects, and energy consumption.

\subsection{Experimental Setup}
\label{sec:eval-setup}

\input{Tables/evaluation-workloads.tex}

All designs use a VCK5000/Versal VC1902~\cite{amd2022am004}, Vitis/Vivado 2022.1, and the same platform shell with one DDR channel. We evaluate 17 of OverGen's 19 workloads: five MachSuite, five DSP, and seven Vision workloads. We omit \texttt{derivative} and \texttt{convert-bit} from the Vision suite: the released OverGen compiler fails on both, and the released \texttt{derivative} source does not match the published dimensions and channel layout.

We evaluate nine cluster images. Three suite cluster images are defined by the suite name, and six cluster images are defined by the resulting cosine similarity clustering. Within an image, switching modes selects the workload. Every image grows from its workloads' DFGs, adding components only as they are needed and sharing what the workloads have in common. For Vision, these components are FUs and connections. MachSuite and DSP need 64-bit and floating-point types that the 16-bit fabric does not express, so their compute is replaced by coarser components, a \emph{fixed-function engine} per MachSuite workload and a shared \emph{micro-programmed executor} for DSP (\S~\ref{sec:flow-scheduling}), while each cluster still shares one copy of every floating-point core and one memory system. Together they exercise the mechanisms of \S~\ref{sec:automatic-flow}. The typed static pipeline is validated separately.

For the measurements, we run one warm-up and five timed invocations, checking every output, and we report the medians. The speedup is defined as OverGen's median time over ours, and aggregates are the geometric means across all 17 workloads, weighted equally.

\subsection{Constructing the OverGen Baseline}
\label{sec:eval-baseline}

OverGen's artifact targets another board and toolchain, so quoting its paper numbers would compare different hardware; we instead rebuilt it on the VCK5000, aiming for its best performance. This took three kinds of repair. \emph{Toolchain:} we reconstructed the build environment, reconciled architecture-description formats, and compiled with deterministic, routing-checked mappings. \emph{Hardware:} we rebuilt missing queues, memories and floating-point blocks, coordinated operand handshakes, and replaced the original board's hard-wired capacities and interfaces. Generated overlays also needed fixes to backpressure, indirect accesses, port numbering, vector widths, switch paths and accumulator and output selection, and the searched 5-core MachSuite design, too large for the device, became a 3-core one. \emph{Workloads:} we corrected initialization, timing, work partitioning and unsupported source constructs. Every retained result matches the reference word for word, and each workload uses the fastest correct median.

\subsection{Execution Performance}
\label{sec:eval-overgen}

Suite images run 9.52$\times$ faster than OverGen in geometric mean and similarity cluster images 10.33$\times$, and all 17 workloads are faster under both, by 4.75--26.43$\times$ and 5.33--36.97$\times$ respectively (Figure~\ref{fig:eval-overgen-comparison}).
Similarity clustering helps DSP the most, lifting it from 10.59$\times$ to 12.56$\times$, because only the smaller DSP similarity-cluster images close timing with double pumping (\S~\ref{sec:generation-memory}). \texttt{solver} is the exception, as each step waits for the previous one, so the extra multiplier throughput is unused, and the 250\,MHz core clock used as the slow clock in double pumping makes it 27\% slower.

\subsection{Effects of Workload Clustering}
\label{sec:eval-clustering}

Clustering mostly preserves workload cycle count. Each cluster image runs its workloads for the same number of cycles as the suite image, to within 0.13\%. However, before dual-porting, clustering enabled two Vision cluster images to close timing at 359 and 400\,MHz, compared to 321\,MHz for the suite image, which made Vision 1.17$\times$ faster and all workloads 1.07$\times$ faster.
Dual porting then raised the Vision suite image to 360\,MHz, which narrowed the gains to 1.07$\times$. Double-pumping also hid the clustering effect, as only the DSP cluster images close timing at twice the clock, and the frequencies of cluster images were reduced to 250\,MHz to close timing. This means that clustering can benefit designs that double-pumping and dual-porting may not improve.

\subsection{Resource Consumption}
\label{sec:eval-resources}

\input{Figures/resources-and-clocks.tex}

Routed compute-unit counts cover execution, storage, control and transfer logic. We evaluate the board resource usage of our images and OverGen on the benchmark suite. We do not show the breakdown per component due to space limits. The gap is large, with our images using 3,509--31,665 LUTs against OverGen's 586,223--788,479 (Figure~\ref{fig:eval-resources}). The clocks differ as much: even on the larger, older UltraScale+ VU9P, OverGen's own paper reports only 92.87\,MHz for a quad-core design~\cite{liu2022overgen}, against 250--400\,MHz for our images. FF, DSP, and BRAM usage stays below 2.1\% of the device, while only URAM, which holds the scratchpads, reaches 6.91\%. The new memory techniques do cost logic, especially double pumping, as the pumped DSP cluster $C_3$ needs 31,665 LUTs, and runs its core at 250\,MHz, with its RAMs and multipliers at 500\,MHz. Besides, the interconnects explain much of the gap. OverGen's reconfigurable switches alone take 47,430--96,040 LUTs, more than our largest image, whereas our flow builds only the connections its workloads use.
In summary, in geometric mean, OverGen's image uses 58.92$\times$ more LUTs than our suite image and 73.22$\times$ more than our cluster image, and our images use 19.4–22.1$\times$ less total fabric area than OverGen.

\subsection{Ablation}
\label{sec:eval-implementation}
\label{sec:eval-ablations}

Our flow applies several optimization stages to the generated overlay (\S\ref{sec:flow-pipeline}--\ref{sec:flow-scheduling}). Figure~\ref{fig:eval-implementation-speedup} retraces them, reporting each earlier configuration's runtime relative to the final suite images ($V_5$). The steps are cumulative, and the ablation disables one at a time.

\input{Figures/implementation-speedup.tex}

Across all 17 workloads, the initial 100\,MHz DDR images ($V_0$) run 92.51$\times$ slower than $V_5$. Registering the DDR interfaces and raising the clock ($V_1$) only narrows this to 60.30$\times$, because serialized DDR requests still dominate. Bulk transfers into an UltraRAM scratchpad ($V_2$) bring it to 16.84$\times$, but engines still wait for each request and operation in turn. \emph{Pipelined engines} ($V_3$) eliminate that wait by keeping many requests in flight, pipelining operators, and interleaving up to 16 independent DSP iterations, which cuts the gap to 4.46$\times$. \emph{Engines with local reuse} ($V_4$) then keep sparse-row and matrix accumulations local, reuse stencil data, split DSP issue into load, multiply/divide and add paths, and overlap Vision frames through ordered queues, all while preserving arithmetic order. This leaves 1.41$\times$ and shrinks Vision's gap from 10.88$\times$ to 1.94$\times$.

The final step uses the hard blocks' spare speed (\S~\ref{sec:generation-memory}). Dual porting removes the remaining 1.28$\times$ for MachSuite and 1.94$\times$ for Vision, whose clock also rises from 321 to 360\,MHz. DSP stays at 1.00$\times$: its dual-port variant brought no gain and its double-pumped suite image did not close timing, so pumping helps only its smaller cluster images (Figure~\ref{fig:eval-overgen-comparison}).

\subsection{Generation and Implementation Cost}
\label{sec:eval-generation}

A new image is quick to generate. However, it is slow to implement. A suite image takes 43.5--80.0 minutes, with 96.6--99.6\% of that time spent on FPGA implementation, and the remaining time is allocated to functional simulation and RTL generation. This makes our suite image 7.1--10.89$\times$ quicker to reach a runnable overlay than OverGen. Even the two cluster images covering a suite’s workloads are still 2.5--4$\times$ faster at reaching a runnable overlay than OverGen. Its implementation alone runs 4.9--6.6 hours. Besides, its architecture search adds 0.5--3.5 hours compared to 5.97 seconds for all clustering cost combined. Compiling workloads remains cheap in both our flow and OverGen.

\subsection{Board Power and Energy}
\label{sec:eval-power}

\input{Figures/energy-estimates.tex}

We evaluate energy per kernel call from measured board power.
Idle power barely separates the designs: 33.01--33.39\,W for our images and 33.51--34.12\,W for OverGen. Thus, the energy consumption follows runtime (Figure~\ref{fig:eval-energy}): OverGen needs 9.67$\times$ the suite images' product, from 6.58$\times$ for MachSuite to 11.85$\times$ for Vision, and cluster images need 0.92$\times$ that of the suite images. This demonstrates that growing overlay out of workload requirements outperforms optimizing initial hardware organization in energy consumption. It also shows that clustering saves a little more energy as their hardware is more specialized.

%% file: Figures/overgen-comparison.tex
\begin{figure}[t]
  \centering
  \begingroup%
  \newcommand{\GRcompactRuntimeRow}[4]{%
    \begin{scope}[shift={(0,#1)}]
      \pgfmathsetmacro{\GRsuiteEnd}{2.15+4.05*ln(#3)/ln(64)}
      \pgfmathsetmacro{\GRclusterEnd}{2.15+4.05*ln(#4)/ln(64)}
      \path[grcSuite] (2.15,.025) rectangle (\GRsuiteEnd,.115);
      \path[grcCluster] (2.15,-.115) rectangle (\GRclusterEnd,-.025);
      \node[anchor=east,inner sep=0pt] at (2.0,0) {#2};
      \node[grcNumber,anchor=east] at (7.15,0) {\pgfmathprintnumber[fixed,precision=2,fixed zerofill]{#3}};
      \node[grcNumber,anchor=east] at (8.25,0) {\pgfmathprintnumber[fixed,precision=2,fixed zerofill]{#4}};
    \end{scope}
  }%
  \resizebox{\columnwidth}{!}{%
  \begin{tikzpicture}[grcFigure]

    \path[grcSuite] (.25,6.96) rectangle (.55,7.16);
    \node[anchor=west,inner sep=1pt] at (.65,7.06) {Our suite image};
    \path[grcCluster] (4.45,6.96) rectangle (4.75,7.16);
    \node[anchor=west,inner sep=1pt] at (4.85,7.06) {Our cluster images};
    \node[grcTitle] at (6.85,6.72) {Suite};
    \node[grcTitle] at (7.95,6.72) {Cluster};

    \foreach \yy in {4.82,2.80,.22,-.16} {
      \path[fill=black!5] (0,\yy-.135) rectangle (8.4,\yy+.135);
    }

    \foreach \tick in {1,2,4,8,16,32,64} {
      \pgfmathsetmacro{\tickX}{2.15+4.05*ln(\tick)/ln(64)}
      \draw[black!14,line width=.3pt] (\tickX,-.32) -- (\tickX,6.38);
      \draw[black!65,line width=.3pt] (\tickX,-.46) -- (\tickX,-.53);
      \node[grcNumber,anchor=north] at (\tickX,-.58) {$\tick\times$};
    }
    \draw[black!75,line width=.5pt] (2.15,-.46) -- (2.15,6.38);
    \draw[black!65,line width=.3pt] (2.15,-.46) -- (6.2,-.46);

    \node[grcTitle,anchor=west] at (0,6.5) {MachSuite};
    \GRcompactRuntimeRow{6.22}{\texttt{stencil-3d}}{6.966957559}{7.035261922}
    \GRcompactRuntimeRow{5.94}{\texttt{stencil-2d}}{8.900833109}{8.989379278}
    \GRcompactRuntimeRow{5.66}{\texttt{gemm}}{5.812408305}{5.870873512}
    \GRcompactRuntimeRow{5.38}{\texttt{crs}}{5.739091318}{5.796982915}
    \GRcompactRuntimeRow{5.10}{\texttt{ellpack}}{5.274369948}{5.329361457}
    \GRcompactRuntimeRow{4.82}{\textbf{MachSuite GM}}{6.420498254}{6.485119441}

    \node[grcTitle,anchor=west] at (0,4.48) {DSP};
    \GRcompactRuntimeRow{4.20}{\texttt{cholesky}}{14.067221570}{17.206893228}
    \GRcompactRuntimeRow{3.92}{\texttt{fft}}{8.405158285}{12.178945217}
    \GRcompactRuntimeRow{3.64}{\texttt{fir}}{26.434761975}{36.971540174}
    \GRcompactRuntimeRow{3.36}{\texttt{solver}}{8.967106790}{6.576158710}
    \GRcompactRuntimeRow{3.08}{\texttt{mm}}{4.752714268}{6.134190360}
    \GRcompactRuntimeRow{2.80}{\textbf{DSP GM}}{10.590213977}{12.559774549}

    \node[grcTitle,anchor=west] at (0,2.46) {Vision};
    \GRcompactRuntimeRow{2.18}{\texttt{acc-sqr}}{9.637835910}{10.061660962}
    \GRcompactRuntimeRow{1.90}{\texttt{acc-weight}}{10.196469087}{11.315448107}
    \GRcompactRuntimeRow{1.62}{\texttt{accumulate}}{9.881155372}{10.315678364}
    \GRcompactRuntimeRow{1.34}{\texttt{bgr2grey}}{15.609173311}{16.295432669}
    \GRcompactRuntimeRow{1.06}{\texttt{blur}}{13.999338568}{14.616534030}
    \GRcompactRuntimeRow{.78}{\texttt{channel-ext}}{14.689336855}{16.305025374}
    \GRcompactRuntimeRow{.50}{\texttt{vecmax}}{9.579609186}{10.630448086}
    \GRcompactRuntimeRow{.22}{\textbf{Vision GM}}{11.691433954}{12.529856075}

    \GRcompactRuntimeRow{-.16}{\textbf{All 17 GM}}{9.520791340}{10.330582383}

    \node[grcNote] at (4.2,-.98) {OverGen time / our time (log; higher is better)};

  \end{tikzpicture}%
  }
  \endgroup
  \caption{Speedup over the correct OverGen configuration.}
  \label{fig:eval-overgen-comparison}
\end{figure}
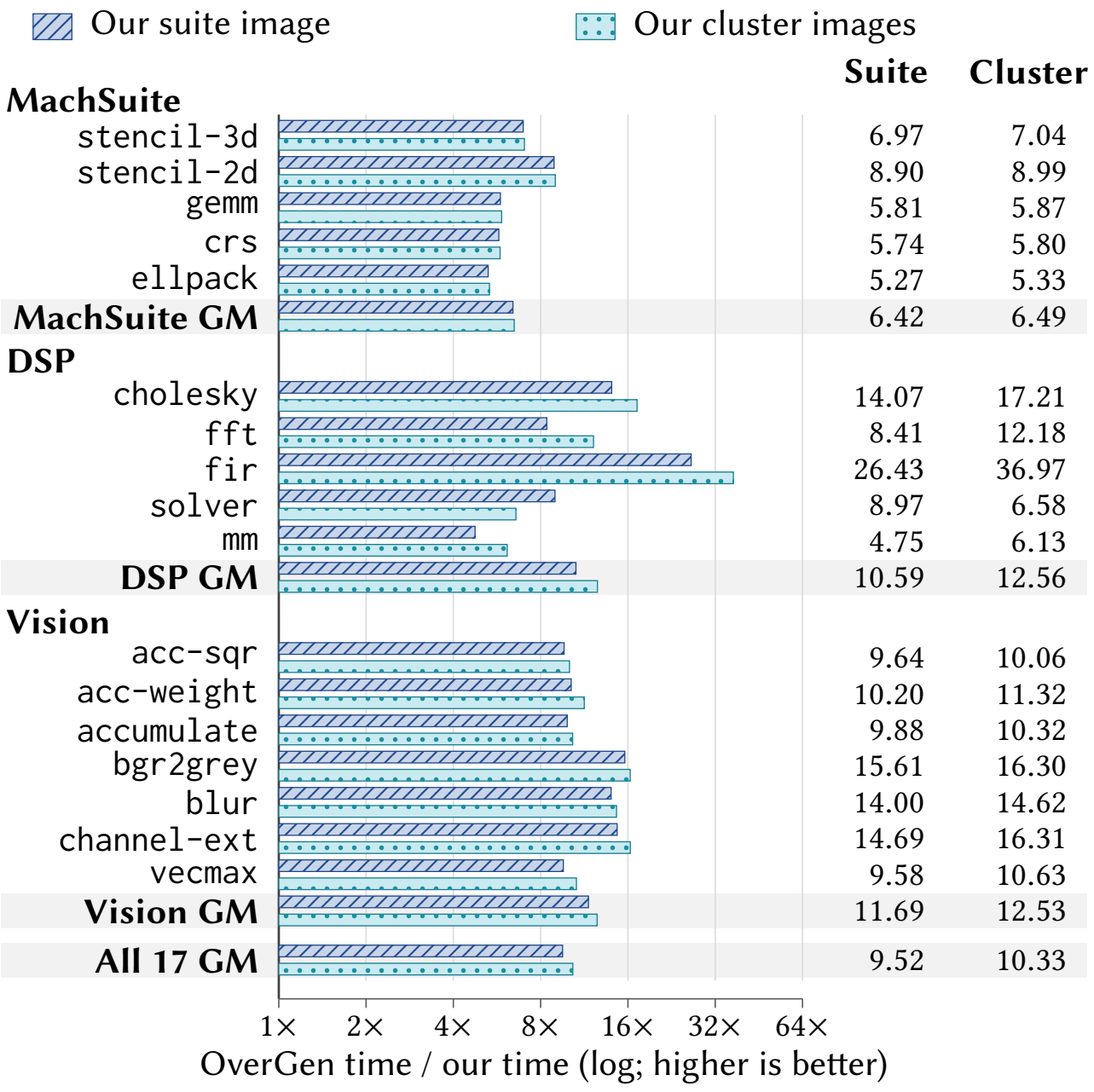

%% file: Tables/evaluation-workloads.tex
\begin{table}[t]
\centering
\footnotesize
\setlength{\tabcolsep}{3.5pt}
\caption{Evaluated workloads and their OverGen baselines.}
\label{tab:eval-workloads}
\begin{tabular}{@{}lllllc@{}}
\toprule
\textbf{Suite} & \textbf{Workload} & \textbf{Type} & \textbf{Input size} & \textbf{Cluster} & \textbf{OverGen} \\
\midrule
MachSuite & \texttt{stencil-3d} & \texttt{i64} & $34^3$ ($32^3$ interior) & $C_6$ & $R_2$, 50 \\
& \texttt{stencil-2d} & \texttt{i64} & $66^2$, $3{\times}3$ filter & $C_6$ & $G_3$, 50 \\
& \texttt{gemm} & \texttt{i64} & $64^2$ matrices & $C_6$ & $R_2$, 50 \\
& \texttt{crs} & \texttt{f64} & 494 rows, 1,638 nonzeros & $C_4$ & $G_3$, 50 \\
& \texttt{ellpack} & \texttt{f64} & 494 rows ${\times}$ 4 slots & $C_4$ & $G_3$, 50 \\
\addlinespace
DSP & \texttt{cholesky} & \texttt{f64} & $48^2$ matrix & $C_3$ & $R_1$, 36 \\
& \texttt{fft} & \texttt{c-f32} & 4,096 points & $C_5$ & $R_1$, 36 \\
& \texttt{fir} & \texttt{f64} & 1,024 samples, 199 taps & $C_3$ & $R_1$, 36 \\
& \texttt{solver} & \texttt{f64} & $48^2$ matrix, 48 vector & $C_3$ & $R_1$, 36 \\
& \texttt{mm} & \texttt{f64} & $32^2$ matrices & $C_3$ & $G_3$, 36 \\
\addlinespace
Vision & \texttt{acc-sqr} & \texttt{i16x4} & $128^2{\times}4$ & $C_1$ & $R_1$, 35 \\
& \texttt{acc-weight} & \texttt{i16x4} & $128^2{\times}4$ & $C_2$ & $R_1$, 35 \\
& \texttt{accumulate} & \texttt{i16x4} & $128^2{\times}4$ & $C_1$ & $R_1$, 35 \\
& \texttt{bgr2grey} & \texttt{i16}\textsuperscript{p} & $128^2{\times}4$ & $C_1$ & $R_1$, 35 \\
& \texttt{blur} & \texttt{i16x4} & $130^2{\times}4$ & $C_1$ & $R_1$, 35 \\
& \texttt{channel-ext} & \texttt{i16}\textsuperscript{p} & $128^2{\times}4$ & $C_2$ & $R_1$, 35 \\
& \texttt{vecmax} & \texttt{i16x4} & $128^2{\times}4$ & $C_2$ & $R_1$, 35 \\
\bottomrule
\multicolumn{6}{@{}p{\columnwidth}@{}}{\scriptsize \texttt{c-f32}: complex \texttt{f32}; \textsuperscript{p}packed. Cl.: cluster. $R_k$/$G_k$: released/generated overlay with $k$ cores, at the programmed clock in MHz.}
\end{tabular}
\end{table}

%% file: Figures/resources-and-clocks.tex
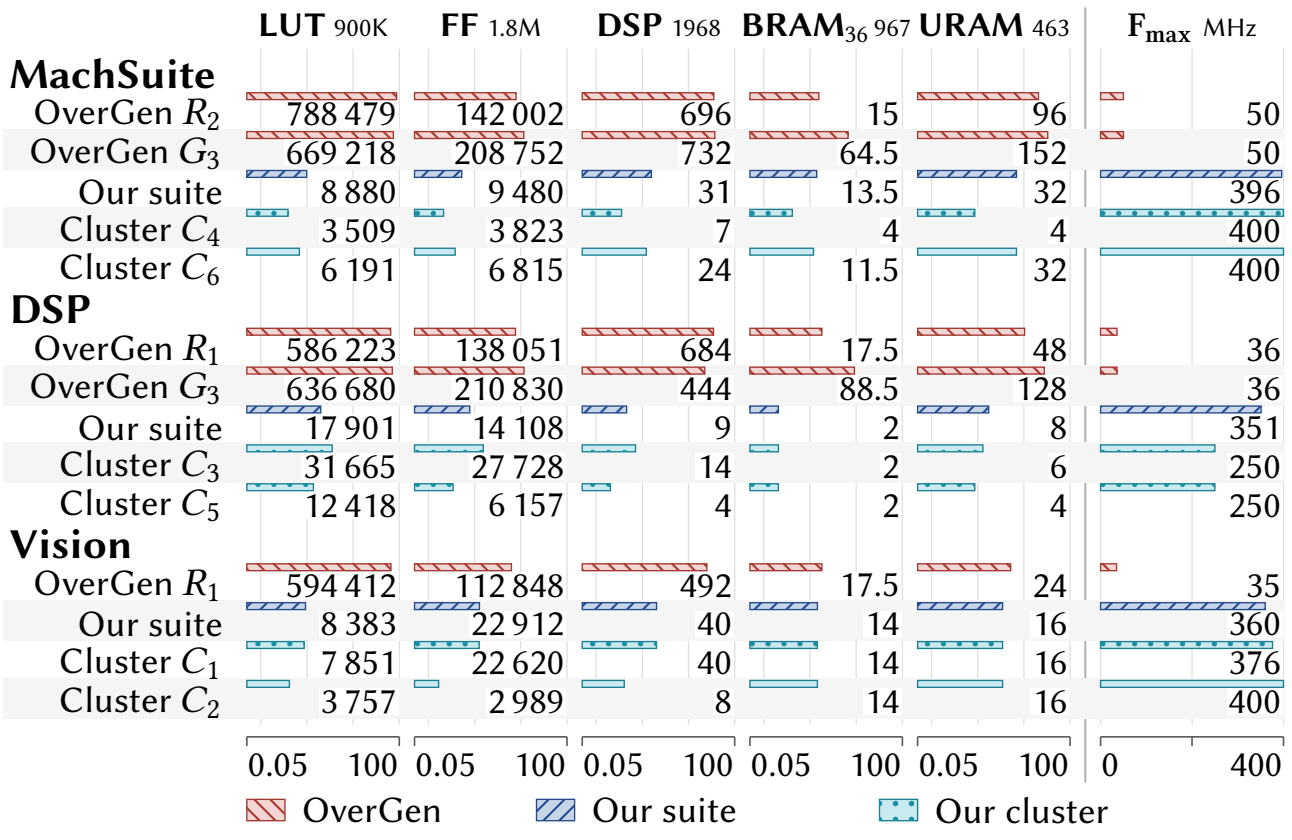
\begin{figure}[t]
  \centering
  \begingroup%
  \newcommand{\GRcompactResourceRow}[9]{%
    \begin{scope}[shift={(0,#1)}]
      \node[anchor=east,inner sep=0pt] at (1.45,-.04) {#2};

      \foreach \GRoffset/\GRpercent/\GRcount in {1.6/#4,2.7/#5,3.8/#6,4.9/#7,6.0/#8} {
        \pgfmathsetmacro{\GRresourceEnd}{\GRoffset+ln(\GRpercent/0.05)/ln(2000)}
        \path[#3] (\GRoffset,.055) rectangle (\GRresourceEnd,.1);
        \node[grcNumber,anchor=east,fill=white,inner sep=.3pt] at (\GRoffset+1,-.04) {\GRcount};
      }

      \pgfmathsetmacro{\GRclockEnd}{7.2+1.2*#9/400}
      \path[#3] (7.2,.055) rectangle (\GRclockEnd,.1);
      \node[grcNumber,anchor=east,fill=white,inner sep=.3pt] at (8.4,-.04) {#9};
    \end{scope}
  }%
  \resizebox{\columnwidth}{!}{%
  \begin{tikzpicture}[
    grcFigure,
    font=\sffamily\fontsize{7}{8}\selectfont,
    colhead/.style={grcTitle,font=\sffamily\fontsize{6.5}{7.5}\selectfont\bfseries}
  ]

    \foreach \GRoffset/\GRname/\GRcapacity in {1.6/LUT/{900K},2.7/FF/{1.8M},3.8/DSP/{1968},4.9/{BRAM$_{36}$\!}/{967},6.0/URAM/{463}} {
      \node[colhead,anchor=base] at (\GRoffset+.5,4.66) {\GRname\ \textmd{\fontsize{5}{6}\selectfont\GRcapacity}};

      \foreach \GRtick in {0.1,1,10,100} {
        \pgfmathsetmacro{\GRtickX}{\GRoffset+ln(\GRtick/0.05)/ln(2000)}
        \draw[black!12,line width=.25pt] (\GRtickX,0.145) -- (\GRtickX,4.58);
      }

      \draw[black!65,line width=.3pt] (\GRoffset,0.065) -- (\GRoffset+1,0.065);
      \draw[black!65,line width=.3pt] (\GRoffset,0.065) -- (\GRoffset,0.005);
      \draw[black!65,line width=.3pt] (\GRoffset+1,0.065) -- (\GRoffset+1,0.005);

      \node[grcNumber,anchor=north west] at (\GRoffset,-0.035) {0.05};
      \node[grcNumber,anchor=north east] at (\GRoffset+1,-0.035) {100};
    }

    \draw[black!30,line width=.4pt] (7.1,4.85) -- (7.1,-0.235);
    \node[colhead,anchor=base] at (7.8,4.66) {$\mathbf{F_{max}}$ \textmd{\fontsize{5}{6}\selectfont MHz}};

    \foreach \GRtick in {0,200,400} {
      \pgfmathsetmacro{\GRtickX}{7.2+1.2*\GRtick/400}
      \draw[black!12,line width=.25pt] (\GRtickX,0.145) -- (\GRtickX,4.58);
      \draw[black!65,line width=.3pt] (\GRtickX,0.065) -- (\GRtickX,0.005);
    }

    \draw[black!65,line width=.3pt] (7.2,0.065) -- (8.4,0.065);
    \node[grcNumber,anchor=north west] at (7.2,-0.035) {0};
    \node[grcNumber,anchor=north east] at (8.4,-0.035) {400};

    \path[fill=black!4] (0,3.785) rectangle (8.4,4.050);
    \path[fill=black!4] (0,3.275) rectangle (8.4,3.540);
    \path[fill=black!4] (0,2.240) rectangle (8.4,2.505);
    \path[fill=black!4] (0,1.730) rectangle (8.4,1.995);
    \path[fill=black!4] (0,0.695) rectangle (8.4,0.960);
    \path[fill=black!4] (0,0.185) rectangle (8.4,0.450);

    \node[grcTitle,anchor=west] at (0,4.42) {MachSuite};
    \GRcompactResourceRow{4.19}{OverGen $R_2$}{grcOvergen}{87.624355/{788\,479}}{7.890403/{142\,002}}{35.365854/{696}}{1.551189/{15}}{20.734341/{96}}{50}
    \GRcompactResourceRow{3.935}{OverGen $G_3$}{grcOvergen}{74.370777/{669\,218}}{11.599395/{208\,752}}{37.195122/{732}}{6.670114/{64.5}}{32.829374/{152}}{50}
    \GRcompactResourceRow{3.68}{Our suite}{grcSuite}{0.986842/{8\,880}}{0.526760/{9\,480}}{1.575203/{31}}{1.396070/{13.5}}{6.911447/{32}}{396}
    \GRcompactResourceRow{3.425}{Cluster $C_4$}{grcCluster}{0.389958/{3\,509}}{0.212427/{3\,823}}{0.355691/{7}}{0.413650/{4}}{0.863931/{4}}{400}
    \GRcompactResourceRow{3.17}{Cluster $C_6$}{grcCluster}{0.688011/{6\,191}}{0.378678/{6\,815}}{1.219512/{24}}{1.189245/{11.5}}{6.911447/{32}}{400}

    \node[grcTitle,anchor=west] at (0,2.875) {DSP};
    \GRcompactResourceRow{2.645}{OverGen $R_1$}{grcOvergen}{65.147471/{586\,223}}{7.670864/{138\,051}}{34.756098/{684}}{1.809721/{17.5}}{10.367171/{48}}{36}
    \GRcompactResourceRow{2.39}{OverGen $G_3$}{grcOvergen}{70.754801/{636\,680}}{11.714860/{210\,830}}{22.560976/{444}}{9.152017/{88.5}}{27.645788/{128}}{36}
    \GRcompactResourceRow{2.135}{Our suite}{grcSuite}{1.989354/{17\,901}}{0.783917/{14\,108}}{0.457317/{9}}{0.206825/{2}}{1.727862/{8}}{351}
    \GRcompactResourceRow{1.88}{Cluster $C_3$}{grcCluster}{3.518959/{31\,665}}{1.540718/{27\,728}}{0.711382/{14}}{0.206825/{2}}{1.295896/{6}}{250}
    \GRcompactResourceRow{1.625}{Cluster $C_5$}{grcCluster}{1.380023/{12\,418}}{0.342116/{6\,157}}{0.203252/{4}}{0.206825/{2}}{0.863931/{4}}{250}

    \node[grcTitle,anchor=west] at (0,1.33) {Vision};
    \GRcompactResourceRow{1.1}{OverGen $R_1$}{grcOvergen}{66.057521/{594\,412}}{6.270448/{112\,848}}{25.000000/{492}}{1.809721/{17.5}}{5.183585/{24}}{35}
    \GRcompactResourceRow{0.845}{Our suite}{grcSuite}{0.931610/{8\,383}}{1.273115/{22\,912}}{2.032520/{40}}{1.447777/{14}}{3.455724/{16}}{360}
    \GRcompactResourceRow{0.59}{Cluster $C_1$}{grcCluster}{0.872488/{7\,851}}{1.256890/{22\,620}}{2.032520/{40}}{1.447777/{14}}{3.455724/{16}}{376}
    \GRcompactResourceRow{0.335}{Cluster $C_2$}{grcCluster}{0.417519/{3\,757}}{0.166085/{2\,989}}{0.406504/{8}}{1.447777/{14}}{3.455724/{16}}{400}

    \path[grcOvergen] (1.6,-0.485) rectangle (1.85,-0.345);
    \node[anchor=west,inner sep=1pt] at (1.92,-0.415) {OverGen};
    \path[grcSuite] (3.5,-0.485) rectangle (3.75,-0.345);
    \node[anchor=west,inner sep=1pt] at (3.82,-0.415) {Our suite};
    \path[grcCluster] (5.75,-0.485) rectangle (6.0,-0.345);
    \node[anchor=west,inner sep=1pt] at (6.07,-0.415) {Our cluster};

  \end{tikzpicture}%
  }
  \endgroup
  \caption{Compute-unit resources (\% of device, log) and $\mathbf{F_{max}}$.}
  \label{fig:eval-resources}
\end{figure}

%% file: Figures/implementation-speedup.tex
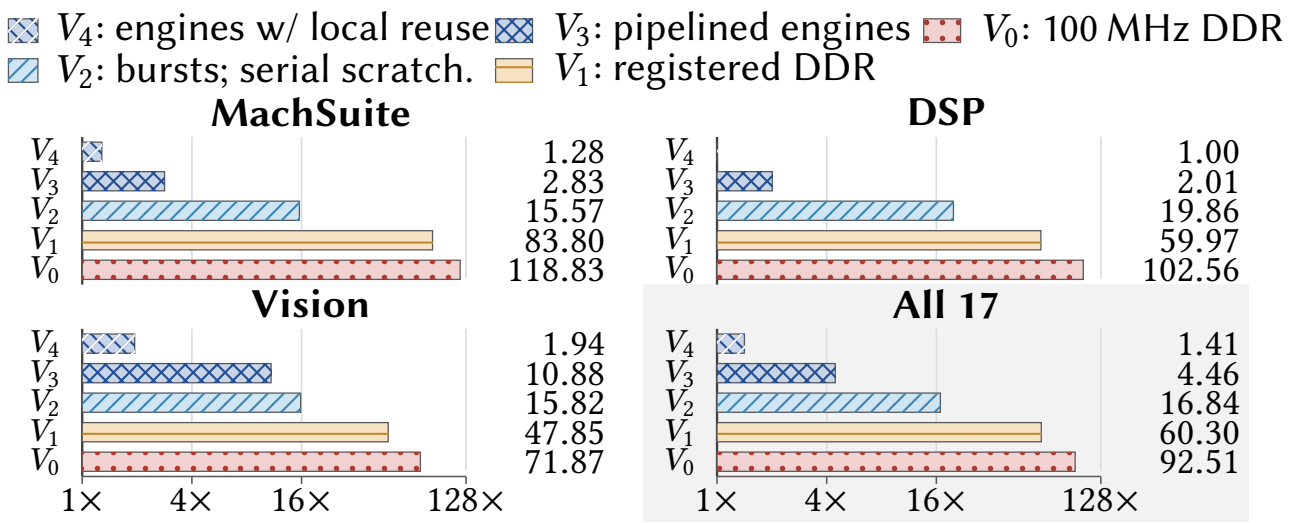
\begin{figure}[t]
  \centering
  \begingroup%
  \newcommand{\GRcompactImplementationPanel}[9]{%
    \begin{scope}[shift={(#1,#2)}]
      \node[grcTitle] at (2.05,.27) {#3};

      \foreach \tick in {1,4,16,128} {
        \pgfmathsetmacro{\tickX}{.5+2.6*ln(\tick)/ln(128)}
        \draw[black!14,line width=.3pt] (\tickX,-.86) -- (\tickX,.1);
        \ifnum#9=1
          \draw[black!65,line width=.3pt] (\tickX,-.9) -- (\tickX,-.95);
          \node[grcNumber,anchor=north] at (\tickX,-.97) {$\tick\times$};
        \fi
      }
      \draw[black!75,line width=.45pt] (.5,-.9) -- (.5,.1);
      \ifnum#9=1 \draw[black!65,line width=.3pt] (.5,-.9) -- (3.1,-.9); \fi

      \foreach \yy/\variant/\value/\style in {0/4/#4/grcVfour,-.2/3/#5/grcVthree,-.4/2/#6/grcVtwo,-.6/1/#7/grcVone,-.8/0/#8/grcVzero} {
        \pgfmathsetmacro{\barEnd}{.5+2.6*ln(\value)/ln(128)}
        \path[\style] (.5,\yy-.065) rectangle (\barEnd,\yy+.065);
        \node[anchor=east,inner sep=0pt,font=\sffamily\fontsize{6.5}{7}\selectfont] at (.37,\yy) {$V_{\variant}$};
        \node[grcNumber,anchor=east] at (4.05,\yy) {\pgfmathprintnumber[fixed,precision=2,fixed zerofill]{\value}};
      }
    \end{scope}
  }%
  \resizebox{\columnwidth}{!}{%
  \begin{tikzpicture}[grcFigure,
    grcVfour/.style={grcVthree,postaction={pattern=north east lines,pattern color=white}}]

    \path[grcVfour] (0,2.89) rectangle (.2,3.05);
    \node[anchor=west,inner sep=1pt] at (.28,2.97) {$V_4$: engines w/ local reuse};

    \path[grcVthree] (3.3,2.89) rectangle (3.55,3.05);
    \node[anchor=west,inner sep=1pt] at (3.65,2.97) {$V_3$: pipelined engines};

    \path[grcVzero] (6.2,2.89) rectangle (6.45,3.05);
    \node[anchor=west,inner sep=1pt] at (6.55,2.97) {$V_0$: 100\,MHz DDR};

    \path[grcVtwo] (0,2.62) rectangle (.2,2.78);
    \node[anchor=west,inner sep=1pt] at (.28,2.7) {$V_2$: bursts; serial scratch.};

    \path[grcVone] (3.3,2.62) rectangle (3.55,2.78);
    \node[anchor=west,inner sep=1pt] at (3.65,2.7) {$V_1$: registered DDR};

    \path[fill=black!5] (4.3,-.36) rectangle (8.4,1.25);

    \GRcompactImplementationPanel{0}{2.15}{MachSuite}{1.281239695}{2.831856715}{15.568747486}{83.804584285}{118.826426735}{0}
    \GRcompactImplementationPanel{4.3}{2.15}{DSP}{1.000000000}{2.009338778}{19.855614522}{59.969881278}{102.558729446}{0}
    \GRcompactImplementationPanel{0}{.85}{Vision}{1.941807346}{10.882583598}{15.823050108}{47.849645306}{71.872088184}{1}
    \GRcompactImplementationPanel{4.3}{.85}{All 17}{1.413608383}{4.456409341}{16.835212997}{60.298052753}{92.511786973}{1}

  \end{tikzpicture}%
  }
  \endgroup
  \caption{Effect of each optimization on suite images ($V_5=1$).}
  \label{fig:eval-implementation-speedup}
\end{figure}

%% file: Figures/energy-estimates.tex
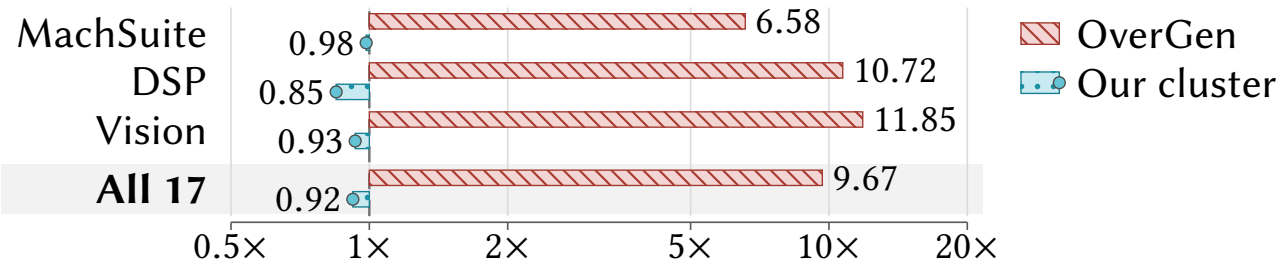
\begin{figure}[t]
  \centering
  \begingroup%
  \pgfmathsetmacro{\GRcompactParity}{1.5+4.8*ln(1/0.5)/ln(40)}%
  \newcommand{\GRcompactEnergyRow}[4]{%
    \begin{scope}[shift={(0,#1)}]
      \pgfmathsetmacro{\GRovergenEnd}{1.5+4.8*ln(#3/0.5)/ln(40)}
      \pgfmathsetmacro{\GRclusterEnd}{1.5+4.8*ln(#4/0.5)/ln(40)}

      \path[grcOvergen] (\GRcompactParity,.02) rectangle (\GRovergenEnd,.12);
      \path[grcCluster] (\GRcompactParity,-.12) rectangle (\GRclusterEnd,-.02);
      \node[anchor=east,inner sep=0pt] at (1.35,0) {#2};
      \node[grcNumber,anchor=west] at (\GRovergenEnd+.06,.07) {\pgfmathprintnumber[fixed,precision=2,fixed zerofill]{#3}};

      \path[fill=grcTeal!70,draw=black!70,line width=.25pt] (\GRclusterEnd,-.07) circle[radius=1.1pt];
      \node[grcNumber,anchor=east] at (\GRclusterEnd-.07,-.07) {\pgfmathprintnumber[fixed,precision=2,fixed zerofill]{#4}};
    \end{scope}
  }%
  \resizebox{\columnwidth}{!}{%
  \begin{tikzpicture}[grcFigure]

    \path[grcOvergen] (6.65,.82) rectangle (6.9,.96);
    \node[anchor=west,inner sep=1pt] at (6.97,.89) {OverGen};
    \path[grcCluster] (6.65,.5) rectangle (6.9,.64);
    \path[fill=grcTeal!70,draw=black!70,line width=.25pt] (6.9,.57) circle[radius=1.1pt];
    \node[anchor=west,inner sep=1pt] at (6.97,.57) {Our cluster};

    \path[fill=black!5] (0,-.28) rectangle (6.4,.04);
    \foreach \tick in {0.5,1,2,5,10,20} {
      \pgfmathsetmacro{\tickX}{1.5+4.8*ln(\tick/0.5)/ln(40)}
      \draw[black!14,line width=.3pt] (\tickX,-.3) -- (\tickX,1.06);
      \draw[black!65,line width=.3pt] (\tickX,-.34) -- (\tickX,-.39);
      \node[grcNumber,anchor=north] at (\tickX,-.41) {$\tick\times$};
    }
    \draw[black!55,densely dashed,line width=.45pt] (\GRcompactParity,-.3) -- (\GRcompactParity,1.06);
    \draw[black!65,line width=.3pt] (1.5,-.34) -- (6.3,-.34);

    \GRcompactEnergyRow{.9}{MachSuite}{6.577020394}{0.984329629}
    \GRcompactEnergyRow{.58}{DSP}{10.716211016}{0.846740235}
    \GRcompactEnergyRow{.26}{Vision}{11.846135691}{0.931493730}
    \GRcompactEnergyRow{-.12}{\textbf{All 17}}{9.674112340}{0.920538507}

  \end{tikzpicture}%
  }
  \endgroup
  \caption{Energy consumption.}
  \label{fig:eval-energy}
\end{figure}

%% file: Sections/6-Related-Work.tex
\section{Related Work}
\label{sec:related-work}

Our work touches on three aspects: overlays that make FPGAs programmable, generators that tailor an architecture to its workloads, and custom compilers that map programs onto the resulting architecture.

\paragraph{Overlays.}\label{sec:related-overlays}
Overlays trade some implementation efficiency for fast compilation and portability~\cite{coole2010intermediate,brant2012zuma,li2019overlaysurvey,podobas2020cgrasurvey}, and also serve virtualization, HLS debugging, and fault recovery~\cite{najem2017extended,jamal2018debug,kourfali2019faulttolerance}. They now target neural networks and language models~\cite{yu2020opu,abdelfattah2018dla,choudhury2022cnn,ioannou2022streaming,qian2025autocompiler,wu2025flightopu,he2026skipopu}, image and signal processing~\cite{choudhury2023flowpix,ioannou2020convolution,bai2025dspopu}, packet parsing and encrypted computation~\cite{mashreghi2023parser,mashreghi2024prismparser,fang2017secure,yang2025ola}, on execution models ranging from time-multiplexed pipelines~\cite{li2018timemultiplexed,ioannou2019lightweight} to SIMT and many-core processors that scale across dies, and HBM and clusters~\cite{alkadi2016fgpu,khanzadi2017datadriven,benabdelhamid2019mitraca,benabdelhamid2021dragon,benabdelhamid2023thesis,benabdelhamid2025kilocore,benabdelhamid2023scalable,elshimy2023nearmemory,naylor2019tinsel}. This diversity is precisely why the architecture underneath should be specialized.

\paragraph{Workload-driven generation and merging.}\label{sec:related-generation}\label{sec:related-merging}
Generators tune FUs, precision, connectivity, and storage from workload information and compiler feedback~\cite{weng2020dsagen,liu2022overgen,bandara2022revamp,tan2021aurora,tan2022asap}, building on dataflow templates, reconfigurable dataflow generators, and specialized processors~\cite{drewes2024template,silva2025hpcgra,duarte2017scratch,boutros2021specializing,ma2021pdlfgpu,ma2024specializedoverlays,sun2025labidus}. OverGen, our baseline, varies compute, network, and scratchpad resources within an architecture graph~\cite{liu2022overgen}, and Plasticine and MAERI show the value of pairing computation with memory and reduction structures~\cite{prabhakar2017plasticine,kwon2018maeri}. Datapath merging instead reuses blocks and connections across DFGs~\cite{moreano2005merging,stojilovic2012selective,coole2015adjustable}, found through frequent subgraphs and patterns~\cite{willsey2019radish,melchert2023apex,mbongue2019asoverlays} or shaped around DFG stages~\cite{jain2016deco,wilson2018recurrent}. These approaches work top-down by having generators prune or tune a predefined template, and merging combines complete data paths. Our flow instead grows bottom-up, starting from the workloads' operations and adding only the units and connections they need, with workload grouping and spectral descriptors deciding what to share.

\paragraph{Native resources and mapping.}\label{sec:related-native}\label{sec:related-graphs}
DSP-based overlays use native modes, internal pipelining, and forwarding~\cite{cheah2014idea,cheah2015forwarding,jain2015efficient,jain2016throughput,jain2022jit}, with compilers that map, share, cascade, and multipump DSPs~\cite{ronak2016mapping,ronak2016sharing,ronak2017multipumping,shi2020ftdl}; memory-centric overlays instead act around distributed storage~\cite{basalama2020spar2,panahi2021memorycentric,panahi2022thesis,kabir2025davinci}. Overlay compilers place, schedule, and route DFGs~\cite{capalija2013overlay,capalija2017thesis,silva2019ready,vieira2021reshape}, helped by graph-aware mapping, equivalent expressions, and communication motifs~\cite{li2022lisa,woodruff2023flexc,li2025plaid}, and extend to control overlays and runtime compilation~\cite{wilson2017fsmoverlay,dickerson2026pyjit2fpga}. Our flow makes operations, memory interfaces, and pipeline timing explicit during construction, uses spectral matching~\cite{umeyama1988matching,shuman2013emerging} only to rank candidates, and leaves dependencies, ports, and compatibility rules to decide what is built.

\paragraph{HLS and infrastructure.}\label{sec:related-hls}\label{sec:related-runtime}\label{sec:related-infrastructure}
HLS~\cite{cong2011hls} gains from templates, stencil and systolic transformations~\cite{cong2018autoaccel,chi2018soda,wang2021autosa}, guided or learned exploration~\cite{sohrabizadeh2022autodse,sohrabizadeh2022gnndse,sohrabizadeh2023harp} and joint code optimization~\cite{pouget2025sisyphus,pouget2025prometheus}; floorplanning, pipelining, stream networks and overlay interfaces raise physical quality~\cite{guo2021autobridge,guo2023tapa,basalama2025streamhls,li2020overlayinterface,wang2025rsn,he2025intar}, and modular or partially reconfigurable flows cut implementation effort~\cite{liu2019rapidroute,aklah2017thesis,park2024refine}. Although they provide automation, these still build a single accelerator or rebuild hardware as needs change, whereas each of our images switches workloads by configuration alone.
Similar to flows that use MLIR, custom accelerator IRs, and architecture-description toolchains~\cite{sharifian2019muir,lattner2021mlir,ye2022scalehls,chen2024allo,lelann2018argen,juneja2025morpher,bouaziz2024mlir}, our flow uses MLIR only for its frontend and grows the overlay and emits RTL in its own stages.

%% file: Sections/7-Conclusion.tex
\section{Conclusion}
\label{sec:conclusion}

We presented a framework that grows programmable FPGA overlays from the dataflow graphs of the workloads they must run. It groups workloads by structural similarity, uses spectral descriptors to match their operations and connections, and builds shared hardware that respects operation semantics, data types, operand roles and execution constraints. The supported workloads then switch modes on one implemented image, with no new synthesis or place and route.

To further accelerate the resulting overlays, we leverage the hard blocks' native capabilities. We use DSP controls that provide programmable arithmetic, internal registers that set the pipeline, and dual-ported or double-pumped memories and multipliers that leverage the hard blocks' spare speed. We also use an MLIR-assisted frontend, graph analysis, and RTL generation to keep the hardware and its configurations consistent.

On a VCK5000, with Vision running on grown overlays and MachSuite and DSP on fixed-function engines and a micro-pro\-grammed executor, our images run all 17 workloads faster than OverGen rebuilt for the same board: 9.52$\times$ in geometric mean with one image per suite and 10.33$\times$ with similarity clusters, and up to 36.97$\times$. It does so while using 19.4--22.1$\times$ less FPGA fabric usage and 9.67--10.51$\times$ less estimated board energy. Clustering alone cuts logic by up to 60\% without changing cycle counts, and double pumping helps most where arithmetic dominates, on the DSP clusters. 

Future work will extend the frontend to support complex control flow, additional operations and data types, and dynamic memory access patterns that dataflow execution supports.

%% file: main.bbl

\begin{thebibliography}{118}


\ifx \showCODEN    \undefined \def \showCODEN     #1{\unskip}     \fi
\ifx \showISBNx    \undefined \def \showISBNx     #1{\unskip}     \fi
\ifx \showISBNxiii \undefined \def \showISBNxiii  #1{\unskip}     \fi
\ifx \showISSN     \undefined \def \showISSN      #1{\unskip}     \fi
\ifx \showLCCN     \undefined \def \showLCCN      #1{\unskip}     \fi
\ifx \shownote     \undefined \def \shownote      #1{#1}          \fi
\ifx \showarticletitle \undefined \def \showarticletitle #1{#1}   \fi
\ifx \showURL      \undefined \def \showURL       {\relax}        \fi
\providecommand\bibfield[2]{#2}
\providecommand\bibinfo[2]{#2}
\providecommand\natexlab[1]{#1}
\providecommand\showeprint[2][]{arXiv:#2}
\makeatletter
\@ifundefined{NAT@parse@date}{}{\let\NAT@parse@date@orig\NAT@parse@date}
\@ifundefined{NAT@parse@date}{}{\def\NAT@parse@date#1#2#3#4#5#6@@{\NAT@parse@date@orig#1#2#3#4#5#6@@\def\NAT@tempyear{0000}\def\NAT@tempexlab{{?}}\ifx\NAT@year\NAT@tempyear\ifx\NAT@exlab\NAT@tempexlab\def\NAT@date{[n.\,d.]}\else\edef\NAT@date{[n.\,d.]\NAT@exlab}\fi\fi}}
\makeatother

\bibitem[Abdelfattah et~al\mbox{.}(2018)]%
        {abdelfattah2018dla}
\bibfield{author}{\bibinfo{person}{Mohamed~S. Abdelfattah},
  \bibinfo{person}{David Han}, \bibinfo{person}{Andrew Bitar},
  \bibinfo{person}{Roberto DiCecco}, \bibinfo{person}{Shane O'Connell},
  \bibinfo{person}{Nitika Shanker}, \bibinfo{person}{Joseph Chu},
  \bibinfo{person}{Ian Prins}, \bibinfo{person}{Joshua Fender},
  \bibinfo{person}{Andrew~C. Ling}, {and} \bibinfo{person}{Gordon~R. Chiu}.}
  \bibinfo{year}{2018}\natexlab{}.
\newblock \showarticletitle{{DLA}: Compiler and {FPGA} Overlay for Neural
  Network Inference Acceleration}. In \bibinfo{booktitle}{\emph{2018 28th
  International Conference on Field Programmable Logic and Applications
  (FPL)}}. \bibinfo{pages}{411--418}.
\newblock
\href{https://doi.org/10.1109/FPL.2018.00077}{doi:\nolinkurl{10.1109/FPL.2018.00077}}


\bibitem[{Advanced Micro Devices, Inc.}(2022)]%
        {amd_pg065}
{Advanced Micro Devices, Inc.} \bibinfo{year}{2022}\natexlab{}.
\newblock \bibinfo{booktitle}{\emph{{Clocking Wizard LogiCORE IP Product Guide
  (PG065)}}}.
\newblock {Advanced Micro Devices, Inc.}
\newblock
\urldef\tempurl%
\url{https://docs.amd.com/r/en-US/pg065-clk-wiz}
\showURL{%
\tempurl}
\newblock
\shownote{Version 6.0}.


\bibitem[Advanced Micro Devices, Inc.(2022)]%
        {amd2022am004}
Advanced Micro Devices, Inc. \bibinfo{year}{2022}\natexlab{}.
\newblock \bibinfo{booktitle}{\emph{{Versal ACAP DSP Engine Architecture
  Manual}}}.
\newblock Advanced Micro Devices, Inc.
\newblock
\urldef\tempurl%
\url{https://docs.amd.com/r/en-US/am004-versal-dsp-engine}
\showURL{%
\tempurl}
\newblock
\shownote{Version 1.2.1}.


\bibitem[Advanced Micro Devices, Inc.(2026)]%
        {amd2026am007}
Advanced Micro Devices, Inc. \bibinfo{year}{2026}\natexlab{}.
\newblock \bibinfo{booktitle}{\emph{{Versal Adaptive SoC Memory Resources
  Architecture Manual}}}.
\newblock Advanced Micro Devices, Inc.
\newblock
\urldef\tempurl%
\url{https://docs.amd.com/r/en-US/am007-versal-memory}
\showURL{%
\tempurl}
\newblock
\shownote{Version 1.2.1}.


\bibitem[{Advanced Micro Devices, Inc.}(2026)]%
        {amd2026ug1353}
\bibfield{author}{\bibinfo{person}{{Advanced Micro Devices, Inc.}}}
  \bibinfo{year}{2026}\natexlab{}.
\newblock \bibinfo{booktitle}{\emph{Versal Architecture {AI} Core Series
  Libraries Guide ({UG1353})}}.
\newblock
\urldef\tempurl%
\url{https://docs.amd.com/r/en-US/ug1353-versal-architecture-ai-libraries}
\showURL{%
\tempurl}
\newblock
\shownote{Version 2026.1, June 23, 2026}.


\bibitem[Aklah(2017)]%
        {aklah2017thesis}
\bibfield{author}{\bibinfo{person}{Zeyad~Tariq Aklah}.}
  \bibinfo{year}{2017}\natexlab{}.
\newblock \emph{\bibinfo{title}{A Hybrid Partially Reconfigurable Overlay
  Supporting Just-In-Time Assembly of Custom Accelerators on {FPGAs}}}.
\newblock \bibinfo{thesistype}{Ph.\,D. Dissertation}.
  \bibinfo{school}{University of Arkansas, Fayetteville}.
\newblock
\urldef\tempurl%
\url{https://scholarworks.uark.edu/etd/1928/}
\showURL{%
\tempurl}


\bibitem[Al~Kadi et~al\mbox{.}(2016)]%
        {alkadi2016fgpu}
\bibfield{author}{\bibinfo{person}{Muhammed Al~Kadi}, \bibinfo{person}{Benedikt
  Janssen}, {and} \bibinfo{person}{Michael Huebner}.}
  \bibinfo{year}{2016}\natexlab{}.
\newblock \showarticletitle{{FGPU}: An {SIMT}-Architecture for {FPGAs}}. In
  \bibinfo{booktitle}{\emph{Proceedings of the ACM/SIGDA International
  Symposium on Field-Programmable Gate Arrays}}
  \emph{(\bibinfo{series}{FPGA})}. \bibinfo{pages}{254--263}.
\newblock
\href{https://doi.org/10.1145/2847263.2847273}{doi:\nolinkurl{10.1145/2847263.2847273}}


\bibitem[Bai et~al\mbox{.}(2025)]%
        {bai2025dspopu}
\bibfield{author}{\bibinfo{person}{Yueyin Bai}, \bibinfo{person}{Song Zhang},
  \bibinfo{person}{Zhiyuan Ma}, \bibinfo{person}{Enhao Tang}, {and}
  \bibinfo{person}{Jun Yu}.} \bibinfo{year}{2025}\natexlab{}.
\newblock \showarticletitle{{DSP-OPU}: An {FPGA}-based overlay processor for
  digital signal processing}.
\newblock \bibinfo{journal}{\emph{Electronic Research Archive}}
  \bibinfo{volume}{33}, \bibinfo{number}{5} (\bibinfo{year}{2025}),
  \bibinfo{pages}{2698--2718}.
\newblock
\href{https://doi.org/10.3934/era.2025119}{doi:\nolinkurl{10.3934/era.2025119}}


\bibitem[Bandara et~al\mbox{.}(2022)]%
        {bandara2022revamp}
\bibfield{author}{\bibinfo{person}{Thilini~Kaushalya Bandara},
  \bibinfo{person}{Dhananjaya Wijerathne}, \bibinfo{person}{Tulika Mitra},
  {and} \bibinfo{person}{Li-Shiuan Peh}.} \bibinfo{year}{2022}\natexlab{}.
\newblock \showarticletitle{{REVAMP}: A Systematic Framework for Heterogeneous
  {CGRA} Realization}. In \bibinfo{booktitle}{\emph{Proceedings of the 27th ACM
  International Conference on Architectural Support for Programming Languages
  and Operating Systems}}. \bibinfo{pages}{918--932}.
\newblock
\href{https://doi.org/10.1145/3503222.3507772}{doi:\nolinkurl{10.1145/3503222.3507772}}


\bibitem[Basalama and Cong(2025)]%
        {basalama2025streamhls}
\bibfield{author}{\bibinfo{person}{Suhail Basalama} {and}
  \bibinfo{person}{Jason Cong}.} \bibinfo{year}{2025}\natexlab{}.
\newblock \showarticletitle{{Stream-HLS}: Towards Automatic Dataflow
  Acceleration}. In \bibinfo{booktitle}{\emph{Proceedings of the 2025 ACM/SIGDA
  International Symposium on Field Programmable Gate Arrays}}.
  \bibinfo{pages}{103--114}.
\newblock
\href{https://doi.org/10.1145/3706628.3708878}{doi:\nolinkurl{10.1145/3706628.3708878}}


\bibitem[Basalama et~al\mbox{.}(2020)]%
        {basalama2020spar2}
\bibfield{author}{\bibinfo{person}{Suhail Basalama},
  \bibinfo{person}{Atiyehsadat Panahi}, \bibinfo{person}{Ange-Thierry Ishimwe},
  {and} \bibinfo{person}{David Andrews}.} \bibinfo{year}{2020}\natexlab{}.
\newblock \showarticletitle{{SPAR-2}: A {SIMD} Processor Array for Machine
  Learning in {IoT} Devices}. In \bibinfo{booktitle}{\emph{2020 3rd
  International Conference on Data Intelligence and Security (ICDIS)}}.
  \bibinfo{pages}{141--147}.
\newblock
\href{https://doi.org/10.1109/ICDIS50059.2020.00025}{doi:\nolinkurl{10.1109/ICDIS50059.2020.00025}}


\bibitem[Ben~Abdelhamid(2023)]%
        {benabdelhamid2023thesis}
\bibfield{author}{\bibinfo{person}{Riadh Ben~Abdelhamid}.}
  \bibinfo{year}{2023}\natexlab{}.
\newblock \emph{\bibinfo{title}{Energy-efficient many-core overlay architecture
  for reconfigurable chips}}.
\newblock \bibinfo{thesistype}{Ph.\,D. Dissertation}.
  \bibinfo{school}{University of Tsukuba}.
\newblock
\href{https://doi.org/10.15068/0002008220}{doi:\nolinkurl{10.15068/0002008220}}


\bibitem[Ben~Abdelhamid et~al\mbox{.}(2025)]%
        {benabdelhamid2025kilocore}
\bibfield{author}{\bibinfo{person}{Riadh Ben~Abdelhamid},
  \bibinfo{person}{Vladislav Valek}, \bibinfo{person}{Kevin Klein}, {and}
  \bibinfo{person}{Dirk Koch}.} \bibinfo{year}{2025}\natexlab{}.
\newblock \showarticletitle{Design Space Exploration of Fast {RISC-V}
  Processors for Scalable Kilo-Core {FPGA} Systems}. In
  \bibinfo{booktitle}{\emph{2025 35th International Conference on
  Field-Programmable Logic and Applications (FPL)}}. \bibinfo{pages}{282--290}.
\newblock
\href{https://doi.org/10.1109/FPL68686.2025.00046}{doi:\nolinkurl{10.1109/FPL68686.2025.00046}}


\bibitem[Ben~Abdelhamid et~al\mbox{.}(2019)]%
        {benabdelhamid2019mitraca}
\bibfield{author}{\bibinfo{person}{Riadh Ben~Abdelhamid},
  \bibinfo{person}{Yoshiki Yamaguchi}, {and} \bibinfo{person}{Taisuke Boku}.}
  \bibinfo{year}{2019}\natexlab{}.
\newblock \showarticletitle{{MITRACA}: A Next-Gen Heterogeneous Architecture}.
  In \bibinfo{booktitle}{\emph{2019 IEEE 13th International Symposium on
  Embedded Multicore/Many-core Systems-on-Chip (MCSoC)}}.
  \bibinfo{pages}{304--311}.
\newblock
\href{https://doi.org/10.1109/MCSoC.2019.00050}{doi:\nolinkurl{10.1109/MCSoC.2019.00050}}


\bibitem[Ben~Abdelhamid et~al\mbox{.}(2021)]%
        {benabdelhamid2021dragon}
\bibfield{author}{\bibinfo{person}{Riadh Ben~Abdelhamid},
  \bibinfo{person}{Yoshiki Yamaguchi}, {and} \bibinfo{person}{Taisuke Boku}.}
  \bibinfo{year}{2021}\natexlab{}.
\newblock \showarticletitle{A Highly-Efficient and Tightly-Connected Many-Core
  Overlay Architecture}.
\newblock \bibinfo{journal}{\emph{IEEE Access}}  \bibinfo{volume}{9}
  (\bibinfo{year}{2021}), \bibinfo{pages}{65277--65292}.
\newblock
\href{https://doi.org/10.1109/ACCESS.2021.3074171}{doi:\nolinkurl{10.1109/ACCESS.2021.3074171}}


\bibitem[Ben~Abdelhamid et~al\mbox{.}(2023)]%
        {benabdelhamid2023scalable}
\bibfield{author}{\bibinfo{person}{Riadh Ben~Abdelhamid},
  \bibinfo{person}{Yoshiki Yamaguchi}, {and} \bibinfo{person}{Taisuke Boku}.}
  \bibinfo{year}{2023}\natexlab{}.
\newblock \showarticletitle{A Scalable Many-core Overlay Architecture on an
  {HBM2}-enabled Multi-Die {FPGA}}.
\newblock \bibinfo{journal}{\emph{ACM Transactions on Reconfigurable Technology
  and Systems}} \bibinfo{volume}{16}, \bibinfo{number}{1}
  (\bibinfo{year}{2023}), \bibinfo{pages}{15:1--15:33}.
\newblock
\href{https://doi.org/10.1145/3547657}{doi:\nolinkurl{10.1145/3547657}}


\bibitem[Bouaziz and Fahmy(2024)]%
        {bouaziz2024mlir}
\bibfield{author}{\bibinfo{person}{Mohamed Bouaziz} {and}
  \bibinfo{person}{Suhaib~A. Fahmy}.} \bibinfo{year}{2024}\natexlab{}.
\newblock \showarticletitle{Leveraging {MLIR} for Efficient Irregular-Shaped
  {CGRA} Overlay Design: ({PhD} Forum Paper)}. In
  \bibinfo{booktitle}{\emph{2024 IEEE 35th International Conference on
  Application-Specific Systems, Architectures and Processors (ASAP)}}.
  \bibinfo{pages}{204--205}.
\newblock
\href{https://doi.org/10.1109/ASAP61560.2024.00048}{doi:\nolinkurl{10.1109/ASAP61560.2024.00048}}


\bibitem[Boutros et~al\mbox{.}(2021)]%
        {boutros2021specializing}
\bibfield{author}{\bibinfo{person}{Andrew Boutros}, \bibinfo{person}{Eriko
  Nurvitadhi}, {and} \bibinfo{person}{Vaughn Betz}.}
  \bibinfo{year}{2021}\natexlab{}.
\newblock \showarticletitle{Specializing for Efficiency: Customizing {AI}
  Inference Processors on {FPGAs}}. In \bibinfo{booktitle}{\emph{2021
  International Conference on Microelectronics (ICM)}}.
  \bibinfo{pages}{62--65}.
\newblock
\href{https://doi.org/10.1109/ICM52667.2021.9664938}{doi:\nolinkurl{10.1109/ICM52667.2021.9664938}}


\bibitem[Brant and Lemieux(2012)]%
        {brant2012zuma}
\bibfield{author}{\bibinfo{person}{Alexander Brant} {and} \bibinfo{person}{Guy
  G.~F. Lemieux}.} \bibinfo{year}{2012}\natexlab{}.
\newblock \showarticletitle{{ZUMA}: An Open {FPGA} Overlay Architecture}. In
  \bibinfo{booktitle}{\emph{2012 IEEE 20th International Symposium on
  Field-Programmable Custom Computing Machines (FCCM)}}.
  \bibinfo{pages}{93--96}.
\newblock
\href{https://doi.org/10.1109/FCCM.2012.25}{doi:\nolinkurl{10.1109/FCCM.2012.25}}


\bibitem[Capalija(2017)]%
        {capalija2017thesis}
\bibfield{author}{\bibinfo{person}{Davor Capalija}.}
  \bibinfo{year}{2017}\natexlab{}.
\newblock \emph{\bibinfo{title}{Architecture, Mapping Algorithms and Physical
  Design of Mesh-of-Functional-Units {FPGA} Overlays for Pipelined Execution of
  Data Flow Graphs}}.
\newblock \bibinfo{thesistype}{Ph.\,D. Dissertation}.
  \bibinfo{school}{University of Toronto}.
\newblock
\urldef\tempurl%
\url{https://hdl.handle.net/1807/78948}
\showURL{%
\tempurl}


\bibitem[Capalija and Abdelrahman(2013)]%
        {capalija2013overlay}
\bibfield{author}{\bibinfo{person}{Davor Capalija} {and}
  \bibinfo{person}{Tarek~S. Abdelrahman}.} \bibinfo{year}{2013}\natexlab{}.
\newblock \showarticletitle{A High-Performance Overlay Architecture for
  Pipelined Execution of Data Flow Graphs}. In \bibinfo{booktitle}{\emph{2013
  23rd International Conference on Field Programmable Logic and Applications
  (FPL)}}. \bibinfo{pages}{1--8}.
\newblock
\href{https://doi.org/10.1109/FPL.2013.6645515}{doi:\nolinkurl{10.1109/FPL.2013.6645515}}


\bibitem[Cheah et~al\mbox{.}(2014)]%
        {cheah2014idea}
\bibfield{author}{\bibinfo{person}{Hui~Yan Cheah}, \bibinfo{person}{Fredrik
  Brosser}, \bibinfo{person}{Suhaib~A. Fahmy}, {and}
  \bibinfo{person}{Douglas~L. Maskell}.} \bibinfo{year}{2014}\natexlab{}.
\newblock \showarticletitle{The {iDEA DSP} Block-Based Soft Processor for
  {FPGAs}}.
\newblock \bibinfo{journal}{\emph{ACM Transactions on Reconfigurable Technology
  and Systems}} \bibinfo{volume}{7}, \bibinfo{number}{3}, Article
  \bibinfo{articleno}{19} (\bibinfo{year}{2014}), \bibinfo{numpages}{23}~pages.
\newblock
\href{https://doi.org/10.1145/2629443}{doi:\nolinkurl{10.1145/2629443}}


\bibitem[Cheah et~al\mbox{.}(2015)]%
        {cheah2015forwarding}
\bibfield{author}{\bibinfo{person}{Hui~Yan Cheah}, \bibinfo{person}{Suhaib~A.
  Fahmy}, {and} \bibinfo{person}{Nachiket Kapre}.}
  \bibinfo{year}{2015}\natexlab{}.
\newblock \showarticletitle{On Data Forwarding in Deeply Pipelined Soft
  Processors}. In \bibinfo{booktitle}{\emph{Proceedings of the ACM/SIGDA
  International Symposium on Field-Programmable Gate Arrays}}
  \emph{(\bibinfo{series}{FPGA})}. \bibinfo{pages}{181--189}.
\newblock
\href{https://doi.org/10.1145/2684746.2689067}{doi:\nolinkurl{10.1145/2684746.2689067}}


\bibitem[Chen et~al\mbox{.}(2024)]%
        {chen2024allo}
\bibfield{author}{\bibinfo{person}{Hongzheng Chen}, \bibinfo{person}{Niansong
  Zhang}, \bibinfo{person}{Shaojie Xiang}, \bibinfo{person}{Zhichen Zeng},
  \bibinfo{person}{Mengjia Dai}, {and} \bibinfo{person}{Zhiru Zhang}.}
  \bibinfo{year}{2024}\natexlab{}.
\newblock \showarticletitle{{Allo}: A Programming Model for Composable
  Accelerator Design}.
\newblock \bibinfo{journal}{\emph{Proceedings of the ACM on Programming
  Languages}} \bibinfo{volume}{8}, \bibinfo{number}{PLDI}, Article
  \bibinfo{articleno}{171} (\bibinfo{year}{2024}),
  \bibinfo{numpages}{28}~pages.
\newblock
\href{https://doi.org/10.1145/3656401}{doi:\nolinkurl{10.1145/3656401}}


\bibitem[Chi et~al\mbox{.}(2018)]%
        {chi2018soda}
\bibfield{author}{\bibinfo{person}{Yuze Chi}, \bibinfo{person}{Jason Cong},
  \bibinfo{person}{Peng Wei}, {and} \bibinfo{person}{Peipei Zhou}.}
  \bibinfo{year}{2018}\natexlab{}.
\newblock \showarticletitle{{SODA}: Stencil with Optimized Dataflow
  Architecture}. In \bibinfo{booktitle}{\emph{Proceedings of the IEEE/ACM
  International Conference on Computer-Aided Design}}
  \emph{(\bibinfo{series}{ICCAD})}. Article \bibinfo{articleno}{116},
  \bibinfo{numpages}{8}~pages.
\newblock
\href{https://doi.org/10.1145/3240765.3240850}{doi:\nolinkurl{10.1145/3240765.3240850}}


\bibitem[Choudhury et~al\mbox{.}(2023)]%
        {choudhury2023flowpix}
\bibfield{author}{\bibinfo{person}{Ziaul Choudhury}, \bibinfo{person}{Anish
  Gulati}, {and} \bibinfo{person}{Suresh Purini}.}
  \bibinfo{year}{2023}\natexlab{}.
\newblock \showarticletitle{{FlowPix}: Accelerating Image Processing Pipelines
  on an {FPGA} Overlay using a Domain Specific Compiler}.
\newblock \bibinfo{journal}{\emph{ACM Transactions on Architecture and Code
  Optimization}} \bibinfo{volume}{20}, \bibinfo{number}{4}
  (\bibinfo{year}{2023}), \bibinfo{pages}{60:1--60:25}.
\newblock
\href{https://doi.org/10.1145/3629523}{doi:\nolinkurl{10.1145/3629523}}


\bibitem[Choudhury et~al\mbox{.}(2022)]%
        {choudhury2022cnn}
\bibfield{author}{\bibinfo{person}{Ziaul Choudhury}, \bibinfo{person}{Shashwat
  Shrivastava}, \bibinfo{person}{Lavanya Ramapantulu}, {and}
  \bibinfo{person}{Suresh Purini}.} \bibinfo{year}{2022}\natexlab{}.
\newblock \showarticletitle{An {FPGA} Overlay for {CNN} Inference with
  Fine-grained Flexible Parallelism}.
\newblock \bibinfo{journal}{\emph{ACM Transactions on Architecture and Code
  Optimization}} \bibinfo{volume}{19}, \bibinfo{number}{3}
  (\bibinfo{year}{2022}), \bibinfo{pages}{34:1--34:26}.
\newblock
\href{https://doi.org/10.1145/3519598}{doi:\nolinkurl{10.1145/3519598}}


\bibitem[Chung(1997)]%
        {chung1997heat}
\bibfield{author}{\bibinfo{person}{Fan R.~K. Chung}.}
  \bibinfo{year}{1997}\natexlab{}.
\newblock \bibinfo{booktitle}{\emph{Spectral Graph Theory}}.
  \bibinfo{series}{CBMS Regional Conference Series in Mathematics},
  Vol.~\bibinfo{volume}{92}.
\newblock \bibinfo{publisher}{American Mathematical Society},
  \bibinfo{address}{Providence, RI}, Chapter 10: Heat Kernels.
\newblock
\showISBNx{978-0-8218-0315-8}


\bibitem[Cong et~al\mbox{.}(2011)]%
        {cong2011hls}
\bibfield{author}{\bibinfo{person}{Jason Cong}, \bibinfo{person}{Bin Liu},
  \bibinfo{person}{Stephen Neuendorffer}, \bibinfo{person}{Juanjo Noguera},
  \bibinfo{person}{Kees Vissers}, {and} \bibinfo{person}{Zhiru Zhang}.}
  \bibinfo{year}{2011}\natexlab{}.
\newblock \showarticletitle{High-Level Synthesis for {FPGAs}: From Prototyping
  to Deployment}.
\newblock \bibinfo{journal}{\emph{IEEE Transactions on Computer-Aided Design of
  Integrated Circuits and Systems}} \bibinfo{volume}{30}, \bibinfo{number}{4}
  (\bibinfo{year}{2011}), \bibinfo{pages}{473--491}.
\newblock
\href{https://doi.org/10.1109/TCAD.2011.2110592}{doi:\nolinkurl{10.1109/TCAD.2011.2110592}}


\bibitem[Cong et~al\mbox{.}(2018)]%
        {cong2018autoaccel}
\bibfield{author}{\bibinfo{person}{Jason Cong}, \bibinfo{person}{Peng Wei},
  \bibinfo{person}{Cody~Hao Yu}, {and} \bibinfo{person}{Peng Zhang}.}
  \bibinfo{year}{2018}\natexlab{}.
\newblock \showarticletitle{Automated Accelerator Generation and Optimization
  with Composable, Parallel and Pipeline Architecture}. In
  \bibinfo{booktitle}{\emph{Proceedings of the ACM/IEEE Design Automation
  Conference}} \emph{(\bibinfo{series}{DAC})}. \bibinfo{pages}{1--6}.
\newblock
\href{https://doi.org/10.1145/3195970.3195999}{doi:\nolinkurl{10.1145/3195970.3195999}}


\bibitem[Coole and Stitt(2010)]%
        {coole2010intermediate}
\bibfield{author}{\bibinfo{person}{James Coole} {and} \bibinfo{person}{Greg
  Stitt}.} \bibinfo{year}{2010}\natexlab{}.
\newblock \showarticletitle{Intermediate Fabrics: Virtual Architectures for
  Circuit Portability and Fast Placement and Routing}. In
  \bibinfo{booktitle}{\emph{Proceedings of the International Conference on
  Hardware/Software Codesign and System Synthesis}}
  \emph{(\bibinfo{series}{CODES+ISSS})}. \bibinfo{pages}{13--22}.
\newblock
\href{https://doi.org/10.1145/1878961.1878966}{doi:\nolinkurl{10.1145/1878961.1878966}}


\bibitem[Coole and Stitt(2015)]%
        {coole2015adjustable}
\bibfield{author}{\bibinfo{person}{James Coole} {and} \bibinfo{person}{Greg
  Stitt}.} \bibinfo{year}{2015}\natexlab{}.
\newblock \showarticletitle{Adjustable-Cost Overlays for Runtime Compilation}.
  In \bibinfo{booktitle}{\emph{Proceedings of the IEEE International Symposium
  on Field-Programmable Custom Computing Machines}}
  \emph{(\bibinfo{series}{FCCM})}. \bibinfo{pages}{21--24}.
\newblock
\href{https://doi.org/10.1109/FCCM.2015.49}{doi:\nolinkurl{10.1109/FCCM.2015.49}}


\bibitem[Da~Silva et~al\mbox{.}(2019)]%
        {silva2019ready}
\bibfield{author}{\bibinfo{person}{Lucas~Bragan{\c{c}}a Da~Silva},
  \bibinfo{person}{Ricardo Ferreira}, \bibinfo{person}{Michael Canesche},
  \bibinfo{person}{Marcelo~M. Menezes}, \bibinfo{person}{Maria~D. Vieira},
  \bibinfo{person}{Jeronimo Penha}, \bibinfo{person}{Peter Jamieson}, {and}
  \bibinfo{person}{Jos{\'e} Augusto~M. Nacif}.}
  \bibinfo{year}{2019}\natexlab{}.
\newblock \showarticletitle{{READY}: A Fine-Grained Multithreading Overlay
  Framework for Modern {CPU-FPGA} Dataflow Applications}.
\newblock \bibinfo{journal}{\emph{ACM Transactions on Embedded Computing
  Systems}} \bibinfo{volume}{18}, \bibinfo{number}{5s} (\bibinfo{year}{2019}),
  \bibinfo{pages}{56:1--56:20}.
\newblock
\href{https://doi.org/10.1145/3358187}{doi:\nolinkurl{10.1145/3358187}}


\bibitem[da~Silva et~al\mbox{.}(2025)]%
        {silva2025hpcgra}
\bibfield{author}{\bibinfo{person}{Lucas~Bragan{\c{c}}a da Silva},
  \bibinfo{person}{C{\'e}sar Grandis}, \bibinfo{person}{Jeronimo Penha},
  \bibinfo{person}{Jos{\'e} Augusto~M. Nacif}, {and}
  \bibinfo{person}{Ricardo~S. Ferreira}.} \bibinfo{year}{2025}\natexlab{}.
\newblock \showarticletitle{Reconfigurable Domain-Specific Architectures Based
  on Coarse-Grained Operators in High-Performance {FPGAs}}.
\newblock \bibinfo{journal}{\emph{Concurrency and Computation: Practice and
  Experience}} \bibinfo{volume}{37}, \bibinfo{number}{15--17}
  (\bibinfo{year}{2025}), \bibinfo{pages}{e70170}.
\newblock
\href{https://doi.org/10.1002/cpe.70170}{doi:\nolinkurl{10.1002/cpe.70170}}


\bibitem[Dickerson et~al\mbox{.}(2026)]%
        {dickerson2026pyjit2fpga}
\bibfield{author}{\bibinfo{person}{Mitchel Dickerson}, \bibinfo{person}{Anand
  Srinivasan}, {and} \bibinfo{person}{Michael Franz}.}
  \bibinfo{year}{2026}\natexlab{}.
\newblock \showarticletitle{Practical {Python FPGA} Acceleration with Fast
  Just-In-Time Compilation and Configuration}.
\newblock \bibinfo{journal}{\emph{ACM Transactions on Architecture and Code
  Optimization}} \bibinfo{volume}{23}, \bibinfo{number}{1}, Article
  \bibinfo{articleno}{31} (\bibinfo{year}{2026}), \bibinfo{numpages}{25}~pages.
\newblock
\href{https://doi.org/10.1145/3797265}{doi:\nolinkurl{10.1145/3797265}}


\bibitem[Drewes et~al\mbox{.}(2024)]%
        {drewes2024template}
\bibfield{author}{\bibinfo{person}{Anna Drewes}, \bibinfo{person}{Vitalii
  Burtsev}, \bibinfo{person}{Bala Gurumurthy}, \bibinfo{person}{Martin
  Wilhelm}, \bibinfo{person}{David Broneske}, \bibinfo{person}{Gunter Saake},
  {and} \bibinfo{person}{Thilo Pionteck}.} \bibinfo{year}{2024}\natexlab{}.
\newblock \showarticletitle{An Architectural Template for {FPGA} Overlays
  Targeting Data Flow Applications}. In \bibinfo{booktitle}{\emph{2024 IEEE
  International Parallel and Distributed Processing Symposium Workshops
  (IPDPSW)}}. \bibinfo{pages}{162--168}.
\newblock
\href{https://doi.org/10.1109/IPDPSW63119.2024.00042}{doi:\nolinkurl{10.1109/IPDPSW63119.2024.00042}}


\bibitem[Duarte et~al\mbox{.}(2017)]%
        {duarte2017scratch}
\bibfield{author}{\bibinfo{person}{Pedro Duarte}, \bibinfo{person}{Pedro
  Tom{\'a}s}, {and} \bibinfo{person}{Gabriel Falc{\~a}o}.}
  \bibinfo{year}{2017}\natexlab{}.
\newblock \showarticletitle{{SCRATCH}: An End-to-End Application-Aware
  Soft-{GPGPU} Architecture and Trimming Tool}. In
  \bibinfo{booktitle}{\emph{Proceedings of the Annual IEEE/ACM International
  Symposium on Microarchitecture}} \emph{(\bibinfo{series}{MICRO})}.
  \bibinfo{pages}{165--177}.
\newblock
\href{https://doi.org/10.1145/3123939.3123953}{doi:\nolinkurl{10.1145/3123939.3123953}}


\bibitem[Elshimy et~al\mbox{.}(2023)]%
        {elshimy2023nearmemory}
\bibfield{author}{\bibinfo{person}{Mahmoud~Ahmed Elshimy},
  \bibinfo{person}{Veronia Iskandar}, \bibinfo{person}{Diana Goehringer}, {and}
  \bibinfo{person}{Mohamed~A. Abd El~Ghany}.} \bibinfo{year}{2023}\natexlab{}.
\newblock \showarticletitle{A Near-Memory Dynamically Programmable Many-Core
  Overlay}. In \bibinfo{booktitle}{\emph{2023 IEEE 16th International Symposium
  on Embedded Multicore/Many-core Systems-on-Chip (MCSoC)}}.
  \bibinfo{pages}{268--275}.
\newblock
\href{https://doi.org/10.1109/MCSOC60832.2023.00047}{doi:\nolinkurl{10.1109/MCSOC60832.2023.00047}}


\bibitem[Fang et~al\mbox{.}(2017)]%
        {fang2017secure}
\bibfield{author}{\bibinfo{person}{Xin Fang}, \bibinfo{person}{Stratis
  Ioannidis}, {and} \bibinfo{person}{Miriam Leeser}.}
  \bibinfo{year}{2017}\natexlab{}.
\newblock \showarticletitle{Secure Function Evaluation Using an {FPGA} Overlay
  Architecture}. In \bibinfo{booktitle}{\emph{Proceedings of the 2017 ACM/SIGDA
  International Symposium on Field-Programmable Gate Arrays}}.
  \bibinfo{pages}{257--266}.
\newblock
\href{https://doi.org/10.1145/3020078.3021746}{doi:\nolinkurl{10.1145/3020078.3021746}}


\bibitem[Guo et~al\mbox{.}(2023)]%
        {guo2023tapa}
\bibfield{author}{\bibinfo{person}{Licheng Guo}, \bibinfo{person}{Yuze Chi},
  \bibinfo{person}{Jason Lau}, \bibinfo{person}{Linghao Song},
  \bibinfo{person}{Xingyu Tian}, \bibinfo{person}{Moazin Khatti},
  \bibinfo{person}{Weikang Qiao}, \bibinfo{person}{Jie Wang},
  \bibinfo{person}{Ecenur Ustun}, \bibinfo{person}{Zhenman Fang},
  \bibinfo{person}{Zhiru Zhang}, {and} \bibinfo{person}{Jason Cong}.}
  \bibinfo{year}{2023}\natexlab{}.
\newblock \showarticletitle{{TAPA}: A Scalable Task-Parallel Dataflow
  Programming Framework for Modern {FPGAs} with Co-Optimization of {HLS} and
  Physical Design}.
\newblock \bibinfo{journal}{\emph{ACM Transactions on Reconfigurable Technology
  and Systems}} \bibinfo{volume}{16}, \bibinfo{number}{4}, Article
  \bibinfo{articleno}{63} (\bibinfo{year}{2023}), \bibinfo{numpages}{31}~pages.
\newblock
\href{https://doi.org/10.1145/3609335}{doi:\nolinkurl{10.1145/3609335}}


\bibitem[Guo et~al\mbox{.}(2021)]%
        {guo2021autobridge}
\bibfield{author}{\bibinfo{person}{Licheng Guo}, \bibinfo{person}{Yuze Chi},
  \bibinfo{person}{Jie Wang}, \bibinfo{person}{Jason Lau},
  \bibinfo{person}{Weikang Qiao}, \bibinfo{person}{Ecenur Ustun},
  \bibinfo{person}{Zhiru Zhang}, {and} \bibinfo{person}{Jason Cong}.}
  \bibinfo{year}{2021}\natexlab{}.
\newblock \showarticletitle{{AutoBridge}: Coupling Coarse-Grained Floorplanning
  and Pipelining for High-Frequency {HLS} Design on Multi-Die {FPGAs}}. In
  \bibinfo{booktitle}{\emph{The 2021 ACM/SIGDA International Symposium on
  Field-Programmable Gate Arrays}}. \bibinfo{pages}{81--92}.
\newblock
\href{https://doi.org/10.1145/3431920.3439289}{doi:\nolinkurl{10.1145/3431920.3439289}}


\bibitem[Hammond et~al\mbox{.}(2011)]%
        {hammond2011wavelets}
\bibfield{author}{\bibinfo{person}{David~K. Hammond}, \bibinfo{person}{Pierre
  Vandergheynst}, {and} \bibinfo{person}{R{\'e}mi Gribonval}.}
  \bibinfo{year}{2011}\natexlab{}.
\newblock \showarticletitle{Wavelets on Graphs via Spectral Graph Theory}.
\newblock \bibinfo{journal}{\emph{Applied and Computational Harmonic Analysis}}
  \bibinfo{volume}{30}, \bibinfo{number}{2} (\bibinfo{date}{March}
  \bibinfo{year}{2011}), \bibinfo{pages}{129--150}.
\newblock
\href{https://doi.org/10.1016/j.acha.2010.04.005}{doi:\nolinkurl{10.1016/j.acha.2010.04.005}}


\bibitem[He et~al\mbox{.}(2025)]%
        {he2025intar}
\bibfield{author}{\bibinfo{person}{Zifan He}, \bibinfo{person}{Anderson
  Truong}, \bibinfo{person}{Yingqi Cao}, {and} \bibinfo{person}{Jason Cong}.}
  \bibinfo{year}{2025}\natexlab{}.
\newblock \showarticletitle{{InTAR}: Inter-Task Auto-Reconfigurable Accelerator
  Design for High Data Volume Variation in {DNNs}}. In
  \bibinfo{booktitle}{\emph{2025 IEEE 33rd Annual International Symposium on
  Field-Programmable Custom Computing Machines (FCCM)}}.
  \bibinfo{pages}{123--132}.
\newblock
\href{https://doi.org/10.1109/FCCM62733.2025.00047}{doi:\nolinkurl{10.1109/FCCM62733.2025.00047}}


\bibitem[He et~al\mbox{.}(2026)]%
        {he2026skipopu}
\bibfield{author}{\bibinfo{person}{Zicheng He}, \bibinfo{person}{Anhao Zhao},
  \bibinfo{person}{Xiaoyu Shen}, \bibinfo{person}{Chen Wu}, {and}
  \bibinfo{person}{Lei He}.} \bibinfo{year}{2026}\natexlab{}.
\newblock \showarticletitle{{SkipOPU}: An {FPGA}-based Overlay Processor for
  Large Language Models with Dynamically Allocated Computation}.
\newblock \bibinfo{journal}{\emph{ACM Transactions on Reconfigurable Technology
  and Systems}} (\bibinfo{year}{2026}).
\newblock
\href{https://doi.org/10.1145/3837086}{doi:\nolinkurl{10.1145/3837086}}
\newblock
\shownote{Just Accepted}.


\bibitem[Ioannou et~al\mbox{.}(2020)]%
        {ioannou2020convolution}
\bibfield{author}{\bibinfo{person}{Lenos Ioannou}, \bibinfo{person}{Abdullah
  Al-Dujaili}, {and} \bibinfo{person}{Suhaib~A. Fahmy}.}
  \bibinfo{year}{2020}\natexlab{}.
\newblock \showarticletitle{High Throughput Spatial Convolution Filters on
  {FPGAs}}.
\newblock \bibinfo{journal}{\emph{IEEE Transactions on Very Large Scale
  Integration (VLSI) Systems}} \bibinfo{volume}{28}, \bibinfo{number}{6}
  (\bibinfo{year}{2020}), \bibinfo{pages}{1392--1402}.
\newblock
\href{https://doi.org/10.1109/TVLSI.2020.2987202}{doi:\nolinkurl{10.1109/TVLSI.2020.2987202}}


\bibitem[Ioannou and Fahmy(2019)]%
        {ioannou2019lightweight}
\bibfield{author}{\bibinfo{person}{Lenos Ioannou} {and}
  \bibinfo{person}{Suhaib~A. Fahmy}.} \bibinfo{year}{2019}\natexlab{}.
\newblock \showarticletitle{Lightweight Programmable {DSP} Block Overlay for
  Streaming Neural Network Acceleration}. In \bibinfo{booktitle}{\emph{2019
  International Conference on Field-Programmable Technology (ICFPT)}}.
  \bibinfo{pages}{355--358}.
\newblock
\href{https://doi.org/10.1109/ICFPT47387.2019.00066}{doi:\nolinkurl{10.1109/ICFPT47387.2019.00066}}


\bibitem[Ioannou and Fahmy(2022)]%
        {ioannou2022streaming}
\bibfield{author}{\bibinfo{person}{Lenos Ioannou} {and}
  \bibinfo{person}{Suhaib~A. Fahmy}.} \bibinfo{year}{2022}\natexlab{}.
\newblock \showarticletitle{Streaming Overlay Architecture for Lightweight
  {LSTM} Computation on {FPGA SoCs}}.
\newblock \bibinfo{journal}{\emph{ACM Transactions on Reconfigurable Technology
  and Systems}} \bibinfo{volume}{16}, \bibinfo{number}{1}, Article
  \bibinfo{articleno}{8} (\bibinfo{year}{2022}), \bibinfo{numpages}{26}~pages.
\newblock
\href{https://doi.org/10.1145/3543069}{doi:\nolinkurl{10.1145/3543069}}


\bibitem[Jain et~al\mbox{.}(2015)]%
        {jain2015efficient}
\bibfield{author}{\bibinfo{person}{Abhishek~Kumar Jain},
  \bibinfo{person}{Suhaib~A. Fahmy}, {and} \bibinfo{person}{Douglas~L.
  Maskell}.} \bibinfo{year}{2015}\natexlab{}.
\newblock \showarticletitle{Efficient Overlay Architecture Based on {DSP}
  Blocks}. In \bibinfo{booktitle}{\emph{2015 IEEE 23rd Annual International
  Symposium on Field-Programmable Custom Computing Machines (FCCM)}}.
  \bibinfo{publisher}{IEEE}, \bibinfo{pages}{25--28}.
\newblock
\href{https://doi.org/10.1109/FCCM.2015.15}{doi:\nolinkurl{10.1109/FCCM.2015.15}}


\bibitem[Jain et~al\mbox{.}(2016a)]%
        {jain2016deco}
\bibfield{author}{\bibinfo{person}{Abhishek~Kumar Jain},
  \bibinfo{person}{Xiangwei Li}, \bibinfo{person}{Pranjul Singhai},
  \bibinfo{person}{Douglas~L. Maskell}, {and} \bibinfo{person}{Suhaib~A.
  Fahmy}.} \bibinfo{year}{2016}\natexlab{a}.
\newblock \showarticletitle{{DeCO}: A {DSP} Block Based {FPGA} Accelerator
  Overlay With Low Overhead Interconnect}. In \bibinfo{booktitle}{\emph{2016
  IEEE 24th Annual International Symposium on Field-Programmable Custom
  Computing Machines (FCCM)}}. \bibinfo{pages}{1--8}.
\newblock
\href{https://doi.org/10.1109/FCCM.2016.10}{doi:\nolinkurl{10.1109/FCCM.2016.10}}


\bibitem[Jain et~al\mbox{.}(2016b)]%
        {jain2016throughput}
\bibfield{author}{\bibinfo{person}{Abhishek~Kumar Jain},
  \bibinfo{person}{Douglas~L. Maskell}, {and} \bibinfo{person}{Suhaib~A.
  Fahmy}.} \bibinfo{year}{2016}\natexlab{b}.
\newblock \showarticletitle{Throughput Oriented {FPGA} Overlays Using {DSP}
  Blocks}. In \bibinfo{booktitle}{\emph{2016 Design, Automation \& Test in
  Europe Conference \& Exhibition (DATE)}}. \bibinfo{pages}{1628--1633}.
\newblock
\urldef\tempurl%
\url{https://abhishekkumarjain.github.io/files/DATE2016.pdf}
\showURL{%
\tempurl}


\bibitem[Jain et~al\mbox{.}(2022)]%
        {jain2022jit}
\bibfield{author}{\bibinfo{person}{Abhishek~Kumar Jain},
  \bibinfo{person}{Douglas~L. Maskell}, {and} \bibinfo{person}{Suhaib~A.
  Fahmy}.} \bibinfo{year}{2022}\natexlab{}.
\newblock \showarticletitle{Coarse Grained {FPGA} Overlay for Rapid
  Just-In-Time Accelerator Compilation}.
\newblock \bibinfo{journal}{\emph{IEEE Transactions on Parallel and Distributed
  Systems}} \bibinfo{volume}{33}, \bibinfo{number}{6} (\bibinfo{year}{2022}),
  \bibinfo{pages}{1478--1490}.
\newblock
\href{https://doi.org/10.1109/TPDS.2021.3116859}{doi:\nolinkurl{10.1109/TPDS.2021.3116859}}


\bibitem[Jamal(2018)]%
        {jamal2018debug}
\bibfield{author}{\bibinfo{person}{Al-Shahna Jamal}.}
  \bibinfo{year}{2018}\natexlab{}.
\newblock \emph{\bibinfo{title}{An {FPGA} Overlay Architecture Supporting
  Software-like Compile Times During On-chip Debug of High-level Synthesis
  Designs}}.
\newblock Master of Applied Science thesis. \bibinfo{school}{University of
  British Columbia}.
\newblock
\href{https://doi.org/10.14288/1.0370945}{doi:\nolinkurl{10.14288/1.0370945}}


\bibitem[Juneja et~al\mbox{.}(2025)]%
        {juneja2025morpher}
\bibfield{author}{\bibinfo{person}{Rohan Juneja}, \bibinfo{person}{Pranav
  Dangi}, \bibinfo{person}{Thilini~Kaushalya Bandara},
  \bibinfo{person}{Zhaoying Li}, \bibinfo{person}{Dhananjaya Wijerathne},
  \bibinfo{person}{Li-Shiuan Peh}, {and} \bibinfo{person}{Tulika Mitra}.}
  \bibinfo{year}{2025}\natexlab{}.
\newblock \showarticletitle{Building an Open {CGRA} Ecosystem for Agile
  Innovation}. In \bibinfo{booktitle}{\emph{2025 IEEE/ACM International
  Conference on Computer-Aided Design (ICCAD)}}. \bibinfo{pages}{1--9}.
\newblock
\href{https://doi.org/10.1109/ICCAD66269.2025.11240628}{doi:\nolinkurl{10.1109/ICCAD66269.2025.11240628}}
\newblock
\shownote{Invited paper}.


\bibitem[Kabir et~al\mbox{.}(2025)]%
        {kabir2025davinci}
\bibfield{author}{\bibinfo{person}{MD~Arafat Kabir}, \bibinfo{person}{Nathaniel
  Fredricks}, \bibinfo{person}{Tendayi Kamucheka}, \bibinfo{person}{Joel
  Mandebi}, \bibinfo{person}{Miaoqing Huang}, \bibinfo{person}{Jason~D. Bakos},
  {and} \bibinfo{person}{David Andrews}.} \bibinfo{year}{2025}\natexlab{}.
\newblock \showarticletitle{{DA-VinCi}: A Deep-Learning Accelerator Overlay
  Using In-Memory Computing}.
\newblock \bibinfo{journal}{\emph{ACM Transactions on Reconfigurable Technology
  and Systems}} \bibinfo{volume}{18}, \bibinfo{number}{4}
  (\bibinfo{year}{2025}), \bibinfo{pages}{46:1--46:38}.
\newblock
\href{https://doi.org/10.1145/3770756}{doi:\nolinkurl{10.1145/3770756}}


\bibitem[Khanzadi et~al\mbox{.}(2017)]%
        {khanzadi2017datadriven}
\bibfield{author}{\bibinfo{person}{Himan Khanzadi}, \bibinfo{person}{Yvon
  Savaria}, {and} \bibinfo{person}{Jean-Pierre David}.}
  \bibinfo{year}{2017}\natexlab{}.
\newblock \showarticletitle{A data driven {CGRA} Overlay Architecture with
  embedded processors}. In \bibinfo{booktitle}{\emph{2017 15th IEEE
  International New Circuits and Systems Conference (NEWCAS)}}.
  \bibinfo{pages}{269--272}.
\newblock
\href{https://doi.org/10.1109/NEWCAS.2017.8010157}{doi:\nolinkurl{10.1109/NEWCAS.2017.8010157}}


\bibitem[Koeplinger et~al\mbox{.}(2018)]%
        {koeplinger2018spatial}
\bibfield{author}{\bibinfo{person}{David Koeplinger}, \bibinfo{person}{Matthew
  Feldman}, \bibinfo{person}{Raghu Prabhakar}, \bibinfo{person}{Yaqi Zhang},
  \bibinfo{person}{Stefan Hadjis}, \bibinfo{person}{Ruben Fiszel},
  \bibinfo{person}{Tian Zhao}, \bibinfo{person}{Luigi Nardi},
  \bibinfo{person}{Ardavan Pedram}, \bibinfo{person}{Christos Kozyrakis}, {and}
  \bibinfo{person}{Kunle Olukotun}.} \bibinfo{year}{2018}\natexlab{}.
\newblock \showarticletitle{{Spatial}: A Language and Compiler for Application
  Accelerators}. In \bibinfo{booktitle}{\emph{Proceedings of the 39th ACM
  SIGPLAN Conference on Programming Language Design and Implementation}}
  \emph{(\bibinfo{series}{PLDI '18})}. \bibinfo{publisher}{Association for
  Computing Machinery}, \bibinfo{pages}{296--311}.
\newblock
\href{https://doi.org/10.1145/3192366.3192379}{doi:\nolinkurl{10.1145/3192366.3192379}}


\bibitem[Kourfali and Stroobandt(2019)]%
        {kourfali2019faulttolerance}
\bibfield{author}{\bibinfo{person}{Alexandra Kourfali} {and}
  \bibinfo{person}{Dirk Stroobandt}.} \bibinfo{year}{2019}\natexlab{}.
\newblock \showarticletitle{In-circuit fault tolerance for {FPGAs} using
  dynamic reconfiguration and virtual overlays}.
\newblock \bibinfo{journal}{\emph{Microelectronics Reliability}}
  \bibinfo{volume}{102} (\bibinfo{year}{2019}), \bibinfo{pages}{113438}.
\newblock
\href{https://doi.org/10.1016/j.microrel.2019.113438}{doi:\nolinkurl{10.1016/j.microrel.2019.113438}}


\bibitem[Kuhn(1955)]%
        {kuhn1955hungarian}
\bibfield{author}{\bibinfo{person}{H.~W. Kuhn}.}
  \bibinfo{year}{1955}\natexlab{}.
\newblock \showarticletitle{The Hungarian Method for the Assignment Problem}.
\newblock \bibinfo{journal}{\emph{Naval Research Logistics Quarterly}}
  \bibinfo{volume}{2}, \bibinfo{number}{1--2} (\bibinfo{date}{March}
  \bibinfo{year}{1955}), \bibinfo{pages}{83--97}.
\newblock
\href{https://doi.org/10.1002/nav.3800020109}{doi:\nolinkurl{10.1002/nav.3800020109}}


\bibitem[Kuon and Rose(2007)]%
        {kuon2007gap}
\bibfield{author}{\bibinfo{person}{Ian Kuon} {and} \bibinfo{person}{Jonathan
  Rose}.} \bibinfo{year}{2007}\natexlab{}.
\newblock \showarticletitle{Measuring the Gap Between {FPGAs} and {ASICs}}.
\newblock \bibinfo{journal}{\emph{IEEE Transactions on Computer-Aided Design of
  Integrated Circuits and Systems}} \bibinfo{volume}{26}, \bibinfo{number}{2}
  (\bibinfo{year}{2007}), \bibinfo{pages}{203--215}.
\newblock


\bibitem[Kwon et~al\mbox{.}(2018)]%
        {kwon2018maeri}
\bibfield{author}{\bibinfo{person}{Hyoukjun Kwon}, \bibinfo{person}{Ananda
  Samajdar}, {and} \bibinfo{person}{Tushar Krishna}.}
  \bibinfo{year}{2018}\natexlab{}.
\newblock \showarticletitle{{MAERI}: Enabling Flexible Dataflow Mapping over
  {DNN} Accelerators via Reconfigurable Interconnects}. In
  \bibinfo{booktitle}{\emph{Proceedings of the International Conference on
  Architectural Support for Programming Languages and Operating Systems}}
  \emph{(\bibinfo{series}{ASPLOS})}. \bibinfo{pages}{461--475}.
\newblock
\href{https://doi.org/10.1145/3173162.3173176}{doi:\nolinkurl{10.1145/3173162.3173176}}


\bibitem[Lattner et~al\mbox{.}(2021)]%
        {lattner2021mlir}
\bibfield{author}{\bibinfo{person}{Chris Lattner}, \bibinfo{person}{Mehdi
  Amini}, \bibinfo{person}{Uday Bondhugula}, \bibinfo{person}{Albert Cohen},
  \bibinfo{person}{Andy Davis}, \bibinfo{person}{Jacques Pienaar},
  \bibinfo{person}{River Riddle}, \bibinfo{person}{Tatiana Shpeisman},
  \bibinfo{person}{Nicolas Vasilache}, {and} \bibinfo{person}{Oleksandr
  Zinenko}.} \bibinfo{year}{2021}\natexlab{}.
\newblock \showarticletitle{{MLIR}: Scaling Compiler Infrastructure for Domain
  Specific Computation}. In \bibinfo{booktitle}{\emph{2021 IEEE/ACM
  International Symposium on Code Generation and Optimization (CGO)}}.
  \bibinfo{publisher}{IEEE}, \bibinfo{pages}{2--14}.
\newblock
\href{https://doi.org/10.1109/CGO51591.2021.9370308}{doi:\nolinkurl{10.1109/CGO51591.2021.9370308}}


\bibitem[Le~Lann et~al\mbox{.}(2018)]%
        {lelann2018argen}
\bibfield{author}{\bibinfo{person}{Jean-Christophe Le~Lann},
  \bibinfo{person}{Th{\'e}otime Bollengier}, \bibinfo{person}{Mohamad Najem},
  {and} \bibinfo{person}{Lo{\"i}c Lagadec}.} \bibinfo{year}{2018}\natexlab{}.
\newblock \showarticletitle{An Integrated Toolchain for Overlay-centric
  System-on-chip}. In \bibinfo{booktitle}{\emph{2018 13th International
  Symposium on Reconfigurable Communication-centric Systems-on-Chip
  (ReCoSoC)}}. \bibinfo{pages}{1--8}.
\newblock
\href{https://doi.org/10.1109/ReCoSoC.2018.8449388}{doi:\nolinkurl{10.1109/ReCoSoC.2018.8449388}}


\bibitem[Li et~al\mbox{.}(2018)]%
        {li2018timemultiplexed}
\bibfield{author}{\bibinfo{person}{Xiangwei Li},
  \bibinfo{person}{Abhishek~Kumar Jain}, \bibinfo{person}{Douglas~L. Maskell},
  {and} \bibinfo{person}{Suhaib~A. Fahmy}.} \bibinfo{year}{2018}\natexlab{}.
\newblock \showarticletitle{A Time-Multiplexed {FPGA} Overlay with Linear
  Interconnect}. In \bibinfo{booktitle}{\emph{2018 Design, Automation \& Test
  in Europe Conference \& Exhibition (DATE)}}. \bibinfo{pages}{1075--1080}.
\newblock
\href{https://doi.org/10.23919/DATE.2018.8342171}{doi:\nolinkurl{10.23919/DATE.2018.8342171}}


\bibitem[Li and Maskell(2019)]%
        {li2019overlaysurvey}
\bibfield{author}{\bibinfo{person}{Xiangwei Li} {and}
  \bibinfo{person}{Douglas~L. Maskell}.} \bibinfo{year}{2019}\natexlab{}.
\newblock \showarticletitle{Time-Multiplexed {FPGA} Overlay Architectures: A
  Survey}.
\newblock \bibinfo{journal}{\emph{ACM Transactions on Design Automation of
  Electronic Systems}} \bibinfo{volume}{24}, \bibinfo{number}{5}, Article
  \bibinfo{articleno}{54} (\bibinfo{year}{2019}), \bibinfo{numpages}{19}~pages.
\newblock
\href{https://doi.org/10.1145/3339861}{doi:\nolinkurl{10.1145/3339861}}


\bibitem[Li et~al\mbox{.}(2020)]%
        {li2020overlayinterface}
\bibfield{author}{\bibinfo{person}{Xiangwei Li}, \bibinfo{person}{Kizheppatt
  Vipin}, \bibinfo{person}{Douglas~L. Maskell}, \bibinfo{person}{Suhaib~A.
  Fahmy}, {and} \bibinfo{person}{Abhishek~Kumar Jain}.}
  \bibinfo{year}{2020}\natexlab{}.
\newblock \showarticletitle{High Throughput Accelerator Interface Framework for
  a Linear Time-Multiplexed {FPGA} Overlay}. In
  \bibinfo{booktitle}{\emph{Proceedings of the IEEE International Symposium on
  Circuits and Systems}} \emph{(\bibinfo{series}{ISCAS})}.
  \bibinfo{pages}{1--5}.
\newblock
\href{https://doi.org/10.1109/ISCAS45731.2020.9181072}{doi:\nolinkurl{10.1109/ISCAS45731.2020.9181072}}


\bibitem[Li et~al\mbox{.}(2025)]%
        {li2025plaid}
\bibfield{author}{\bibinfo{person}{Zhaoying Li}, \bibinfo{person}{Pranav
  Dangi}, \bibinfo{person}{Chenyang Yin}, \bibinfo{person}{Thilini~Kaushalya
  Bandara}, \bibinfo{person}{Rohan Juneja}, \bibinfo{person}{Cheng Tan},
  \bibinfo{person}{Zhenyu Bai}, {and} \bibinfo{person}{Tulika Mitra}.}
  \bibinfo{year}{2025}\natexlab{}.
\newblock \showarticletitle{Enhancing {CGRA} Efficiency Through Aligned Compute
  and Communication Provisioning}. In \bibinfo{booktitle}{\emph{Proceedings of
  the 30th ACM International Conference on Architectural Support for
  Programming Languages and Operating Systems, Volume 1}}.
  \bibinfo{pages}{410--425}.
\newblock
\href{https://doi.org/10.1145/3669940.3707230}{doi:\nolinkurl{10.1145/3669940.3707230}}


\bibitem[Li et~al\mbox{.}(2022)]%
        {li2022lisa}
\bibfield{author}{\bibinfo{person}{Zhaoying Li}, \bibinfo{person}{Dan Wu},
  \bibinfo{person}{Dhananjaya Wijerathne}, {and} \bibinfo{person}{Tulika
  Mitra}.} \bibinfo{year}{2022}\natexlab{}.
\newblock \showarticletitle{{LISA}: Graph Neural Network based Portable Mapping
  on Spatial Accelerators}. In \bibinfo{booktitle}{\emph{2022 IEEE
  International Symposium on High-Performance Computer Architecture (HPCA)}}.
  \bibinfo{pages}{444--459}.
\newblock
\urldef\tempurl%
\url{https://www.comp.nus.edu.sg/~tulika/HPCA_LISA_2022.pdf}
\showURL{%
\tempurl}


\bibitem[Liu et~al\mbox{.}(2019)]%
        {liu2019rapidroute}
\bibfield{author}{\bibinfo{person}{Leo Liu}, \bibinfo{person}{Jay Weng}, {and}
  \bibinfo{person}{Nachiket Kapre}.} \bibinfo{year}{2019}\natexlab{}.
\newblock \showarticletitle{{RapidRoute}: Fast Assembly of Communication
  Structures for {FPGA} Overlays}. In \bibinfo{booktitle}{\emph{2019 IEEE 27th
  Annual International Symposium on Field-Programmable Custom Computing
  Machines (FCCM)}}. \bibinfo{pages}{61--64}.
\newblock
\href{https://doi.org/10.1109/FCCM.2019.00018}{doi:\nolinkurl{10.1109/FCCM.2019.00018}}


\bibitem[Liu et~al\mbox{.}(2022)]%
        {liu2022overgen}
\bibfield{author}{\bibinfo{person}{Sihao Liu}, \bibinfo{person}{Jian Weng},
  \bibinfo{person}{Dylan Kupsh}, \bibinfo{person}{Atefeh Sohrabizadeh},
  \bibinfo{person}{Zhengrong Wang}, \bibinfo{person}{Licheng Guo},
  \bibinfo{person}{Jiuyang Liu}, \bibinfo{person}{Maxim Zhulin},
  \bibinfo{person}{Rishabh Mani}, \bibinfo{person}{Lucheng Zhang},
  \bibinfo{person}{Jason Cong}, {and} \bibinfo{person}{Tony Nowatzki}.}
  \bibinfo{year}{2022}\natexlab{}.
\newblock \showarticletitle{{OverGen}: Improving {FPGA} Usability through
  Domain-specific Overlay Generation}. In \bibinfo{booktitle}{\emph{2022 55th
  IEEE/ACM International Symposium on Microarchitecture (MICRO)}}.
  \bibinfo{publisher}{IEEE}, \bibinfo{pages}{35--56}.
\newblock
\href{https://doi.org/10.1109/MICRO56248.2022.00018}{doi:\nolinkurl{10.1109/MICRO56248.2022.00018}}


\bibitem[{LLVM Project}(0000a)]%
        {llvmClangIR}
\bibfield{author}{\bibinfo{person}{{LLVM Project}}.}
  \bibinfo{year}{[n.\,d.]}\natexlab{a}.
\newblock \bibinfo{title}{{ClangIR}}.
\newblock \bibinfo{howpublished}{Clang documentation}.
\newblock
\urldef\tempurl%
\url{https://clang.llvm.org/docs/CIR/index.html}
\showURL{%
\tempurl}
\newblock
\shownote{Accessed November 12, 2026}.


\bibitem[{LLVM Project}(0000b)]%
        {llvmMLIRLangRef}
\bibfield{author}{\bibinfo{person}{{LLVM Project}}.}
  \bibinfo{year}{[n.\,d.]}\natexlab{b}.
\newblock \bibinfo{title}{{MLIR} Language Reference}.
\newblock \bibinfo{howpublished}{Online documentation}.
\newblock
\urldef\tempurl%
\url{https://mlir.llvm.org/docs/LangRef/}
\showURL{%
\tempurl}
\newblock
\shownote{Accessed November 12, 2026}.


\bibitem[{LLVM Project}(0000c)]%
        {llvmMLIRPasses}
\bibfield{author}{\bibinfo{person}{{LLVM Project}}.}
  \bibinfo{year}{[n.\,d.]}\natexlab{c}.
\newblock \bibinfo{title}{{MLIR} Passes}.
\newblock \bibinfo{howpublished}{Online documentation}.
\newblock
\urldef\tempurl%
\url{https://mlir.llvm.org/docs/Passes/}
\showURL{%
\tempurl}
\newblock
\shownote{Accessed November 12, 2026}.


\bibitem[Lysecky et~al\mbox{.}(2005)]%
        {lysecky2005firm}
\bibfield{author}{\bibinfo{person}{Roman Lysecky}, \bibinfo{person}{Kris
  Miller}, \bibinfo{person}{Frank Vahid}, {and} \bibinfo{person}{Kees
  Vissers}.} \bibinfo{year}{2005}\natexlab{}.
\newblock \showarticletitle{Firm-Core Virtual {FPGA} for Just-in-Time {FPGA}
  Compilation}. In \bibinfo{booktitle}{\emph{Proceedings of the 2005 ACM/SIGDA
  13th International Symposium on Field-Programmable Gate Arrays (FPGA)}}.
  \bibinfo{publisher}{ACM}, \bibinfo{pages}{271}.
\newblock
\href{https://doi.org/10.1145/1046192.1046233}{doi:\nolinkurl{10.1145/1046192.1046233}}


\bibitem[Ma(2024)]%
        {ma2024specializedoverlays}
\bibfield{author}{\bibinfo{person}{Rui Ma}.} \bibinfo{year}{2024}\natexlab{}.
\newblock \emph{\bibinfo{title}{Automatically specialized {FPGA} overlays: a
  trade-off between programmability, performance, and area}}.
\newblock \bibinfo{thesistype}{Ph.\,D. Dissertation}. \bibinfo{school}{The
  University of Texas at Austin}.
\newblock
\href{https://doi.org/10.26153/tsw/54309}{doi:\nolinkurl{10.26153/tsw/54309}}


\bibitem[Ma et~al\mbox{.}(2021)]%
        {ma2021pdlfgpu}
\bibfield{author}{\bibinfo{person}{Rui Ma}, \bibinfo{person}{Jia-Ching Hsu},
  \bibinfo{person}{Tian Tan}, \bibinfo{person}{Eriko Nurvitadhi},
  \bibinfo{person}{David Sheffield}, \bibinfo{person}{Rob Pelt},
  \bibinfo{person}{Martin Langhammer}, \bibinfo{person}{Jaewoong Sim},
  \bibinfo{person}{Aravind Dasu}, {and} \bibinfo{person}{Derek Chiou}.}
  \bibinfo{year}{2021}\natexlab{}.
\newblock \showarticletitle{Specializing {FGPU} for Persistent Deep Learning}.
\newblock \bibinfo{journal}{\emph{ACM Transactions on Reconfigurable Technology
  and Systems}} \bibinfo{volume}{14}, \bibinfo{number}{2}
  (\bibinfo{year}{2021}), \bibinfo{pages}{10:1--10:23}.
\newblock
\href{https://doi.org/10.1145/3457886}{doi:\nolinkurl{10.1145/3457886}}


\bibitem[Mandebi~Mbongue et~al\mbox{.}(2019)]%
        {mbongue2019asoverlays}
\bibfield{author}{\bibinfo{person}{Joel Mandebi~Mbongue},
  \bibinfo{person}{Danielle Tchuinkou~Kwadjo}, {and}
  \bibinfo{person}{Christophe Bobda}.} \bibinfo{year}{2019}\natexlab{}.
\newblock \showarticletitle{Automatic Generation of Application-Specific {FPGA}
  Overlays with {RapidWright}}. In \bibinfo{booktitle}{\emph{2019 International
  Conference on Field-Programmable Technology (ICFPT)}}.
  \bibinfo{pages}{303--306}.
\newblock
\href{https://doi.org/10.1109/ICFPT47387.2019.00053}{doi:\nolinkurl{10.1109/ICFPT47387.2019.00053}}


\bibitem[Mashreghi-Moghadam et~al\mbox{.}(2023)]%
        {mashreghi2023parser}
\bibfield{author}{\bibinfo{person}{Parisa Mashreghi-Moghadam},
  \bibinfo{person}{Tarek Ould-Bachir}, {and} \bibinfo{person}{Yvon Savaria}.}
  \bibinfo{year}{2023}\natexlab{}.
\newblock \showarticletitle{An Area-efficient Memory-based Architecture for
  {P4}-programmable Streaming Parsers in {FPGAs}}. In
  \bibinfo{booktitle}{\emph{2023 IEEE International Symposium on Circuits and
  Systems (ISCAS)}}. \bibinfo{pages}{1--5}.
\newblock
\href{https://doi.org/10.1109/ISCAS46773.2023.10182176}{doi:\nolinkurl{10.1109/ISCAS46773.2023.10182176}}


\bibitem[Mashreghi-Moghadam et~al\mbox{.}(2024)]%
        {mashreghi2024prismparser}
\bibfield{author}{\bibinfo{person}{Parisa Mashreghi-Moghadam},
  \bibinfo{person}{Tarek Ould-Bachir}, {and} \bibinfo{person}{Yvon Savaria}.}
  \bibinfo{year}{2024}\natexlab{}.
\newblock \showarticletitle{{PrismParser}: A Framework for Implementing
  Efficient {P4}-Programmable Packet Parsers on {FPGA}}.
\newblock \bibinfo{journal}{\emph{Future Internet}} \bibinfo{volume}{16},
  \bibinfo{number}{9} (\bibinfo{year}{2024}), \bibinfo{pages}{307}.
\newblock
\href{https://doi.org/10.3390/fi16090307}{doi:\nolinkurl{10.3390/fi16090307}}


\bibitem[Melchert et~al\mbox{.}(2023)]%
        {melchert2023apex}
\bibfield{author}{\bibinfo{person}{Jackson Melchert}, \bibinfo{person}{Kathleen
  Feng}, \bibinfo{person}{Caleb Donovick}, \bibinfo{person}{Ross Daly},
  \bibinfo{person}{Ritvik Sharma}, \bibinfo{person}{Clark Barrett},
  \bibinfo{person}{Mark~A. Horowitz}, \bibinfo{person}{Pat Hanrahan}, {and}
  \bibinfo{person}{Priyanka Raina}.} \bibinfo{year}{2023}\natexlab{}.
\newblock \showarticletitle{{APEX}: A Framework for Automated Processing
  Element Design Space Exploration using Frequent Subgraph Analysis}. In
  \bibinfo{booktitle}{\emph{Proceedings of the International Conference on
  Architectural Support for Programming Languages and Operating Systems, Volume
  3}} \emph{(\bibinfo{series}{ASPLOS})}. \bibinfo{pages}{33--45}.
\newblock
\href{https://doi.org/10.1145/3582016.3582070}{doi:\nolinkurl{10.1145/3582016.3582070}}


\bibitem[Moreano et~al\mbox{.}(2005)]%
        {moreano2005merging}
\bibfield{author}{\bibinfo{person}{Nahri Moreano}, \bibinfo{person}{Edson
  Borin}, \bibinfo{person}{Cid de Souza}, {and} \bibinfo{person}{Guido
  Araujo}.} \bibinfo{year}{2005}\natexlab{}.
\newblock \showarticletitle{Efficient Datapath Merging for Partially
  Reconfigurable Architectures}.
\newblock \bibinfo{journal}{\emph{IEEE Transactions on Computer-Aided Design of
  Integrated Circuits and Systems}} \bibinfo{volume}{24}, \bibinfo{number}{7}
  (\bibinfo{year}{2005}), \bibinfo{pages}{969--980}.
\newblock
\href{https://doi.org/10.1109/TCAD.2005.850844}{doi:\nolinkurl{10.1109/TCAD.2005.850844}}


\bibitem[Najem et~al\mbox{.}(2017)]%
        {najem2017extended}
\bibfield{author}{\bibinfo{person}{Mohamad Najem},
  \bibinfo{person}{Th{\'e}otime Bollengier}, \bibinfo{person}{Jean-Christophe
  Le~Lann}, {and} \bibinfo{person}{Lo{\"i}c Lagadec}.}
  \bibinfo{year}{2017}\natexlab{}.
\newblock \showarticletitle{Extended overlay architectures for heterogeneous
  {FPGA} cluster management}.
\newblock \bibinfo{journal}{\emph{Journal of Systems Architecture}}
  \bibinfo{volume}{78} (\bibinfo{year}{2017}), \bibinfo{pages}{1--14}.
\newblock
\href{https://doi.org/10.1016/j.sysarc.2017.06.001}{doi:\nolinkurl{10.1016/j.sysarc.2017.06.001}}


\bibitem[Naylor et~al\mbox{.}(2019)]%
        {naylor2019tinsel}
\bibfield{author}{\bibinfo{person}{Matthew Naylor}, \bibinfo{person}{Simon~W.
  Moore}, {and} \bibinfo{person}{David Thomas}.}
  \bibinfo{year}{2019}\natexlab{}.
\newblock \showarticletitle{{Tinsel}: A Manythread Overlay for {FPGA}
  Clusters}. In \bibinfo{booktitle}{\emph{2019 29th International Conference on
  Field Programmable Logic and Applications (FPL)}}. \bibinfo{pages}{375--383}.
\newblock
\href{https://doi.org/10.1109/FPL.2019.00066}{doi:\nolinkurl{10.1109/FPL.2019.00066}}


\bibitem[Nowatzki et~al\mbox{.}(2017)]%
        {nowatzki2017streamdataflow}
\bibfield{author}{\bibinfo{person}{Tony Nowatzki}, \bibinfo{person}{Vinay
  Gangadhar}, \bibinfo{person}{Newsha Ardalani}, {and}
  \bibinfo{person}{Karthikeyan Sankaralingam}.}
  \bibinfo{year}{2017}\natexlab{}.
\newblock \showarticletitle{Stream-Dataflow Acceleration}. In
  \bibinfo{booktitle}{\emph{Proceedings of the 44th Annual International
  Symposium on Computer Architecture}} \emph{(\bibinfo{series}{ISCA '17})}.
  \bibinfo{publisher}{Association for Computing Machinery},
  \bibinfo{pages}{416--429}.
\newblock
\href{https://doi.org/10.1145/3079856.3080255}{doi:\nolinkurl{10.1145/3079856.3080255}}


\bibitem[Panahi(2022)]%
        {panahi2022thesis}
\bibfield{author}{\bibinfo{person}{Atiyehsadat Panahi}.}
  \bibinfo{year}{2022}\natexlab{}.
\newblock \emph{\bibinfo{title}{A Memory-Centric Customizable Domain-Specific
  {FPGA} Overlay for Accelerating Machine Learning Applications}}.
\newblock \bibinfo{thesistype}{Ph.\,D. Dissertation}.
  \bibinfo{school}{University of Arkansas, Fayetteville}.
\newblock
\urldef\tempurl%
\url{https://scholarworks.uark.edu/etd/4618/}
\showURL{%
\tempurl}


\bibitem[Panahi et~al\mbox{.}(2021)]%
        {panahi2021memorycentric}
\bibfield{author}{\bibinfo{person}{Atiyehsadat Panahi}, \bibinfo{person}{Suhail
  Balsalama}, \bibinfo{person}{Ange-Thierry Ishimwe}, \bibinfo{person}{Joel
  Mandebi~Mbongue}, {and} \bibinfo{person}{David Andrews}.}
  \bibinfo{year}{2021}\natexlab{}.
\newblock \showarticletitle{A Customizable Domain-Specific Memory-Centric
  {FPGA} Overlay for Machine Learning Applications}. In
  \bibinfo{booktitle}{\emph{2021 31st International Conference on
  Field-Programmable Logic and Applications (FPL)}}. \bibinfo{pages}{24--27}.
\newblock
\href{https://doi.org/10.1109/FPL53798.2021.00012}{doi:\nolinkurl{10.1109/FPL53798.2021.00012}}


\bibitem[Park and DeHon(2024)]%
        {park2024refine}
\bibfield{author}{\bibinfo{person}{Dongjoon Park} {and}
  \bibinfo{person}{Andr{\'e} DeHon}.} \bibinfo{year}{2024}\natexlab{}.
\newblock \showarticletitle{{REFINE}: Runtime Execution Feedback for
  {INcremental Evolution} on {FPGA} Designs}. In
  \bibinfo{booktitle}{\emph{Proceedings of the ACM/SIGDA International
  Symposium on Field-Programmable Gate Arrays}}
  \emph{(\bibinfo{series}{FPGA})}. \bibinfo{pages}{108--118}.
\newblock
\href{https://doi.org/10.1145/3626202.3637560}{doi:\nolinkurl{10.1145/3626202.3637560}}


\bibitem[Podobas et~al\mbox{.}(2020)]%
        {podobas2020cgrasurvey}
\bibfield{author}{\bibinfo{person}{Artur Podobas}, \bibinfo{person}{Kentaro
  Sano}, {and} \bibinfo{person}{Satoshi Matsuoka}.}
  \bibinfo{year}{2020}\natexlab{}.
\newblock \showarticletitle{A Survey on Coarse-Grained Reconfigurable
  Architectures From a Performance Perspective}.
\newblock \bibinfo{journal}{\emph{IEEE Access}}  \bibinfo{volume}{8}
  (\bibinfo{year}{2020}), \bibinfo{pages}{146719--146743}.
\newblock
\href{https://doi.org/10.1109/ACCESS.2020.3012084}{doi:\nolinkurl{10.1109/ACCESS.2020.3012084}}


\bibitem[Pouget et~al\mbox{.}(2025a)]%
        {pouget2025prometheus}
\bibfield{author}{\bibinfo{person}{St{\'e}phane Pouget},
  \bibinfo{person}{Michael Lo}, \bibinfo{person}{Louis-No{\"e}l Pouchet}, {and}
  \bibinfo{person}{Jason Cong}.} \bibinfo{year}{2025}\natexlab{a}.
\newblock \showarticletitle{Holistic Optimization Framework for {FPGA}
  Accelerators}.
\newblock \bibinfo{journal}{\emph{ACM Transactions on Design Automation of
  Electronic Systems}} \bibinfo{volume}{31}, \bibinfo{number}{1}, Article
  \bibinfo{articleno}{7} (\bibinfo{year}{2025}), \bibinfo{numpages}{37}~pages.
\newblock
\href{https://doi.org/10.1145/3769307}{doi:\nolinkurl{10.1145/3769307}}


\bibitem[Pouget et~al\mbox{.}(2025b)]%
        {pouget2025sisyphus}
\bibfield{author}{\bibinfo{person}{St{\'e}phane Pouget},
  \bibinfo{person}{Louis-No{\"e}l Pouchet}, {and} \bibinfo{person}{Jason
  Cong}.} \bibinfo{year}{2025}\natexlab{b}.
\newblock \showarticletitle{A Unified Framework for Automated Code
  Transformation and Pragma Insertion}. In
  \bibinfo{booktitle}{\emph{Proceedings of the ACM/SIGDA International
  Symposium on Field-Programmable Gate Arrays}}
  \emph{(\bibinfo{series}{FPGA})}. \bibinfo{pages}{187--198}.
\newblock
\href{https://doi.org/10.1145/3706628.3708873}{doi:\nolinkurl{10.1145/3706628.3708873}}


\bibitem[Prabhakar et~al\mbox{.}(2017)]%
        {prabhakar2017plasticine}
\bibfield{author}{\bibinfo{person}{Raghu Prabhakar}, \bibinfo{person}{Yaqi
  Zhang}, \bibinfo{person}{David Koeplinger}, \bibinfo{person}{Matt Feldman},
  \bibinfo{person}{Tian Zhao}, \bibinfo{person}{Stefan Hadjis},
  \bibinfo{person}{Ardavan Pedram}, \bibinfo{person}{Christos Kozyrakis}, {and}
  \bibinfo{person}{Kunle Olukotun}.} \bibinfo{year}{2017}\natexlab{}.
\newblock \showarticletitle{{Plasticine}: A Reconfigurable Architecture for
  Parallel Patterns}. In \bibinfo{booktitle}{\emph{Proceedings of the
  International Symposium on Computer Architecture}}
  \emph{(\bibinfo{series}{ISCA})}. \bibinfo{pages}{389--402}.
\newblock
\href{https://doi.org/10.1145/3079856.3080256}{doi:\nolinkurl{10.1145/3079856.3080256}}


\bibitem[Qian et~al\mbox{.}(2025)]%
        {qian2025autocompiler}
\bibfield{author}{\bibinfo{person}{Kai Qian}, \bibinfo{person}{Haodong Lu},
  \bibinfo{person}{Yinqiu Liu}, \bibinfo{person}{Zexu Zhang}, {and}
  \bibinfo{person}{Kun Wang}.} \bibinfo{year}{2025}\natexlab{}.
\newblock \showarticletitle{End-to-end Compilation is All {FPGAs} Need: A
  Unified Overlay-based {FPGA} Compiler for Deep Learning}. In
  \bibinfo{booktitle}{\emph{Proceedings of the 30th Asia and South Pacific
  Design Automation Conference}}. \bibinfo{pages}{258--264}.
\newblock
\href{https://doi.org/10.1145/3658617.3697562}{doi:\nolinkurl{10.1145/3658617.3697562}}


\bibitem[Ronak and Fahmy(2016a)]%
        {ronak2016sharing}
\bibfield{author}{\bibinfo{person}{Bajaj Ronak} {and}
  \bibinfo{person}{Suhaib~A. Fahmy}.} \bibinfo{year}{2016}\natexlab{a}.
\newblock \showarticletitle{Improved Resource Sharing for {FPGA DSP} Blocks}.
  In \bibinfo{booktitle}{\emph{Proceedings of the International Conference on
  Field Programmable Logic and Applications}} \emph{(\bibinfo{series}{FPL})}.
  \bibinfo{pages}{1--4}.
\newblock
\href{https://doi.org/10.1109/FPL.2016.7577373}{doi:\nolinkurl{10.1109/FPL.2016.7577373}}


\bibitem[Ronak and Fahmy(2016b)]%
        {ronak2016mapping}
\bibfield{author}{\bibinfo{person}{Bajaj Ronak} {and}
  \bibinfo{person}{Suhaib~A. Fahmy}.} \bibinfo{year}{2016}\natexlab{b}.
\newblock \showarticletitle{Mapping for Maximum Performance on {FPGA DSP}
  Blocks}.
\newblock \bibinfo{journal}{\emph{IEEE Transactions on Computer-Aided Design of
  Integrated Circuits and Systems}} \bibinfo{volume}{35}, \bibinfo{number}{4}
  (\bibinfo{year}{2016}), \bibinfo{pages}{573--585}.
\newblock
\href{https://doi.org/10.1109/TCAD.2015.2474363}{doi:\nolinkurl{10.1109/TCAD.2015.2474363}}


\bibitem[Ronak and Fahmy(2017)]%
        {ronak2017multipumping}
\bibfield{author}{\bibinfo{person}{Bajaj Ronak} {and}
  \bibinfo{person}{Suhaib~A. Fahmy}.} \bibinfo{year}{2017}\natexlab{}.
\newblock \showarticletitle{Multipumping Flexible {DSP} Blocks for Resource
  Reduction on {Xilinx FPGAs}}.
\newblock \bibinfo{journal}{\emph{IEEE Transactions on Computer-Aided Design of
  Integrated Circuits and Systems}} \bibinfo{volume}{36}, \bibinfo{number}{9}
  (\bibinfo{year}{2017}), \bibinfo{pages}{1471--1482}.
\newblock
\href{https://doi.org/10.1109/TCAD.2016.2629421}{doi:\nolinkurl{10.1109/TCAD.2016.2629421}}


\bibitem[Sharifian et~al\mbox{.}(2019)]%
        {sharifian2019muir}
\bibfield{author}{\bibinfo{person}{Amirali Sharifian}, \bibinfo{person}{Reza
  Hojabr}, \bibinfo{person}{Navid Rahimi}, \bibinfo{person}{Sihao Liu},
  \bibinfo{person}{Apala Guha}, \bibinfo{person}{Tony Nowatzki}, {and}
  \bibinfo{person}{Arrvindh Shriraman}.} \bibinfo{year}{2019}\natexlab{}.
\newblock \showarticletitle{{$\mu$IR}: An Intermediate Representation for
  Transforming and Optimizing the Microarchitecture of Application
  Accelerators}. In \bibinfo{booktitle}{\emph{Proceedings of the Annual
  IEEE/ACM International Symposium on Microarchitecture}}
  \emph{(\bibinfo{series}{MICRO})}. \bibinfo{pages}{940--953}.
\newblock
\href{https://doi.org/10.1145/3352460.3358292}{doi:\nolinkurl{10.1145/3352460.3358292}}


\bibitem[Shi et~al\mbox{.}(2020)]%
        {shi2020ftdl}
\bibfield{author}{\bibinfo{person}{Runbin Shi}, \bibinfo{person}{Yuhao Ding},
  \bibinfo{person}{Xuechao Wei}, \bibinfo{person}{He Li}, \bibinfo{person}{Hang
  Liu}, \bibinfo{person}{Hayden Kwok-Hay So}, {and} \bibinfo{person}{Caiwen
  Ding}.} \bibinfo{year}{2020}\natexlab{}.
\newblock \showarticletitle{{FTDL}: A Tailored {FPGA}-Overlay for Deep Learning
  with High Scalability}. In \bibinfo{booktitle}{\emph{2020 57th ACM/IEEE
  Design Automation Conference (DAC)}}. \bibinfo{pages}{1--6}.
\newblock
\href{https://doi.org/10.1109/DAC18072.2020.9218581}{doi:\nolinkurl{10.1109/DAC18072.2020.9218581}}


\bibitem[Shuman et~al\mbox{.}(2013)]%
        {shuman2013emerging}
\bibfield{author}{\bibinfo{person}{David~I. Shuman}, \bibinfo{person}{Sunil~K.
  Narang}, \bibinfo{person}{Pascal Frossard}, \bibinfo{person}{Antonio Ortega},
  {and} \bibinfo{person}{Pierre Vandergheynst}.}
  \bibinfo{year}{2013}\natexlab{}.
\newblock \showarticletitle{The Emerging Field of Signal Processing on Graphs:
  Extending High-Dimensional Data Analysis to Networks and Other Irregular
  Domains}.
\newblock \bibinfo{journal}{\emph{IEEE Signal Processing Magazine}}
  \bibinfo{volume}{30}, \bibinfo{number}{3} (\bibinfo{date}{May}
  \bibinfo{year}{2013}), \bibinfo{pages}{83--98}.
\newblock
\href{https://doi.org/10.1109/MSP.2012.2235192}{doi:\nolinkurl{10.1109/MSP.2012.2235192}}


\bibitem[Sohrabizadeh et~al\mbox{.}(2022a)]%
        {sohrabizadeh2022gnndse}
\bibfield{author}{\bibinfo{person}{Atefeh Sohrabizadeh},
  \bibinfo{person}{Yunsheng Bai}, \bibinfo{person}{Yizhou Sun}, {and}
  \bibinfo{person}{Jason Cong}.} \bibinfo{year}{2022}\natexlab{a}.
\newblock \showarticletitle{Automated Accelerator Optimization Aided by Graph
  Neural Networks}. In \bibinfo{booktitle}{\emph{Proceedings of the ACM/IEEE
  Design Automation Conference}} \emph{(\bibinfo{series}{DAC})}.
  \bibinfo{pages}{55--60}.
\newblock
\href{https://doi.org/10.1145/3489517.3530409}{doi:\nolinkurl{10.1145/3489517.3530409}}


\bibitem[Sohrabizadeh et~al\mbox{.}(2023)]%
        {sohrabizadeh2023harp}
\bibfield{author}{\bibinfo{person}{Atefeh Sohrabizadeh},
  \bibinfo{person}{Yunsheng Bai}, \bibinfo{person}{Yizhou Sun}, {and}
  \bibinfo{person}{Jason Cong}.} \bibinfo{year}{2023}\natexlab{}.
\newblock \showarticletitle{Robust {GNN}-Based Representation Learning for
  {HLS}}. In \bibinfo{booktitle}{\emph{Proceedings of the IEEE/ACM
  International Conference on Computer-Aided Design}}
  \emph{(\bibinfo{series}{ICCAD})}. \bibinfo{pages}{1--9}.
\newblock
\href{https://doi.org/10.1109/ICCAD57390.2023.10323853}{doi:\nolinkurl{10.1109/ICCAD57390.2023.10323853}}


\bibitem[Sohrabizadeh et~al\mbox{.}(2022b)]%
        {sohrabizadeh2022autodse}
\bibfield{author}{\bibinfo{person}{Atefeh Sohrabizadeh},
  \bibinfo{person}{Cody~Hao Yu}, \bibinfo{person}{Min Gao}, {and}
  \bibinfo{person}{Jason Cong}.} \bibinfo{year}{2022}\natexlab{b}.
\newblock \showarticletitle{{AutoDSE}: Enabling Software Programmers to Design
  Efficient {FPGA} Accelerators}.
\newblock \bibinfo{journal}{\emph{ACM Transactions on Design Automation of
  Electronic Systems}} \bibinfo{volume}{27}, \bibinfo{number}{4}, Article
  \bibinfo{articleno}{32} (\bibinfo{year}{2022}), \bibinfo{numpages}{27}~pages.
\newblock
\href{https://doi.org/10.1145/3494534}{doi:\nolinkurl{10.1145/3494534}}


\bibitem[Stojilovi{\'c} et~al\mbox{.}(2012)]%
        {stojilovic2012selective}
\bibfield{author}{\bibinfo{person}{Mirjana Stojilovi{\'c}},
  \bibinfo{person}{David Novo}, \bibinfo{person}{Lazar Saranovac},
  \bibinfo{person}{Philip Brisk}, {and} \bibinfo{person}{Paolo Ienne}.}
  \bibinfo{year}{2012}\natexlab{}.
\newblock \showarticletitle{Selective Flexibility: Breaking the Rigidity of
  Datapath Merging}. In \bibinfo{booktitle}{\emph{Proceedings of the Design,
  Automation and Test in Europe Conference and Exhibition}}
  \emph{(\bibinfo{series}{DATE})}. \bibinfo{pages}{1543--1548}.
\newblock
\href{https://doi.org/10.1109/DATE.2012.6176718}{doi:\nolinkurl{10.1109/DATE.2012.6176718}}


\bibitem[Sun et~al\mbox{.}(2025)]%
        {sun2025labidus}
\bibfield{author}{\bibinfo{person}{Gongjin Sun}, \bibinfo{person}{Seongyoung
  Kang}, \bibinfo{person}{Jane He}, \bibinfo{person}{Se-Min Lim}, {and}
  \bibinfo{person}{Sang-Woo Jun}.} \bibinfo{year}{2025}\natexlab{}.
\newblock \showarticletitle{{Labidus}: {RISC-V} Overlay with Streaming
  Asynchronous Custom Instructions}. In \bibinfo{booktitle}{\emph{2025 IEEE
  36th International Conference on Application-specific Systems, Architectures
  and Processors (ASAP)}}. \bibinfo{pages}{57--64}.
\newblock
\href{https://doi.org/10.1109/ASAP65064.2025.00018}{doi:\nolinkurl{10.1109/ASAP65064.2025.00018}}


\bibitem[Tan et~al\mbox{.}(2022)]%
        {tan2022asap}
\bibfield{author}{\bibinfo{person}{Cheng Tan}, \bibinfo{person}{Thierry Tambe},
  \bibinfo{person}{Jeff~(Jun) Zhang}, \bibinfo{person}{Bo Fang},
  \bibinfo{person}{Tong Geng}, \bibinfo{person}{Gu-Yeon Wei},
  \bibinfo{person}{David Brooks}, \bibinfo{person}{Antonino Tumeo},
  \bibinfo{person}{Ganesh Gopalakrishnan}, {and} \bibinfo{person}{Ang Li}.}
  \bibinfo{year}{2022}\natexlab{}.
\newblock \showarticletitle{{ASAP}: Automatic Synthesis of Area-Efficient and
  Precision-Aware {CGRAs}}. In \bibinfo{booktitle}{\emph{Proceedings of the ACM
  International Conference on Supercomputing}} \emph{(\bibinfo{series}{ICS})}.
  Article \bibinfo{articleno}{4}, \bibinfo{numpages}{13}~pages.
\newblock
\href{https://doi.org/10.1145/3524059.3532359}{doi:\nolinkurl{10.1145/3524059.3532359}}


\bibitem[Tan et~al\mbox{.}(2021)]%
        {tan2021aurora}
\bibfield{author}{\bibinfo{person}{Cheng Tan}, \bibinfo{person}{Chenhao Xie},
  \bibinfo{person}{Ang Li}, \bibinfo{person}{Kevin~J. Barker}, {and}
  \bibinfo{person}{Antonino Tumeo}.} \bibinfo{year}{2021}\natexlab{}.
\newblock \showarticletitle{{AURORA}: Automated Refinement of Coarse-Grained
  Reconfigurable Accelerators}. In \bibinfo{booktitle}{\emph{Proceedings of the
  Design, Automation and Test in Europe Conference and Exhibition}}
  \emph{(\bibinfo{series}{DATE})}. \bibinfo{pages}{1388--1393}.
\newblock
\href{https://doi.org/10.23919/DATE51398.2021.9473955}{doi:\nolinkurl{10.23919/DATE51398.2021.9473955}}


\bibitem[Umeyama(1988)]%
        {umeyama1988matching}
\bibfield{author}{\bibinfo{person}{Shinji Umeyama}.}
  \bibinfo{year}{1988}\natexlab{}.
\newblock \showarticletitle{An Eigendecomposition Approach to Weighted Graph
  Matching Problems}.
\newblock \bibinfo{journal}{\emph{IEEE Transactions on Pattern Analysis and
  Machine Intelligence}} \bibinfo{volume}{10}, \bibinfo{number}{5}
  (\bibinfo{year}{1988}), \bibinfo{pages}{695--703}.
\newblock
\href{https://doi.org/10.1109/34.6778}{doi:\nolinkurl{10.1109/34.6778}}


\bibitem[Vaishnav et~al\mbox{.}(2018)]%
        {vaishnav2018survey}
\bibfield{author}{\bibinfo{person}{Anuj Vaishnav}, \bibinfo{person}{Khoa~Dang
  Pham}, {and} \bibinfo{person}{Dirk Koch}.} \bibinfo{year}{2018}\natexlab{}.
\newblock \showarticletitle{A Survey on {FPGA} Virtualization}. In
  \bibinfo{booktitle}{\emph{2018 28th International Conference on Field
  Programmable Logic and Applications (FPL)}}. \bibinfo{publisher}{IEEE},
  \bibinfo{pages}{131--138}.
\newblock
\href{https://doi.org/10.1109/FPL.2018.00031}{doi:\nolinkurl{10.1109/FPL.2018.00031}}


\bibitem[Vieira et~al\mbox{.}(2021)]%
        {vieira2021reshape}
\bibfield{author}{\bibinfo{person}{Maria Vieira}, \bibinfo{person}{Michael
  Canesche}, \bibinfo{person}{Lucas Bragan{\c c}a}, \bibinfo{person}{Josu{\'e}
  Campos}, \bibinfo{person}{Mateus Silva}, \bibinfo{person}{Ricardo Ferreira},
  {and} \bibinfo{person}{Jos{\'e}~A. Nacif}.} \bibinfo{year}{2021}\natexlab{}.
\newblock \showarticletitle{{RESHAPE}: A Run-Time Dataflow Hardware-Based
  Mapping for {CGRA} Overlays}. In \bibinfo{booktitle}{\emph{2021 IEEE
  International Symposium on Circuits and Systems (ISCAS)}}.
  \bibinfo{pages}{1--5}.
\newblock
\href{https://doi.org/10.1109/ISCAS51556.2021.9401168}{doi:\nolinkurl{10.1109/ISCAS51556.2021.9401168}}


\bibitem[Wang et~al\mbox{.}(2025)]%
        {wang2025rsn}
\bibfield{author}{\bibinfo{person}{Chengyue Wang}, \bibinfo{person}{Xiaofan
  Zhang}, \bibinfo{person}{Jason Cong}, {and} \bibinfo{person}{James~C. Hoe}.}
  \bibinfo{year}{2025}\natexlab{}.
\newblock \showarticletitle{Reconfigurable Stream Network Architecture}. In
  \bibinfo{booktitle}{\emph{Proceedings of the 52nd Annual International
  Symposium on Computer Architecture}}. \bibinfo{pages}{1848--1866}.
\newblock
\href{https://doi.org/10.1145/3695053.3731088}{doi:\nolinkurl{10.1145/3695053.3731088}}


\bibitem[Wang et~al\mbox{.}(2021)]%
        {wang2021autosa}
\bibfield{author}{\bibinfo{person}{Jie Wang}, \bibinfo{person}{Licheng Guo},
  {and} \bibinfo{person}{Jason Cong}.} \bibinfo{year}{2021}\natexlab{}.
\newblock \showarticletitle{{AutoSA}: A Polyhedral Compiler for
  High-Performance Systolic Arrays on {FPGA}}. In \bibinfo{booktitle}{\emph{The
  2021 ACM/SIGDA International Symposium on Field-Programmable Gate Arrays}}.
  \bibinfo{pages}{93--104}.
\newblock
\href{https://doi.org/10.1145/3431920.3439292}{doi:\nolinkurl{10.1145/3431920.3439292}}


\bibitem[Weng et~al\mbox{.}(2020)]%
        {weng2020dsagen}
\bibfield{author}{\bibinfo{person}{Jian Weng}, \bibinfo{person}{Sihao Liu},
  \bibinfo{person}{Vidushi Dadu}, \bibinfo{person}{Zhengrong Wang},
  \bibinfo{person}{Preyas Shah}, {and} \bibinfo{person}{Tony Nowatzki}.}
  \bibinfo{year}{2020}\natexlab{}.
\newblock \showarticletitle{{DSAGEN}: Synthesizing Programmable Spatial
  Accelerators}. In \bibinfo{booktitle}{\emph{2020 ACM/IEEE 47th Annual
  International Symposium on Computer Architecture (ISCA)}}.
  \bibinfo{pages}{268--281}.
\newblock
\href{https://doi.org/10.1109/ISCA45697.2020.00032}{doi:\nolinkurl{10.1109/ISCA45697.2020.00032}}


\bibitem[Willsey et~al\mbox{.}(2019)]%
        {willsey2019radish}
\bibfield{author}{\bibinfo{person}{Max Willsey}, \bibinfo{person}{Vincent~T.
  Lee}, \bibinfo{person}{Alvin Cheung}, \bibinfo{person}{Rastislav Bod{\'i}k},
  {and} \bibinfo{person}{Luis Ceze}.} \bibinfo{year}{2019}\natexlab{}.
\newblock \showarticletitle{Iterative Search for Reconfigurable Accelerator
  Blocks With a Compiler in the Loop}.
\newblock \bibinfo{journal}{\emph{IEEE Transactions on Computer-Aided Design of
  Integrated Circuits and Systems}} \bibinfo{volume}{38}, \bibinfo{number}{3}
  (\bibinfo{year}{2019}), \bibinfo{pages}{407--418}.
\newblock
\href{https://doi.org/10.1109/TCAD.2018.2878194}{doi:\nolinkurl{10.1109/TCAD.2018.2878194}}


\bibitem[Wilson and Stitt(2017)]%
        {wilson2017fsmoverlay}
\bibfield{author}{\bibinfo{person}{David Wilson} {and} \bibinfo{person}{Greg
  Stitt}.} \bibinfo{year}{2017}\natexlab{}.
\newblock \bibinfo{title}{A Scalable, Low-Overhead Finite-State Machine Overlay
  for Rapid {FPGA} Application Development}.
\newblock
\showeprint[arxiv]{1705.02732}~[cs.AR]
\href{https://doi.org/10.48550/arXiv.1705.02732}{doi:\nolinkurl{10.48550/arXiv.1705.02732}}
\newblock
\shownote{Presented at the 3rd International Workshop on Overlay Architectures
  for FPGAs (OLAF 2017)}.


\bibitem[Wilson et~al\mbox{.}(2018)]%
        {wilson2018recurrent}
\bibfield{author}{\bibinfo{person}{David Wilson}, \bibinfo{person}{Greg Stitt},
  {and} \bibinfo{person}{James Coole}.} \bibinfo{year}{2018}\natexlab{}.
\newblock \showarticletitle{A Recurrently Generated Overlay Architecture for
  Rapid {FPGA} Application Development}. In
  \bibinfo{booktitle}{\emph{Proceedings of the 9th International Symposium on
  Highly-Efficient Accelerators and Reconfigurable Technologies}}.
  \bibinfo{pages}{4:1--4:6}.
\newblock
\href{https://doi.org/10.1145/3241793.3241797}{doi:\nolinkurl{10.1145/3241793.3241797}}


\bibitem[Woodruff et~al\mbox{.}(2023)]%
        {woodruff2023flexc}
\bibfield{author}{\bibinfo{person}{Jackson Woodruff}, \bibinfo{person}{Thomas
  K{\oe}hler}, \bibinfo{person}{Alexander Brauckmann}, \bibinfo{person}{Chris
  Cummins}, \bibinfo{person}{Sam Ainsworth}, {and} \bibinfo{person}{Michael
  F.~P. O'Boyle}.} \bibinfo{year}{2023}\natexlab{}.
\newblock \bibinfo{title}{Rewriting History: Repurposing Domain-Specific
  {CGRAs}}.
\newblock
\showeprint[arxiv]{2309.09112}~[cs.AR]
\urldef\tempurl%
\url{https://arxiv.org/abs/2309.09112}
\showURL{%
\tempurl}


\bibitem[Wu et~al\mbox{.}(2025)]%
        {wu2025flightopu}
\bibfield{author}{\bibinfo{person}{Chen Wu}, \bibinfo{person}{Shaoqiang Lu},
  \bibinfo{person}{Yangbo Wei}, \bibinfo{person}{Junhong Qian},
  \bibinfo{person}{Jinlong Yan}, \bibinfo{person}{Zhanfei Chen},
  \bibinfo{person}{Rumin Zhang}, \bibinfo{person}{Xiao Shi}, {and}
  \bibinfo{person}{Lei He}.} \bibinfo{year}{2025}\natexlab{}.
\newblock \showarticletitle{{FlightOPU}: An {FPGA} Overlay Processor for {LLM}
  with {HBM}-Aware Multi-Die Architecture}. In \bibinfo{booktitle}{\emph{2025
  International Conference on Field Programmable Technology (ICFPT)}}.
  \bibinfo{pages}{248--251}.
\newblock
\href{https://doi.org/10.1109/ICFPT67023.2025.00048}{doi:\nolinkurl{10.1109/ICFPT67023.2025.00048}}


\bibitem[Yang et~al\mbox{.}(2025)]%
        {yang2025ola}
\bibfield{author}{\bibinfo{person}{Yang Yang}, \bibinfo{person}{Rajgopal
  Kannan}, {and} \bibinfo{person}{Viktor~K. Prasanna}.}
  \bibinfo{year}{2025}\natexlab{}.
\newblock \showarticletitle{{OLA}: An {FPGA}-based Overlay Accelerator for
  Privacy Preserving Machine Learning with Homomorphic Encryption}. In
  \bibinfo{booktitle}{\emph{Proceedings of the 2025 ACM/SIGDA International
  Symposium on Field Programmable Gate Arrays}}. \bibinfo{pages}{127--138}.
\newblock
\href{https://doi.org/10.1145/3706628.3708868}{doi:\nolinkurl{10.1145/3706628.3708868}}


\bibitem[Ye et~al\mbox{.}(2022)]%
        {ye2022scalehls}
\bibfield{author}{\bibinfo{person}{Hanchen Ye}, \bibinfo{person}{Cong Hao},
  \bibinfo{person}{Jianyi Cheng}, \bibinfo{person}{Hyunmin Jeong},
  \bibinfo{person}{Jack Huang}, \bibinfo{person}{Stephen Neuendorffer}, {and}
  \bibinfo{person}{Deming Chen}.} \bibinfo{year}{2022}\natexlab{}.
\newblock \showarticletitle{{ScaleHLS}: A New Scalable High-Level Synthesis
  Framework on Multi-Level Intermediate Representation}. In
  \bibinfo{booktitle}{\emph{Proceedings of the IEEE International Symposium on
  High-Performance Computer Architecture}} \emph{(\bibinfo{series}{HPCA})}.
  \bibinfo{pages}{741--755}.
\newblock
\href{https://doi.org/10.1109/HPCA53966.2022.00060}{doi:\nolinkurl{10.1109/HPCA53966.2022.00060}}


\bibitem[Yu et~al\mbox{.}(2020)]%
        {yu2020opu}
\bibfield{author}{\bibinfo{person}{Yunxuan Yu}, \bibinfo{person}{Chen Wu},
  \bibinfo{person}{Tiandong Zhao}, \bibinfo{person}{Kun Wang}, {and}
  \bibinfo{person}{Lei He}.} \bibinfo{year}{2020}\natexlab{}.
\newblock \showarticletitle{{OPU}: An {FPGA}-Based Overlay Processor for
  Convolutional Neural Networks}.
\newblock \bibinfo{journal}{\emph{IEEE Transactions on Very Large Scale
  Integration (VLSI) Systems}} \bibinfo{volume}{28}, \bibinfo{number}{1}
  (\bibinfo{year}{2020}), \bibinfo{pages}{35--47}.
\newblock
\href{https://doi.org/10.1109/TVLSI.2019.2939726}{doi:\nolinkurl{10.1109/TVLSI.2019.2939726}}


\end{thebibliography}
